# Twisted DNA origami-based chiral monolayers for spin filtering

Haozhi Wang[1†], Fangfei Yin[2†], Linyun Li[1], Mingqiang Li[1], Zheng Fang[2], Chenyun Sun[1], Bochen Li[1], Jiye Shi[3], Jiang Li[4,5], Lihua Wang[3,4], Shiping Song[3,4], Xiaolei Zuo[2*], Xiaoguo Liu[1*], Chunhai Fan[1]

1. School of Chemistry and Chemical Engineering, New Cornerstone Science Laboratory, Frontiers Science Center for Transformative Molecules, Zhangjiang Institute for Advanced Study and National Center for Translational Medicine, Shanghai Jiao Tong University, Shanghai 200240, China.

2. Institute of Molecular Medicine, Shanghai Key Laboratory for Nucleic Acid Chemistry and Nanomedicine, Renji Hospital, School of Medicine, Shanghai Jiao Tong University, Shanghai, China.

3. CAS Key Laboratory of Interfacial Physics and Technology, Shanghai Institute of Applied Physics, Chinese Academy of Sciences, Shanghai 201800, China.

4. The Interdisciplinary Research Center, Shanghai Synchrotron Radiation Facility, Shanghai Advanced Research Institute, Chinese Academy of Sciences, Shanghai 201210, China.

5. Institute of Materiobiology, Department of Chemistry, College of Science, Shanghai University, Shanghai 200444, China.

**ABSTRACT**

DNA monolayers with inherent chirality play a pivotal role across various domains, including biosensors, DNA chips, and bioelectronics. Nonetheless, conventional DNA chiral monolayers, typically constructed from single-stranded DNA (ssDNA) or double-stranded DNA (dsDNA), often lack structural orderliness and design flexibility at the interface. Structural DNA nanotechnology emerges as a promising solution to tackle these challenges. In this study, we present a strategy for crafting highly adaptable twisted DNA origami-based chiral monolayers. These structures exhibit distinct interfacial assembly characteristics and effectively mitigate the structural disorder of dsDNA monolayers, which is constrained by a limited persistence length of ~50 nm of dsDNA. We highlight the spin-filtering capabilities of four representative DNA origami-based chiral monolayers, demonstrating a maximal one-order-of-magnitude increase in spin-filtering efficiency per unit area compared to conventional dsDNA chiral monolayers. Intriguingly, our findings reveal that the higher-order, tertiary, chiral structure of twisted DNA origami further enhances the spin-filtering efficiency. This work paves the way for the rational design of DNA chiral monolayers.

## Introduction

DNA monolayers represent a category of membrane materials widely applied in many fields, including electronic[1-3], Raman[4, 5], acoustic[6] and biosensor[7-9]. The performance of a DNA chiral monolayer can be improved through the selection of a reasonable design strategy and immobilization method[10]. Taking electrochemical detection as an example, the conformational stability and modification density of DNA chiral monolayer directly affect the sensitivity and speed of electro-chemical identification. Ever since Tarlov and colleagues initially obtained orientation-controlled and relatively dense DNA monolayer using thiolated DNA on the surface of gold, this strategy has become one of the most popular immobilization methods[11]. Subsequently, researchers have employed various design and assembly strategies, such as single-stranded DNA[12, 13] (ssDNA), double-stranded DNA[14] (dsDNA) and tetrahedral DNA nanostructure[15, 16] (TDN), leading to the gradual improvement of the structure and function of DNA monolayers and the expansion of their application range. Among these developments, chirality stands out as a particularly important feature applicable in the field of spin-filtering[17, 18].

When electrons transmit through a chiral monolayer, they induce a magnetic moment and are influenced by an applied magnetic field, selectively permitting electrons with a specific spin direction to pass through[19]. Materials exhibiting such properties are referred to as spin transport materials, and the dsDNA chiral monolayer services as a representative example. In the initial experiment illustrating the spin selectivity of dsDNA chiral monolayer, it was employed to induce spin polarization in photoelectrons emitted from gold surfaces under vacuum conditions[19]. This revealed a linear correlation between the chirality-induced spin selectivity (CISS) efficiency and the length of dsDNA. Subsequent electro-chemical experiments using probes such as methylene blue (MB) and Nile blue (NB) indicated that the secondary structure of dsDNA, rather than its primary structure, determines the spin direction of spin selectivity[20]. To elucidate the CISS phenomenon mediated by DNA chiral monolayers, several ideal DNA models were proposed[21, 22]. These theoretical models predicted how the degree of coupling between the direction of the applied magnetic field and the electron transport pathway guided by dsDNA conformation influences the CISS effect, underscoring the significance of aligning the magnetic field line and dsDNA helix axis. However, traditional methods for constructing dsDNA chiral monolayers introduced substantial deviations from ideal models due to DNA-Au surface interactions, interactions among densely packed DNA strands, and electrostatic and spatial effects within the monolayer[23]. Additionally, the limited persistence length of dsDNA[24-26] (~50 nm) and the absence of direct structural evidence for dsDNA chiral monolayers constrained the available design space and hampered the precise quantification of CISS efficiency.

The twisted DNA origami (TDO) folding strategy, first reported by Dietz and colleagues[27], has been widely applied in nanodevices[28, 29], chiral materials[30], and bioengineering[31, 32]. TDO is designed to strictly obey the Watson-Crick base pairing rules, preserving the primary and secondary structure of dsDNA. Therefore, the local chirality of the sugar-phosphate backbone, the chiral π-stacking of base aromatic heterocycles, and the right-handed helical nature of B-DNA are maintained. In this way, TDO represents a unique multilevel chiral model system for constructing novel DNA chiral monolayer. Moreover, the persistence length of dsDNA bundles within multi-helical DNA nanostructures is orders of magnitude greater than that of individual dsDNA, providing a more extensive design space for the construction of DNA chiral monolayer.

In this study, we reported the construction of four distinct DNA chiral monolayers using TDO nanostructures at the Au electrode interface. We provided comprehensive analyses of their structures and interfacial assembly parameters, and observed exceptional performances of TDO-based chiral monolayers as highly efficient spin filters. By employing atomic fore microscopy (AFM) and scanning electron microscopy (SEM) morphological characterization, we elucidated the conformation and quantity of dsDNA responsible for mediating the CISS effect within TDO-based chiral monolayers. To quantify the charge transport[33] (CT) yield under various experimental conditions, we introduced MB reduction probes into the electrolyte environment and monitored the total number of reduced probe molecules facilitated by the TDO-based chiral monolayer. Our results demonstrate a significant enhancement in the unit area CISS effect compared to traditional dsDNA chiral monolayers when employing TDO-based chiral monolayers. Interestingly, we found that the tertiary chiral structure of TDO further enhances the efficiency of CISS.

## Results and Discussion

### Designing principle of TDO-based chiral monolayers

We first designed three TDOs with achiral (V-TDO), right-handed (R-TDO), and left-handed (L-TDO) tertiary chiral structures, respectively. Each TDO consists of 60 dsDNA strands immobilized by crossovers[34] under a honeycomb lattice arrangement[35, 36], with a combination of 6 × 10 dsDNAs, where each dsDNA strand is comprised of 120 base pairs. A deletion or insertion of a base pair would induce left- or right-handed torque and a push or pull force on the adjacent base pairs, resulting in local stress deformation. This deformation can be relieved by compensatory global left- or right-handed twisting and bending[37].

TDOs were twisted by periodically inserting or deleting a total of 6 base pairs every 21 base pairs (three turns) to introduce right-handed or left-handed twists, respectively (Figure 1). Highly monodisperse V-TDO, R-TDO, and L-TDOs were folded by programmed annealing and purified by PEG purification (Figure S1), with yields of 69.8%, 65.2%, and 49.6%, respectively. The three TDOs have twist angles of 25°, 50°, and -35°, respectively, which is defined as the angle between the symmetry axis of the top and bottom planes. The twist angle of V-TDO arises from the accumulation of stresses caused by the difference in the number of base pairs per turn between natural DNA (~10.66 bp/turn) and designed DNA[38] (10.5 bp/turn). This accumulation of stresses also results in a lesser deviation of R-TDO and L-TDO from their original designs as mutual chiral enantiomers. To isolate the effect of different chiral structures of TDO-based chiral monolayers on CISS effect, we used scaffold strands of different lengths to maintain an identical height for V-TDO (7560 nt), R-TDO (8064 nt), and L-TDO (7249 nt). Excessive domains of scaffold strands were processed into loop structures with the shortest possible length and were positioned away from the terminus of the TDO (Figure S7-10). In addition, redundant ssDNA loops in these TDOs had little influence on the CISS efficiency of TDO monolayers cause they are unable to maintain structural stability during the electrochemical tests, which is a prerequisite to produce CISS effect using DNA[18].

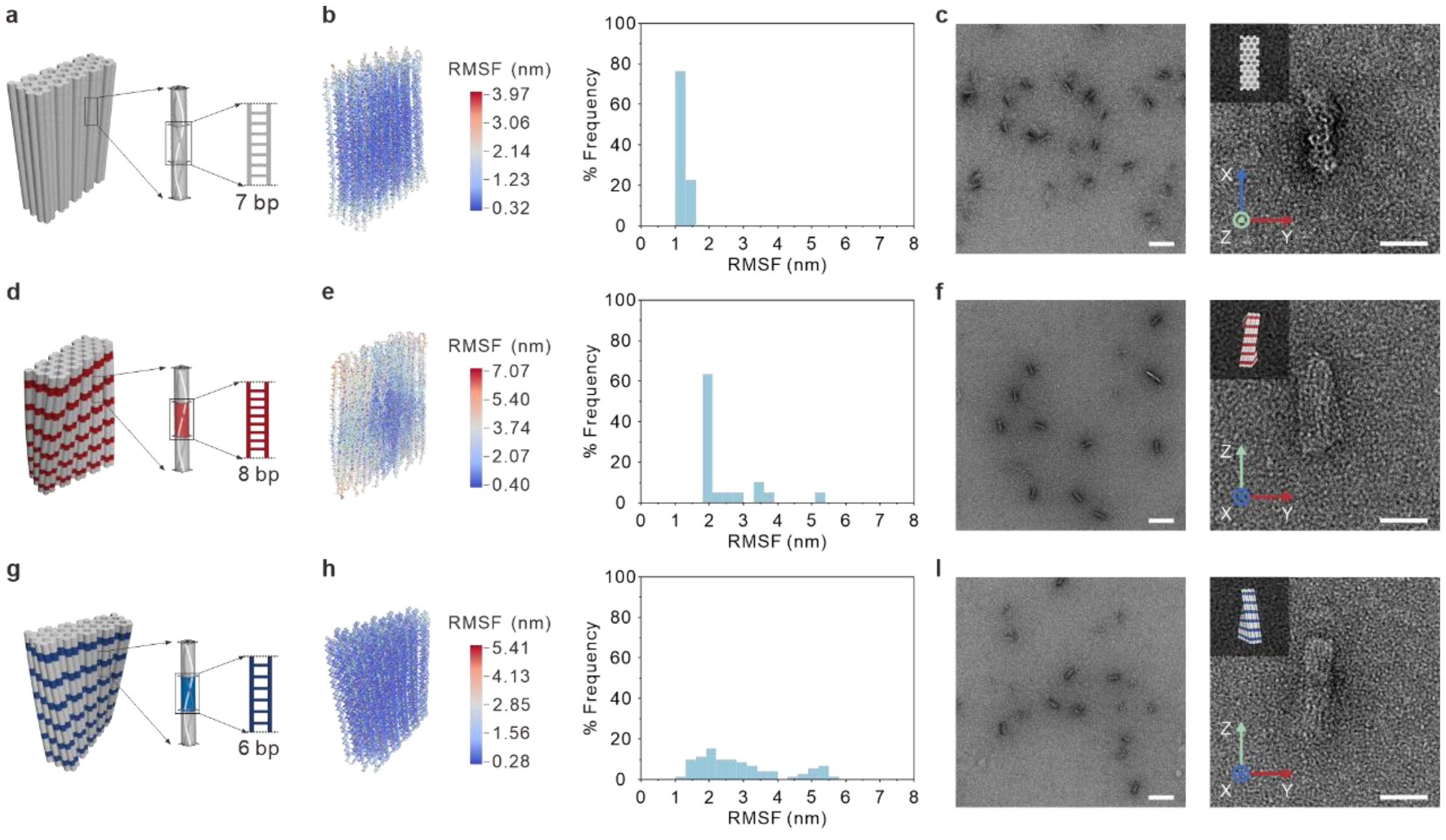

**Figure 1. Design principle, Molecular Dynamics (MD) simulations and structural characterizations of TDOs.** (a, d, g) Illustration of the base pair distributions of inserting (red) or deleting (blue) base pairs in V-TDO, R-TDO and L-TDO, respectively. (b, e, h) The last configuration and corresponding RMSF obtained through OxDNA MD simulation of V-TDO, R-TDO and L-TDO, respectively. (c, f, l) Left: Negative-stain transmission electron microscopy (TEM) images of V-TDO, R-TDO and L-TDO, respectively. Right: zoom-in views of the TDO monomers and corresponding model. Scale bars are 100 nm (left) and 20 nm (right), respectively.

The analysis of coarse-grained models based on OxDNA Molecular Dynamics (MD) simulations[39-41] reveals that dsDNAs within TDOs achieve stable conformations and spatial orientations when compared to ssDNA or dsDNA of equivalent length. Firstly, we established a coarse-grained model for ssDNA and dsDNA, each with a length of 120 base pairs. After relaxation, their conformations of the parts beyond the respective persistence lengths (ssDNA: ~2 base pairs, ~0.75 nm; dsDNA: ~150 base pairs, ~50 nm) exhibited significant deviations, resulting in severe bending and fluctuation (Figure S2). The Root Mean Square Fluctuation (RMSF) exhibited a 90% distribution ranging from 3.45 nm to 6.75 nm for ssDNA and from 1.45 nm to 5.05 nm for dsDNA, respectively. These intrinsic limitations of ssDNA and dsDNA properties hinder the precise control of chiral monolayers constructed with ssDNA and dsDNA. In contrast, owing to the stringent crossover constraints applied to dsDNA within TDOs, the conformations of dsDNA in TDO-based chiral monolayers exhibit greater stability. The 90% distribution of RMSF for V-TDO, R-TDO, and L-TDO spans from 1.15 nm to 1.75 nm, 1.95 nm to 3.75 nm, and 1.45 nm to 5.05 nm, respectively. The persistence length of dsDNA in TDO theoretically depends on the TDO's height. Additionally, there is an angle between the spatial orientation of dsDNA and the interface in traditional dsDNA monolayer, which is affected by the modification density[42]. By contrast, the spatial orientation of dsDNA in the TDO chiral monolayer is consistent with that of TDO on the gold interface. Consequently, TDO forms a highly organized dsDNA cluster comprising 60 dsDNAs, significantly simplifying the structural characterization of chiral monolayers.

To construct the TDO-based chiral monolayers on gold surface, we extended 30 staple strands (15 nt) at the same sites at the bottom of different TDOs, which were oriented from 3' ends to 5' ends. After hybridization with thiolated DNA linker strands at the 3' end, the thiolated TDO assemblies were purified by PEG purification and then assembled onto the gold/nickel electrode.

**Structural properties of TDO-based chiral monolayers**

Next, we confirmed that various TDOs were regularly assembled in a manner approximately perpendicular to the gold/nickel electrode interface. SEM images showed a significant difference between the blank electrode surface and the TDO-based chiral monolayers modified electrode surface, confirming the successful assembly monolayers on electrode (Figure 2a). The surface of the blank electrode showed bright white due to the excellent electrical conductivity of gold, while the areas covered by TDO-based chiral monolayers showed uniformly grey black, indicating poorer conductivity and weaker signals (Figure S5). However, direct observation of the TDO on the electrode surface with SEM was challenging due to the poor electron beam tolerance of DNA. Therefore, we used in-situ silicification method[43] to visualize the details of TDO-based chiral monolayers. The SEM images showed abundant grey rectangles with dimensions of ~37.25 nm in length and ~18.1 nm in width, close to theoretical length-width ratio of corresponding TDO (Figure 2b, right). The slight increase in size was attributed to the thickness of the grown $SiO_2$ layer (Figure S4). In addition, to exclude the influence of silicification on the morphology of the electrode surface, we performed TDOs assembly on specific region of the electrode to form monolayer and then subjected the entire electrode to the silicification. The boundary of TDO based-chiral monolayer revealed significant differences between the TDOs assembled region and the bare electrode region (Figure 2b, middle), where the bare electrode region remained unchanged after silicification. The ratio of monodisperse TDOs with the z-axis perpendicular to and parallel to the electrode surface was determined to be 2.1 ± 0.2, revealing that TDOs in the monolayers were predominantly assembled perpendicular to the electrode surface.

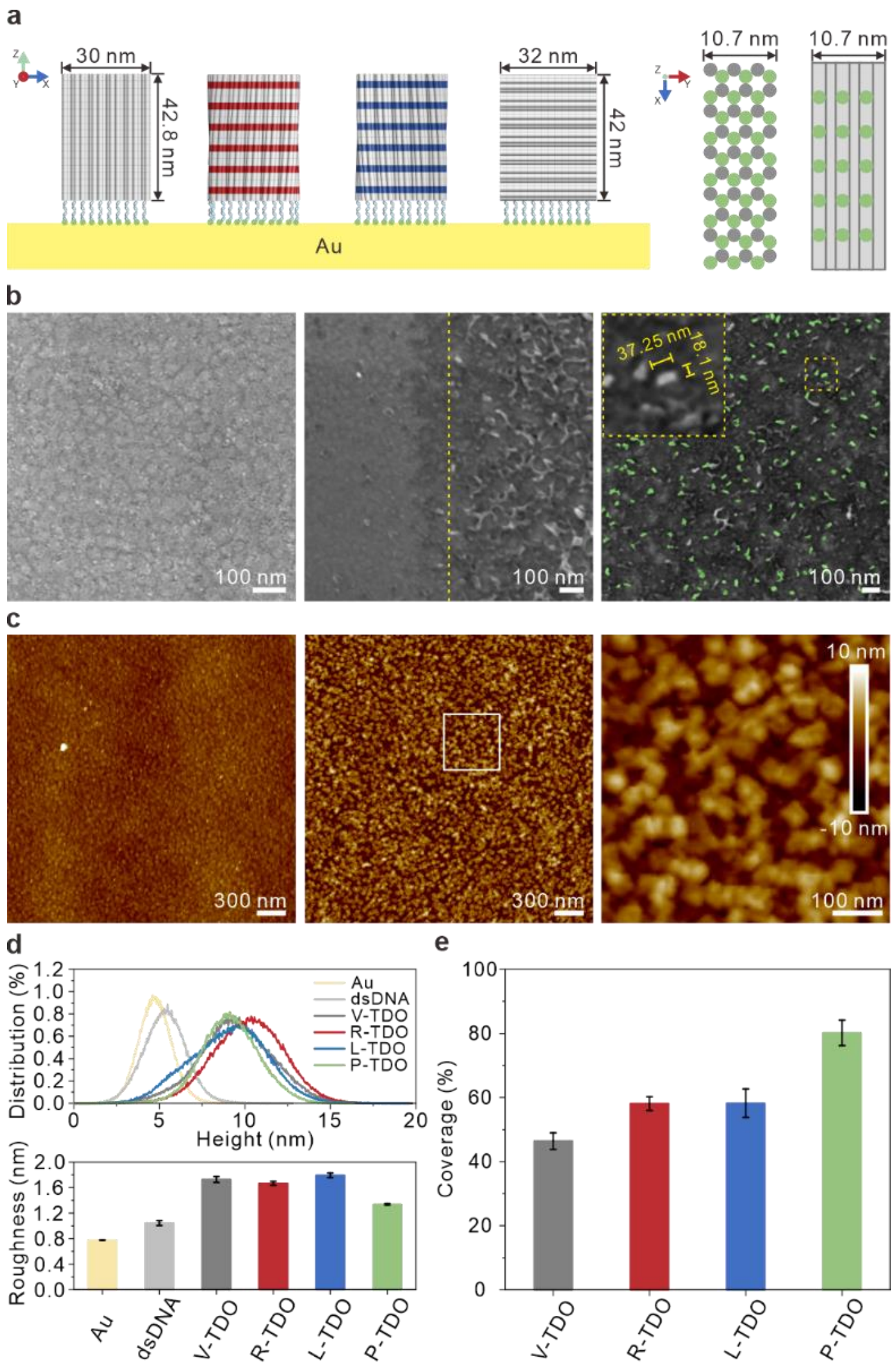

**Figure 2. Schematic and experimental data of the TDO-based chiral monolayers on gold surface.** (a) Left: illustration of thiolated TDOs assembled onto gold surfaces. Right: the sites of extended staple strands for V-TDO, R-TDO, L-TDO (left), and P-TDO (right), respectively. (b) SEM images of blank electrode surface (left), the boundary of silicified R-TDO chiral monolayer-modified electrode surface (middle), and silicified R-TDO chiral monolayer-modified electrode surface with a zoom-in view of a R-TDO coated with $SiO_2$ (right). Scale bars, 100 nm. (c) AFM images of blank electrode surfaces (left), R-TDO chiral monolayer-modified electrode surfaces (middle), and corresponding zoom-in view (right). Scale bars are 300 nm (left and middle) and 100 nm (right). (d) Height distribution (top) and surface roughness (bottom) of blank electrode surfaces, 15 bp dsDNA chiral monolayer-modified electrode surface, and TDO chiral monolayer-modified electrode surfaces, respectively. (e) The coverage of TDO-based chiral monolayers on corresponding modified electrode surfaces.

AFM images further illustrated that the averaged modification density of TDOs on the electrode surface is 93,188 per $cm^2$, and the averaged roughness of the TDO-based chiral monolayers is 1.7 nm. In the case of R-TDO, AFM images showed that R-TDOs were mostly assembled in the z-axis perpendicular to the electrode surface (Figure 2c, middle), with an assembly density of approximately 99,894 R-TDOs per $cm^2$. The zoom-in image of the R-TDO monolayer showed clear observation of rectangles with dimensions of 39.40 nm × 19.70 nm, which were consistent with the theoretical sizes of R-TDO. Some rectangular shapes with dimensions of approximately 43.1 nm × 35.0 nm corresponded to the projection in the y-axis direction of R-TDO, which may result from structural collapse and topple caused by drying during the sample preparation process. All electrodes modified with different TDO-based chiral monolayers exhibited a surface roughness of ~1.0 nm higher than that of the blank electrode, along with an increased range of height distribution, providing statistical evidence for the successful assembly of TDO monolayer on the electrode (Figure 2d). The coverage of V-TDO, R-TDO, and L-TDO chiral monolayer on the electrode surface, after subtracting the blank background (Figure 2e and Figure S3), was approximately 46.4% ± 2.6%, 58.1% ± 2.2%, and 58.2% ± 4.4%, respectively.

**Spin Selectivity studies on TDO-based chiral monolayers**

Figure 3b illustrates the electro-chemical platform to study the CISS effect of TDO-based chiral monolayers. The previously reported DNA-binding reductive probe MB was added to the electrolyte solution[44-46]. MB binds reversibly to dsDNA and undergoes a DNA-mediated proton coupling reaction at a potential of -220 mV versus the AgCl/Ag electrode, followed by a two-electron reduction reaction to produce leucomethylene blue (LB) (Figure 3a). By magnetizing the nickel working electrode using a permanent neodymium magnet (0.66 T), a spin-polarized current is generated at a poised negative potential, and the polarization sign can be adjusted by changing the magnetic field direction[47, 48]. By integrating the Faraday response of the reduced probe molecules using cyclic voltammetry, we can determine the total number of reduced probes, which quantifies the MB reduction yields through different TDO-based chiral monolayers.

We observed that the TDO chiral monolayer exhibited stable spin selectivity and, notably, displayed significantly higher CISS efficiency when compared to the dsDNA monolayer, after establishing the fabrication and characterization method for TDO-based chiral monolayers. A 15bp dsDNA (hybridized with 15bp DNA linker strand and its complementary strand) monolayer and three TDO-based chiral monolayers were used as positive controls. Their CISS effect were studied by cyclic voltammetry in identical electrolyte environment. As shown in Figure 3c-3g, the reduction yields of MB in the dsDNA, V-TDO, R-TDO, and L-TDO monolayers exhibited regular variation with the changing of the magnetic field direction (up and down) beneath the electrode, and remained stable in more than 50 cyclic voltametric measurements. The result for 15 bp dsDNA chiral monolayer was consistent with previously report[20], the ratio of up reduction peak to down reduction peak of MB was 1.98% ± 0.95%. The normalized frequency statistics indicated a higher MB reduction yield for the up direction of the magnetic field (Figure S6). The CISS efficiency of the dsDNA monolayer[49, 50], quantified as monolayer spin polarization (MSP), was determined to be +1.98% ± 0.95%, with the spin-selective direction defined as "up" and denoted by "+". The three chiral monolayers composed of V-TDO, R-TDO and L-TDO showed directional stable spin selectivity, with MSP of +2.23% ± 1.08%, +4.20% ± 1.29%, and +2.31% ± 1.34%, respectively. These findings demonstrated the potential of TDO-based chiral monolayers as stable spin transport materials.

The negative control experiment involving the P-TDO chiral monolayer demonstrated that the variations in CISS efficiency among different TDO-based chiral monolayers were attributed to differences in their tertiary chiral structures. This observation ruled out the influence of dsDNA linker modification quantity and the effects of toppled TDOs within the chiral monolayer on CISS efficiency. The height and contact area of the P-TDO were nearly identical to those of other TDOs, and the polarization current was transmitted from the electrode surface through 15 thiolated dsDNA linker of the same sequence (Figure 2a). To restore the CISS effect of the toppled TDO within the TDO-based chiral monolayer, the orientation of dsDNA in P-TDO was designed to be perpendicular to the magnetic field direction, corresponding to parallel alignment with the electrode surface. The results in Figure 4a illustrate that the reduction yield of MB in the solution above the P-TDO chiral monolayer modified electrode did not exhibit a regular change with alterations in the magnetic field direction. The integral reduction peak ratio of MB was 1.16% ± 2.13%. This error range indicates that P-TDO chiral monolayer does not have stable spin selectivity. Hence, when the TDO chiral monolayer exhibits the CISS effect, the CT mediated by dsDNA must interact with its chiral potential field, giving rise to the spin-orbit coupling (SOC) effect.

To standardize the quantitative criteria for the CISS efficiency of both traditional dsDNA chiral monolayers and TDO-based chiral monolayers, we established a method to calculate the CISS efficiency of TDO-based chiral monolayers. The information on TDO modification density obtained through interface morphology characterization served as the data basis for calculating the total coverage area of dsDNA in different TDO-based chiral monolayers. We counted the total number of dsDNA within the TDO-based chiral monolayer that exhibited the CISS effect and normalized the monolayer spin polarization (MSP) of these DNA chiral monolayers to the CISS efficiency produced by ideal dsDNA under unit area conditions, termed normalized spin polarization (NSP). The calculation formula is as follows:

$$NSP = \frac{MSP}{S \times \rho \times \rho_Z \times \rho_\perp}$$

where $MSP$ represents the CISS efficiency of the DNA chiral monolayer. $S$ represents the working electrode area. $\rho$ represents the percentage of the chiral monolayer modification area relative to the working electrode area. $\rho_Z$ represents the effective area constant, which is 60.43% for the honeycomb lattice arrangement of dsDNA in TDO. $\rho_\perp$ represents the percentage of dsDNA perpendicular to the interface. For dsDNA chiral monolayer, $\rho_Z$ and $\rho_\perp$ are both 100%, as indicated in prior research. Although the dsDNA chiral monolayer was regarded as a dense and uniform ideal structure, various analytical techniques, such as electrochemistry[20], Mott polarimeter[19, 51], and C-AFM measurements[52], have demonstrated consistent NSP results for dsDNA chiral monolayers assembled using the same method. Therefore, in the case of the 15 bp dsDNA chiral monolayer in this study, we have only adjusted the value of $\rho$ to 75%, in accordance with previous findings[11, 20].

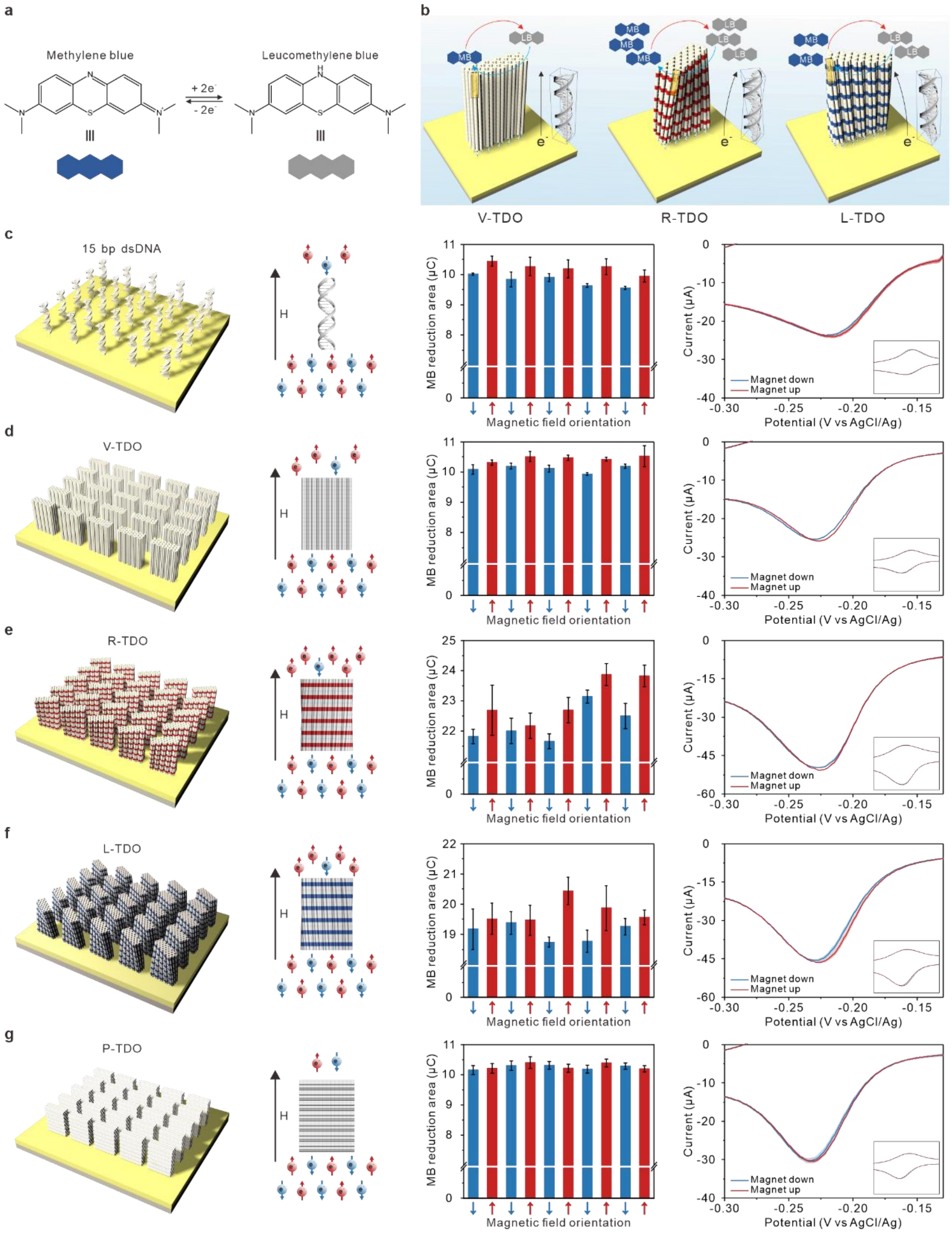

**Figure 3. Representative cyclic voltammetry data for 15 bp dsDNA and TDO-based chiral monolayers.** (a, b) Illustration of DNA-mediated MB redox reactions on the electrochemical platform. (c-g) Schematic models of 15 bp dsDNA, V-TDO, R-TDO, L-TDO and P-TDO chiral monolayers on electrodes, respectively (first column); The second column presents the corresponding spin filtering model when the magnetic field direction is up. The red electrons indicate spin direction up, while the blue electrons indicate spin direction down. The number of total electrons implies the spin filtering efficiency; The third column presents the reduction yield of MB after 9 consecutive conversions in the magnetic field direction after signal stabilization, 5 cyclic voltammetry tests were performed after each conversion. The data were determined based on the first scan with the down direction of magnetic field; The fourth column presents representative cyclic voltammograms with the magnet up (red curve) and magnet down (blue curve). All insertions show the corresponding global cyclic voltammograms, which are -60 to 40 μA in vertical coordinates and -0.30 to -0.13 V in horizontal coordinates.

Based on the calculation method, we determined that the NSP of the TDO-based chiral monolayer increased by ap-proximately an order of magnitude compared to the 15 bp dsDNA chiral monolayer. Specifically, the NSP values for the 15 bp dsDNA, V-TDO, R-TDO, and L-TDO-based chiral monolayers were +7.25% ± 3.49%, +32.80% ± 15.89%, +48.84% ± 15.04%, and +26.91 ± 15.60% (Figure 4b), respectively. Notably, the NSP of the R-TDO chiral monolayer exhibited the most significant enhancement, with a 573% increase over the 15 bp dsDNA chiral monolayer. Following that, the V-TDO chiral monolayer showed a 352% increase, while the L-TDO chiral monolayer exhibited the lowest increase at 271%. Since R-TDO, V-TDO, and L-TDO have the same height, we speculate that the enhanced CISS efficiency per unit area in the R-TDO chiral monolayer arises from its right-handed tertiary chiral structure, while the left-handed tertiary chiral structure in L-TDO reduces its CISS efficiency per unit area. Additionally, the growth rate of CISS efficiency per base pair is approximately 29.41%, slightly lower than the previously reported 49.67%[20]. This reduced growth rate may be due to an increase in the number of dsDNA base pairs or increased dephasing, both of which enhance electron scattering, thereby reducing the average conductivity and leading to decreased growth rate of CISS efficiency[53]. Importantly, under this standard, CISS efficiency calculations exclude measurement errors of up to 30% between different electrodes[54], and surface morphology characterization of the electrodes after electrochemical testing can provide precise parameters for chiral monolayer NSP calculations.

**Effect of tertiary chiral structure of DNA on CISS**

Circular dichroism (CD) spectroscopy results revealed that the dsDNA in R-TDO has a more compact right-handed helical structure (Figure 4c). CD curves of all TDOs exhibited the characteristics of B-DNA, confirming the integrity of the right-handed secondary chiral structure[55-57]. Interestingly, the peak of R-TDO displayed a red-shift of ~5 nm compared to V-TDO, indicating a more compact right-handed helical structure of dsDNA. In contrast, the peak of L-TDO exhibited a blueshift of ~2 nm compared to V-TDO, suggesting a looser right-handed helix structure of dsDNA. The effect of these changes in dsDNA helicity on the CISS effect is consistent with the Hamiltonian model predicted by Ai-Min Guo et al.[53] and R. Gutierrez et al.[58], where the more compact helicity of dsDNA in the R-TDO monolayer further enhances its CISS efficiency, and vice versa.

Further insights into the helicity of dsDNA within each TDO in the testing environment were revealed through the last configuration results obtained from OxDNA MD simulations. We reconstructed coarse-grained models of TDOs in a 12 mM $Mg^{2+}$ solution environment (Figure 4d). Statistical analysis of coaxial stacking distances between adjacent base pairs on the backbone chains showed that the average coaxial stacking distances of dsDNA in V-TDO, R-TDO, and L-TDO within the simulated testing environment were consistent with those in the rigid model of V-TDO (Figure 4e). In other words, the length of dsDNA did not undergo significant changes due to tension or compression forces. Statistical analysis of the rotation per nucleotide between adjacent base pair planes in the outermost dsDNA of TDO revealed values of 34.0° ± 4.4°, 30.9° ± 10.2°, 29.3° ± 7.6°, and 32.9° ± 8.0° for the rigid V-TDO model (Figure 4f), V-TDO, R-TDO, and L-TDO, respectively. This indicates a gradual compaction of the B-DNA helical structure in L-TDO, V-TDO, and R-TDO. In summary, TDO primarily exerts torsional forces, rather than tension or compression, on its dsDNA to balance the local stresses introduced by the tertiary chiral structure. This results in changes in the helicity of the dsDNA within TDO when introducing tertiary chiral structure to the overall TDO structure, thereby influencing the CISS efficiency of the TDO monolayers.

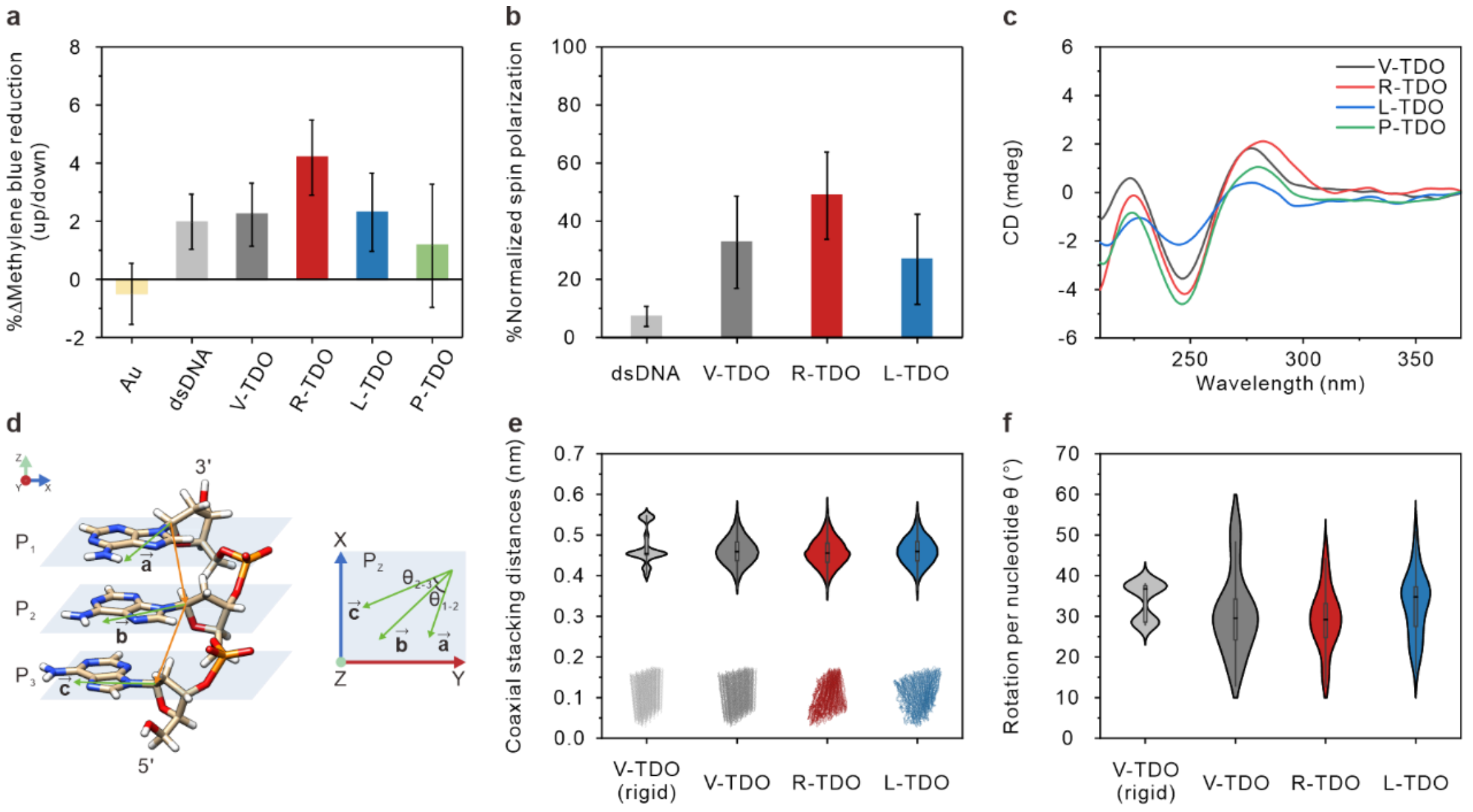

**Figure 4. CISS efficiency and structure analysis of dsDNA or TDO-based chiral monolayers.** (a) Cyclic voltammetry data for the two magnetizations were collected and summarized for Au surface, 15 bp dsDNA chiral monolayer, and TDO-based chiral monolayers, respectively. (b) Normalized spin polarization of 15 bp dsDNA and TDO-based chiral monolayers, respectively. (C) Circular dichroism spectrum of TDOs. (d) Left: the partial all-atom model of DNA. P1, P2, and P3 represent the adjacent base pair planes (pale blue); the solid yellow line with arrow represents the coaxial stacking distance between adjacent base pair planes; the solid green line with arrow represents the space vector of the base. Right: The YZ plane projection of these vectors is normalized to a point, and the angle θ between the vectors in adjacent planes represents the rotation per nucleotide. (e) Coaxial stacking distances of the rigid coarse-grained model of V-TDO and the relaxed coarse-grained models of V-TDO, R-TDO, and L-TDO, respectively. The inserted models are the last configuration of scaffold strand corresponding to TDOs. (f) Rotation per nucleotide of the rigid coarse-grained model of V-TDO and the released coarse-grained models of V-TDO, R-TDO, and L-TDO, respectively. These results were obtained through OxDNA MD simulations at 12.5 mmol/L $Mg^{2+}$ concentration.

## Conclusion

In summary, we have proposed a method for constructing DNA chiral monolayers on gold surfaces using twisted DNA origami. We determined the conformation, spatial orientation, and modification density of individual nanostructures within the TDO-based chiral monolayers, with the assistance of SEM and AFM. The TDO-based chiral monolayers have the following advantages. Firstly, owing to the high programmability of DNA origami, controlling the spatial orientation of dsDNA within the twisted DNA origami-based chiral monolayers became feasible. Secondly, the design principles of DNA origami enabled the effective length of dsDNA to surpass the intrinsic persistence length limit, resulting in the formation of structurally stable and highly order DNA clusters at the interface. Lastly, the twisted DNA origami folding strategy made it possible to finely tune the helicity of dsDNA structures within the TDO chiral monolayer with single-base precision.

Taking electron spin filtering material as an example, we studied the CISS behavior of typical TDO-based chiral monolayers. Our experiments demonstrated that the TDO-based chiral monolayers exhibit stable CISS effects, contingent upon the chiral potential field of dsDNA within the TDO interacting with the orbital motion of the spin electrons. We found that the TDO-based chiral monolayers not only surpass the persistence length limit of dsDNA to enhance their CISS efficiency but also allow further modulation of CISS efficiency through control of the helicity of dsDNA. Specifically, introducing a right-handed tertiary chiral structure compacted the helicity of dsDNA within TDO, leading to improved CISS efficiency, while the left-handed tertiary chiral structure had the opposite result. In conclusion, we have demonstrated the successful construction of TDO-based chiral monolayers and showcased their potential applications as electron spin-filtering membranes. This strategy provides valuable guidance for the development of novel electrochemical and electronic devices based on DNA origami structures.

## ASSOCIATED CONTENT

### Supporting Information

The Supporting Information is available free of charge.

The PDF file includes:

- Materials and methods
- Supplementary text
- Figure S1 to S10
- Table S1 to S4

## AUTHOR INFORMATION

Corresponding Author

**Xiaoguo Liu**-School of Chemistry and Chemical Engineering, New Cornerstone Science Laboratory, Frontiers Science Center for Transformative Molecules, Zhangjiang Institute for Advanced Study and National Center for Translational Medicine, Shanghai Jiao Tong University, Shanghai 200240, China; ORCID: https://orcid.org/0000-0002-7834-399X; Email: *liuxiaoguo@sjtu.edu.cn

**Xiaolei Zuo**-Institute of Molecular Medicine, Shanghai Key Laboratory for Nucleic Acid Chemistry and Nanomedicine, Renji Hospital, School of Medicine, Shanghai Jiao Tong University, Shanghai, China; ORCID: https://orcid.org/0000-0001-7505-2727; Email: *zuoxiaolei@sjtu.edu.cn

### Author Contributions

**H. W. and F. Y. contributed equally to this work.**

### Notes

**The authors declare no competing financial interest.**

## ACKNOWLEDGMENT

Funding Sources

**National Key R&D Program of China (2022YFF0710000).**

**National Natural Science Foundation of China (92056117, 21934007, 22122406, T2188102, 22025404 and 22204098).**

**The New Cornerstone Science Foundation.**

## REFERENCES

(1) Gooding, J. J.; Mearns, F.; Yang, W.; Liu, J. Self-assembled monolayers into the 21st century: recent advances and applications. *Electroanalysis: An International Journal Devoted to Fundamental and Practical Aspects of Electroanalysis* **2003**, *15* (2), 81-96.

(2) Chaki, N. K.; Vijayamohanan, K. Self-assembled monolayers as a tunable platform for biosensor applications. *Biosensors and Bioelectronics* **2002**, *17* (1-2), 1-12.

(3) Willner, I.; Zayats, M. Electronic aptamer-based sensors. *Angewandte Chemie International Edition* **2007**, *46* (34), 6408-6418.

(4) Freeman, R. G.; Grabar, K. C.; Allison, K. J.; Bright, R. M.; Davis, J. A.; Guthrie, A. P.; Hommer, M. B.; Jackson, M. A.; Smith, P. C.; Walter, D. G. Self-assembled metal colloid monolayers: an approach to SERS substrates. *Science* **1995**, *267* (5204), 1629-1632.

(5) Zhao, B.; Shen, J.; Chen, S.; Wang, D.; Li, F.; Mathur, S.; Song, S.; Fan, C. Gold nanostructures encoded by non-fluorescent small molecules in polyA-mediated nanogaps as universal SERS nanotags for recognizing various bioactive molecules. *Chemical Science* **2014**, *5* (11), 4460-4466.

(6) O'sullivan, C.; Guilbault, G. Commercial quartz crystal microbalances–theory and applications. *Biosensors and bioelectronics* **1999**, *14* (8-9), 663-670.

(7) Flink, S.; van Veggel, F. C.; Reinhoudt, D. N. Sensor functionalities in self-assembled monolayers. *Advanced Materials* **2000**, *12* (18), 1315-1328.

(8) Yeung, S. Y.; Mucha, A.; Deshmukh, R.; Boutrus, M.; Arnebrant, T.; Sellergren, B. r. Reversible self-assembled monolayers (rSAMs): Adaptable surfaces for enhanced multivalent interactions and ultrasensitive virus detection. *ACS Central Science* **2017**, *3* (11), 1198-1207.

(9) Malinsky, M. D.; Kelly, K. L.; Schatz, G. C.; Van Duyne, R. P. Chain length dependence and sensing capabilities of the localized surface plasmon resonance of silver nanoparticles chemically modified with alkanethiol self-assembled monolayers. *Journal of the American Chemical Society* **2001**, *123* (7), 1471-1482.

(10) De-Los-Santos-Lvarez, P.; Lobo-CastaN, M. J.; Miranda-Ordieres, A. J.; TuN-Blanco, P. Current strategies for electrochemical detection of DNA with solid electrodes. *Analytical and Bioanalytical Chemistry* **2004**, *378* (1), 104-118.

(11) Steel, A. B.; Herne, T. M.; Tarlov, M. J. Electrochemical quantitation of DNA immobilized on gold. *Analytical Chemistry* **1998**, *70* (22), 4670-4677.

(12) Li, C.; Tao, Y.; Yang, Y.; Xiang, Y.; Li, G. In vitro analysis of DNA–protein interactions in gene transcription using DNAzyme-based electrochemical assay. *Analytical chemistry* **2017**, *89* (9), 5003-5007.

(13) Li, C.; Wu, D.; Hu, X.; Xiang, Y.; Shu, Y.; Li, G. One-step modification of electrode surface for ultrasensitive and highly selective detection of nucleic acids with practical applications. *Analytical chemistry* **2016**, *88* (15), 7583-7590.

(14) Li, C.; Hu, X.; Lu, J.; Mao, X.; Xiang, Y.; Shu, Y.; Li, G. Design of DNA nanostructure-based interfacial probes for the electrochemical detection of nucleic acids directly in whole blood. *Chemical science* **2018**, *9* (4), 979-984.

(15) Zhou, Z.; Sohn, Y. S.; Nechushtai, R.; Willner, I. DNA tetrahedra modules as versatile optical sensing platforms for multiplexed analysis of miRNAs, endonucleases, and aptamer–ligand complexes. *ACS nano* **2020**, *14* (7), 9021-9031.
(16) Pei, H.; Liang, L.; Yao, G.; Li, J.; Huang, Q.; Fan, C. Reconfigurable three-dimensional DNA nanostructures for the construction of intracellular logic sensors. *Angewandte Chemie International Edition* **2012**, *51* (36), 9020-9024.
(17) Ray, S. G.; Daube, S. S.; Leitus, G.; Vager, Z.; Naaman, R. Chirality-induced spin-selective properties of self-assembled monolayers of DNA on gold. *Physical Review Letters* **2006**, *96* (3), 036101.
(18) Naaman, R.; Waldeck, D. H. Spintronics and chirality: Spin selectivity in electron transport through chiral molecules. *Annual Review of Physical Chemistry* **2015**, *66* (1), 263-281.
(19) B. Gohler, V. H., T. Z. Markus, M. Kettner, G. F. Hanne, Z. Vager, R. Naaman, H. Zacharias. Spin selectivity in electron transmission through self-assembled monolayers of double-stranded DNA. *Science* **2011**, *331* (6019), 894-897.
(20) Zwang, T. J.; Hürlimann, S.; Hill, M. G.; Barton, J. K. Helix-dependent spin filtering through the DNA duplex. *Journal of the American Chemical Society* **2016**, *138* (48), 15551-15554.
(21) Yeganeh, S.; Ratner, M. A.; Medina, E.; Mujica, V. Chiral electron transport: Scattering through helical potentials. *The Journal of chemical physics* **2009**, *131* (1), 014707.
(22) Huertas-Hernando, D.; Guinea, F.; Brataas, A. Spin-orbit coupling in curved graphene, fullerenes, nanotubes, and nanotube caps. *Physical Review B* **2006**, *74* (15), 155426.
(23) Lee, C.-Y.; Gong, P.; Harbers, G. M.; Grainger, D. W.; Castner, D. G.; Gamble, L. J. Surface Coverage and Structure of Mixed DNA/Alkylthiol Monolayers on Gold: Characterization by XPS, NEXAFS, and Fluorescence Intensity Measurements. *Analytical Chemistry* **2006**, *78* (10), 3316-3325.
(24) Manning, G. S. The persistence length of DNA is reached from the persistence length of its null isomer through an internal electrostatic stretching force. *Biophysical journal* **2006**, *91* (10), 3607-3616.
(25) Stephanie Geggier, A. K., Alexander Vologodskii. Temperature dependence of DNA persistence length. *Nucleic Acids Research* **2010**, *39* (4), 1419-1426.
(26) Brinkers, S.; Dietrich, H. R. C.; De Groote, F. H.; Young, I. T.; Rieger, B. The persistence length of double stranded DNA determined using dark field tethered particle motion. *The Journal of Chemical Physics* **2009**, *130* (21), 215105.
(27) H. Dietz, S. M. D., W. M. Shih. Folding DNA into twisted and curved nanoscale shapes. *Science* **2009**, *325* (5941), 725-730.
(28) Peil, A.; Xin, L.; Both, S.; Shen, L.; Ke, Y.; Weiss, T.; Zhan, P.; Liu, N. DNA assembly of modular components into a rotary nanodevice. *ACS Nano* **2022**, *16* (4), 5284-5291.
(29) Yao, G.; Zhang, F.; Wang, F.; Peng, T.; Liu, H.; Poppleton, E.; Šulc, P.; Jiang, S.; Liu, L.; Gong, C.; et al. Meta-DNA structures. *Nature Chemistry* **2020**, *12* (11), 1067-1075.
(30) Maier, A. M.; Bae, W.; Schiffels, D.; Emmerig, J. F.; Schiff, M.; Liedl, T. Self-assembled DNA tubes forming helices of controlled diameter and chirality. *ACS Nano* **2017**, *11* (2), 1301-1306.
(31) Grome, M. W.; Zhang, Z.; Pincet, F.; Lin, C. Vesicle tubulation with self-assembling DNA nanosprings. *Angewandte Chemie International Edition* **2018**, *57* (19), 5330-5334.
(32) Perrault, S. D.; Shih, W. M. Virus-inspired membrane encapsulation of DNA nanostructures to achieve in vivo stability. *ACS Nano* **2014**, *8* (5), 5132-5140.
(33) O'Neil, M. A.; Barton, J. K. DNA charge transport: Conformationally gated hopping through stacked domains. *Journal of the American Chemical Society* **2004**, *126* (37), 11471-11483.
(34) Winfree, E.; Liu, F.; Wenzler, L. A.; Seeman, N. C. Design and self-assembly of two-dimensional DNA crystals. *Nature* **1998**, *394* (6693), 539-544.

(35) Ke, Y.; Voigt, N. V.; Gothelf, K. V.; Shih, W. M. Multilayer DNA origami packed on hexagonal and hybrid lattices. *Journal of the American Chemical Society* **2012**, *134* (3), 1770-1774.
(36) Douglas, S. M.; Dietz, H.; Liedl, T.; Högberg, B.; Graf, F.; Shih, W. M. Self-assembly of DNA into nanoscale three-dimensional shapes. *Nature* **2009**, *459* (7245), 414-418.
(37) Ji, J.; Karna, D.; Mao, H. DNA origami nano-mechanics. *Chemical Society Reviews* **2021**, *50* (21), 11966-11978.
(38) Rothemund, P. W. K. Folding DNA to create nanoscale shapes and patterns. *Nature* **2006**, *440* (7082), 297-302.
(39) Fochtman, T. *Molecular dynamics simulations of DNA-functionalized nanoparticle building blocks on GPUs*; University of Arkansas, 2017.
(40) Doye, J. P. K.; Fowler, H.; Prešern, D.; Bohlin, J.; Rovigatti, L.; Romano, F.; Šulc, P.; Wong, C. K.; Louis, A. A.; Schreck, J. S.; et al. The oxDNA coarse-grained model as a tool to simulate DNA origami. Springer US, 2023; pp 93-112.
(41) Pan, K.; Bricker, W. P.; Ratanalert, S.; Bathe, M. Structure and conformational dynamics of scaffolded DNA origami nanoparticles. *Nucleic Acids Research* **2017**, *45* (11), 6284-6298.
(42) West, R. M. Review—Electrical Manipulation of DNA Self-Assembled Monolayers: Electrochemical Melting of Surface-Bound DNA. *Journal of The Electrochemical Society* **2020**, *167*, 037544.
(43) Liu, X.; Zhang, F.; Jing, X.; Pan, M.; Liu, P.; Li, W.; Zhu, B.; Li, J.; Chen, H.; Wang, L.; et al. Complex silica composite nanomaterials templated with DNA origami. *Nature* **2018**, *559* (7715), 593-598.
(44) Kelley, S. O.; Barton, J. K.; Jackson, N. M.; Hill, M. G. Electrochemistry of methylene blue bound to a DNA-modified electrode. *Bioconjugate Chemistry* **1997**, *8* (1), 31-37.
(45) Nano, A.; Furst, A. L.; Hill, M. G.; Barton, J. K. DNA electrochemistry: Charge-transport pathways through DNA films on gold. *Journal of the American Chemical Society* **2021**, *143* (30), 11631-11640.
(46) Pheeney, C. G.; Barton, J. K. DNA electrochemistry with tethered methylene blue. *Langmuir* **2012**, *28* (17), 7063-7070.
(47) Xie, H.; Chen, X.; Zhang, Q.; Mu, Z.; Zhang, X.; Yan, B.; Wu, Y. Magnetization switching in polycrystalline Mn3Sn thin film induced by self-generated spin-polarized current. *Nature Communications* **2022**, *13* (1), 5744.
(48) Zhang, S.; Levy, P.; Fert, A. Mechanisms of spin-polarized current-driven magnetization switching. *Physical review letters* **2002**, *88* (23), 236601.
(49) Gorodetsky, A. A.; Hammond, W. J.; Hill, M. G.; Slowinski, K.; Barton, J. K. Scanning electrochemical microscopy of DNA monolayers modified with nile blue. *Langmuir* **2008**, *24* (24), 14282-14288.
(50) Muren, N. B.; Barton, J. K. Electrochemical assay for the signal-on detection of human DNA methyltransferase activity. *Journal of the American Chemical Society* **2013**, *135* (44), 16632-16640.
(51) Grynko, O.; Lozovski, V.; Ozerov, O.; Repetskii, S.; Tretyak, O.; Vyshyvana, I. Electron spin-polarizer on the base of DNA array. In *2014 IEEE 34th International Scientific Conference on Electronics and Nanotechnology (ELNANO)*, 2014; IEEE: pp 250-253.
(52) Mishra, S.; Mondal, A. K.; Pal, S.; Das, T. K.; Smolinsky, E. Z. B.; Siligardi, G.; Naaman, R. Length-dependent electron spin polarization in oligopeptides and DNA. *The Journal of Physical Chemistry C* **2020**, *124* (19), 10776-10782.
(53) Guo, A.-M.; Sun, Q.-F. Spin-selective transport of electrons in DNA double helix. *Physical Review Letters* **2012**, *108* (21), 218102.
(54) Möllers, P. V.; Ulku, S.; Jayarathna, D.; Tassinari, F.; Nürenberg, D.; Naaman, R.; Achim, C.; Zacharias, H. Spin-selective electron transmission through self-assembled monolayers of double-stranded peptide nucleic acid. *Chirality* **2021**, *33* (2), 93-102.

(55) Baase, W. A.; Johnson Jr, W. C. Circular dichroism and DNA secondary structure. *Nucleic Acids Research* **1979**, *6* (2), 797-814.
(56) Arnott, S. The sequence dependence of circular dichroism spectra of DNA duplexes. *Nucleic acids research* **1975**, *2* (9), 1493-1502.
(57) Yeston, J. DNA circular dichroism in gas phase. *Science* **2020**, *368* (6498), 1443-1445.
(58) Gutierrez, R.; Díaz, E.; Naaman, R.; Cuniberti, G. Spin-selective transport through helical molecular systems. *Physical Review B* **2012**, *85* (8), 081404.

**Supplementary Information for**

# Twisted DNA origami-based chiral monolayers for spin filtering

Haozhi Wang[1†], Fangfei Yin[2†], Linyun Li[1], Mingqiang Li[1], Zheng Fang[2], Chenyun Sun[1], Bochen Li[1], Jiye Shi[3], Jiang Li[4,5], Lihua Wang[3,4], Shiping Song[3,4], Xiaolei Zuo[2*], Xiaoguo Liu[1*], Chunhai Fan[1]

1. School of Chemistry and Chemical Engineering, New Cornerstone Science Laboratory, Frontiers Science Center for Transformative Molecules, Zhangjiang Institute for Advanced Study and National Center for Translational Medicine, Shanghai Jiao Tong University, Shanghai 200240, China.

2. Institute of Molecular Medicine, Shanghai Key Laboratory for Nucleic Acid Chemistry and Nanomedicine, Renji Hospital, School of Medicine, Shanghai Jiao Tong University, Shanghai, China.

3. CAS Key Laboratory of Interfacial Physics and Technology, Shanghai Institute of Applied Physics, Chinese Academy of Sciences, Shanghai 201800, China.

4. The Interdisciplinary Research Center, Shanghai Synchrotron Radiation Facility, Shanghai Advanced Research Institute, Chinese Academy of Sciences, Shanghai 201210, China.

5. Institute of Materiobiology, Department of Chemistry, College of Science, Shanghai University, Shanghai 200444, China.

*Corresponding author:
Xiaolei Zuo. Email: zuoxiaolei@sjtu.edu.cn
Xiaoguo Liu. Email: liuxiaoguo@sjtu.edu.cn

## Materials and Methods

Synthesis of TDOs

TDOs were designed with caDNAno software. Staple DNA strands were purchased from Sangon Biotech (Shanghai, China), and scaffold DNA strands were purchased from Bioruler (cat. No. B3003, B3007 and B3005). All DNA strands were used as received without further purification. The DNA strands were dissolved in ultrapure water and stored at -20 °C. To fold the TDOs, the staple strands were mixed with the scaffold strands at a molar ratio of 5:1 in TE-$MgCl_2$ buffer (10 mM Tris, 1 mM EDTA, 12.5 mM $MgCl_2$, pH = 8.0). The mixture was then annealed in a PCR thermocycler (Eppendorf) using the following protocol: 90 °C for 10 min, then cooled from 65 °C to 25 °C at a rate of -1 °C per 60 min. To remove excess staple strands, a PEG purification method was utilized: the annealed mixture was mixed with PEG buffer (5 mM Tris, 1 mM EDTA, 500 mM NaCl, 15% w/v polyethylene glycol, Mw: 8,000 g/mol, pH = 8.0) at a volumetric ratio of 1:1 and centrifuged at 10,000 rcf for 15 min. The resulting DNA pellet was dissolved in TE-$MgCl_2$ buffer and shaken at 600 rpm and 25 °C for 12 hours. The purified TDOs were diluted to a concentration of 20 nmol/L in TE-$MgCl_2$ buffer and quantified by a UV-2600i UV-visible spectroscopy (Shimadzu Inc.). All purified TDOs were stored at 4 °C before further use.

Post assembly of thiolated DNA linker strands

Thiolated DNA linker strands were synthesized and purified by HPLC (Sangon Biotech, Shanghai, China), with a 3' -end modified with C6 S-S phosphoramidite. The sequence of linker strands is 3'-TATTATTATTATTAT-5'. The linker DNA strands were dissolved in ultrapure water and stored at -20 °C before use. To assemble the linker strands onto the TDOs, disulfide bonds were chemically reduced to sulfhydryl groups using tris(2-carboxyethyl)phosphine hydrochloride (TCEP) at a molar ratio of linker strands to TCEP of 1:250. The reduced mixture was annealed with purified TDOs in a PCR thermocycler, following a protocol of 45 °C for 10 min, and then cooled from 45 °C to 25 °C at a rate of -1 °C per 30 min. The excess staple strands were also removed with the PEG purification method. The resulting DNA pellet was redissolved in TE-$MgCl_2$ buffer and shaken for 12 hours at 600 rpm and 25 °C. The purified TDO assemblies were diluted to a concentration of 20 nmol/L with TE-$MgCl_2$ buffer and quantified by UV-visible spectroscopy. All purified TDO assemblies should be used as soon as possible to ensure their stability.

Acquisition of TEM images

For TEM imaging of TDOs and TDO assemblies, 10 μL of purified TDOs or TDO assemblies (20 nmol/L) solution drop was adsorbed onto glow discharged, carbon-film-coated copper grid for 5 min. The sample was washed with ultrapure water, then stained for 60 s using a 1% aqueous uranyl formate solution. After stained, sample was washed with ultrapure water and dried at 25 °C. All images were acquired using a Talos F200C G2 operated at 200 kV.

Electrode fabrication

The fabrication of gold/nickel working electrode was carried out at Topvendor Tech (Beijing, China). P-type oxidized silicon wafers were initially coated with a 200 nm nickel, followed by deposition of 15 nm gold. The resulting surfaces were then cleaved into 1.2 cm × 1.2 cm rectangles and stored in a dry environment.

Preparation of TDO-based chiral monolayer modified electrode

Gold/nickel working electrode were gently rinsed by ethanol and ultrapure water and dried with argon before plasma cleaning (Harrick Scientific Products Inc.) for 2 minutes. A rubber (3M) gasket was then attached to the surface to create a liquid well, and 50 μL of thiol modified TDO solutions were added to form a monolayer. The TDO-based chiral monolayer modified area of each electrode was 0.6 cm × 0.6 cm. The TDOs were incubated on the surface for 12-18 hours, and once the TDOs were deposited, the surface could not be dried to prevent the structure from affecting the measurement performance of the monolayer. Finally, the surface was rinsed with TE-$MgCl_2$ buffer to remove excess TDOs, and electrochemical experiments were performed immediately afterward.

Acquisition of SEM images

SEM imaging was performed using Zeiss GeminiSEM 360. The TDO-based chiral monolayer modified electrode was washed with ultrapure water and dried. Then scanned with SEM directly without sputtering an additional thin layer of conductive material on the surface of samples. The operation voltage was set to 5 kV to reduce the electron damage to DNA samples.

Silicification of TDO-based monolayers on electrode

A 2% (v/v) TMAPS (50% in methanol) solution was slowly added to 1 mL TAE/$MgCl_2$ buffer (10 mM Tris, 1 mM EDTA, 12.5 mM $MgAc_2$, pH = 8.0) while vigorously stirring. After 40 min, 2% (v/v) TEOS was slowly added and stirred for another 40 minutes to ensure sufficient pre-hydrolysis. Subsequently, the TDO monolayer modified electrodes were gently cleaned with ultrapure water and then transferred into the silicification solution. The samples were allowed to react under static conditions at room temperature for 16 h. Afterwards, they were rinsed with ultrapure water to remove excess reagents and dried.

Acquisition of AFM images

The TDO chiral monolayer modified electrode was washed with ultrapure water and dried. Samples were characterized on a Bruker Multimode 8 Atomic Force Microscope. The samples were imaged in Scan-Asyst mode in Air using Scan-Asyst Air probes. Image analysis was performed by NanoScope Analysis 1.7 software.

Electrochemical experiments

The experimental setup consisted of a conventional three-electrode system, comprising a gold/nickel working electrode, a platinum wire auxiliary electrode, and an Ag/AgCl reference electrode. These electrodes connected to an electrochemical workstation (CHI 660c). Cyclic voltammetry (CV) was used for the experiments with the following parameters: Init E(V) = -0.10, High E(V) = -0.15, Low E(V) = -0.30, and a scan rate of 0.05 V/s. To investigate the effect of a magnetic field, a 6619 Gauss surface strength neodymium magnet was used. Plastic parts were used in the setup to avoid extraneous objects that could influence the magnetic field. Each experiment comparing the yield of reduced MB under magnetic field directed up or down, by comparing the same surface in the same solution to minimize variability caused by other factors. During the experiments, methylene blue (Sangon biotech. CAS # [7220-79-3]) solution was diluted to 10 μM by TE-$MgCl_2$ buffer. The magnetic field direction was switched, scanned, and switched again multiple times, and each electrochemical test that changed the direction of the magnetic field was refilled with a 50 μL MB buffer to ensure signal stability. Cyclic voltammetry tests were performed five times in each magnetic field direction, as continuous testing would cause the signal to decay progressively. This decay was restored by waiting for approximately 1 minute between scans.

Circular dichroism spectroscopy experiments

A Jasco J-1500 spectropolarimeter was used to collect circular dichroism (CD) spectra. Data were obtained from samples containing 20 nM TDOs in TE-$MgCl_2$ buffer using a 1.0 mm path length cell. Data presented in Figure 4c represent the average of three scans.

OxDNA MD simulation

The oxDNA MD simulation treats dsDNA as a rigid model of nucleotides. Interactions such as hydrogen bonding, stacking, and excluded volume are captured by potential functions. While the model allows for specific Watson-Crick base-pairing, it does not capture noncanonical base-pairing interactions. Using long-range molecular dynamics (MD) simulations (~3.03 μs) in a 20 mM cationic environment, we studied the hybridization thermodynamics and mechanical properties of double-stranded and single-stranded DNA in TDOs. The coaxial stacking distances were calculated by measuring the distance between adjacent bases on the scaffold strand, excluding values from crossovers and outliers from single chain segments in TDOs. A Python script that calculates the coaxial stacking distances was provided in materials availability.

Monolayer spin polarization calculation

The spin polarization ($SP$) is defined as:

$$SP = \frac{I_+ - I_-}{I_+ + I_-}$$

in which $I_+$ and $I_-$ are the intensities of the signals corresponding to the spin oriented parallel and antiparallel to the electrons' velocity. The amount of charge transferred to the probe ($Q$), which is determined by integrating the current under the reductive or oxidative peak in the cyclic voltammograms, can be related to the injected spin polarization by the following equation:

$$Q = I_{+}(\eta_{+}) + I_{-}(\eta_{-})$$

where $\eta_{+}$ and $\eta_{-}$ are the yield for injected spin oriented parallel and antiparallel, respectively, to the velocity of the electrons that reduce the probe compared to the total amount injected.

Monolayer spin polarization ($MSP$) can be derived as:

$$MSP = \frac{\eta_{+} - \eta_{-}}{\eta_{+} + \eta_{-}}$$

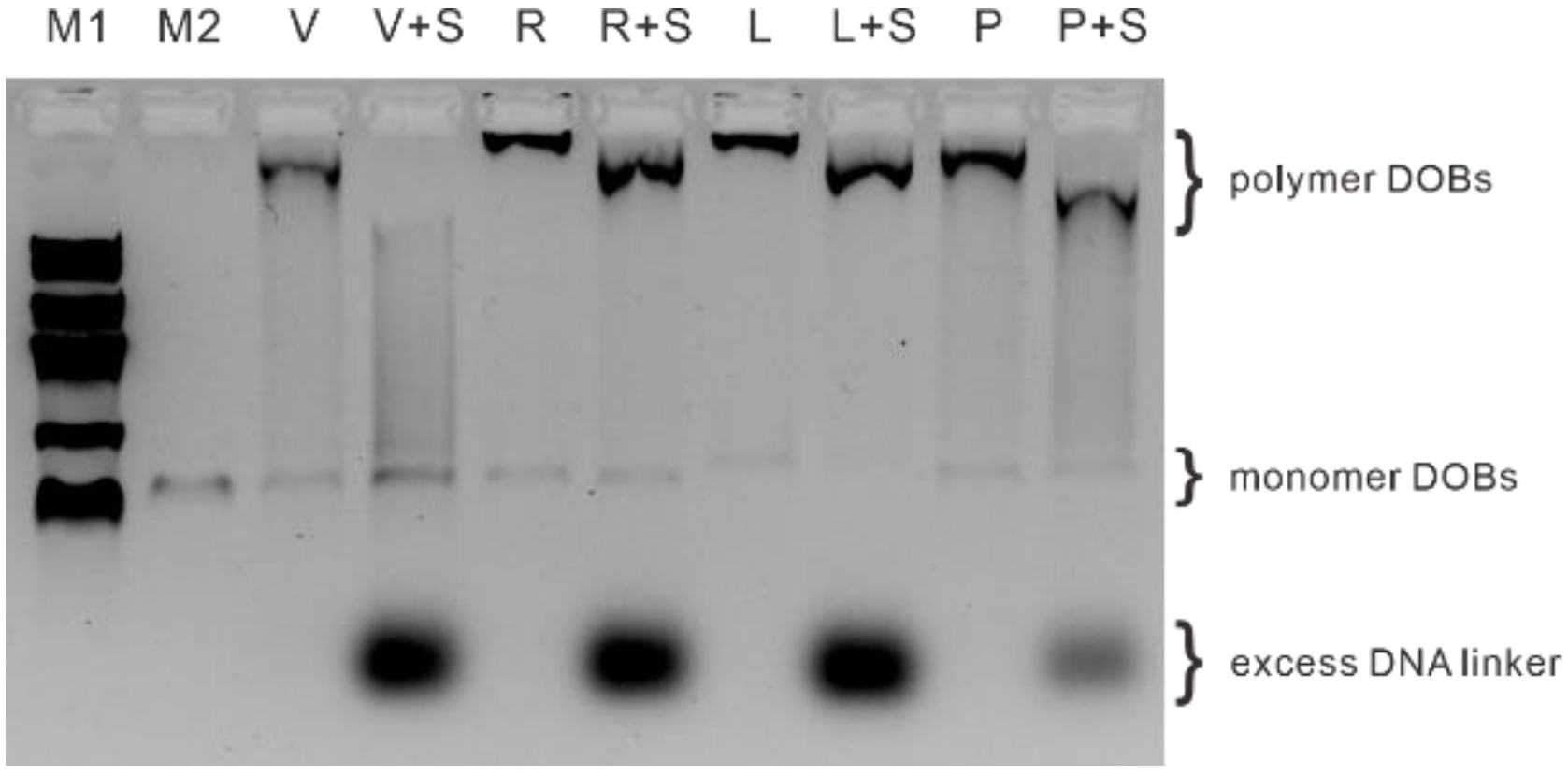

**Figure S1.** 1 % agarose gel of the PEG purified TDOs and unpurified thiolated TDOs. Lane M1: 10 kb DNA ladder; Lane M2: 8064 bp DNA scaffold strand. Lane V, R, L, and P: 20 nmol/L of V-TDO, R-TDO, L-TDO, and P-TDO after PEG purification, respectively. Lanes V+S, R+S, L+S, and P+S: 20 nmol/L of the corresponding thiolated TDOs before PEG purification.

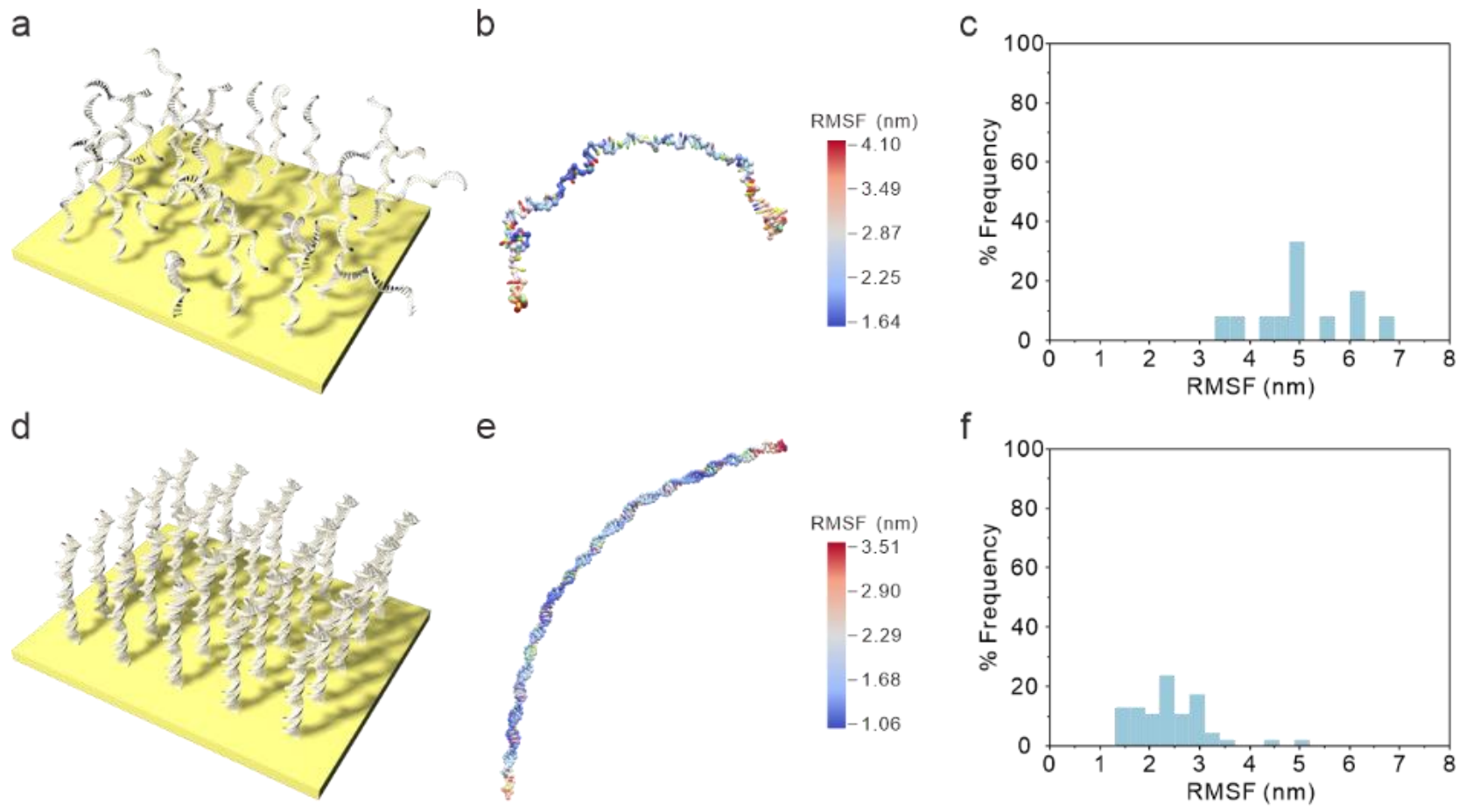

**Figure S2.** Configuration analysis of traditional DNA chiral monolayer through OxDNA MD simulation. (a-c) Illustration of the ssDNA monolayer modified gold interface, the last configuration results outcomes of the 120 bp ssDNA coarse-grained model with corresponding RMSF results and RMSF distribution. (d-f) Illustration of the dsDNA monolayer modified gold interface, the last configuration results outcomes of the 120 bp dsDNA coarse-grained model with corresponding RMSF results and RMSF distribution.

Monolayer coverage of the electrode surface

AFM data showed that the height distribution range of the blank electrode is ± 2 nm, which serves as the background. The coverage of the dsDNA and TDO-based chiral monolayer is determined by subtracting the blank background and is shown in blue in Figure S3.

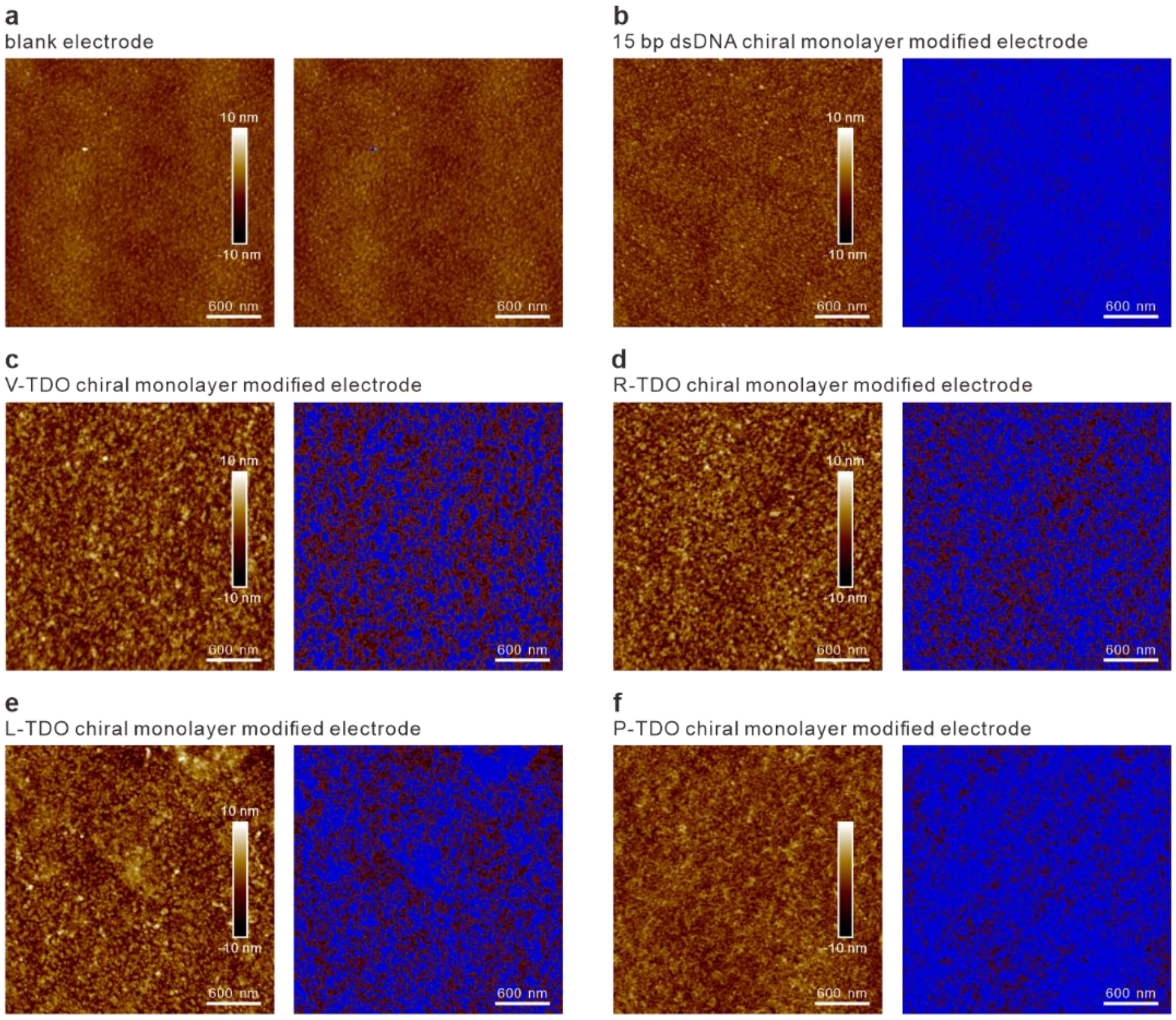

**Figure S3.** The coverage of different chiral monolayers on the electrode. (a-f) are blank electrodes, 15 bp dsDNA chiral monolayer modified electrode, V-TDO chiral monolayer modified electrode, R-TDO chiral monolayer modified electrode, L-TDO chiral monolayer modified electrode, P-TDO chiral monolayer modified electrode (left) and the area covered by corresponding chiral monolayer (right, blue).

Boundary height analysis of TDO-based chiral monolayer before and after silicification

AFM data showed the changes of electrode surface height after silicification with different TDO-based chiral monolayers. The heights of the region modified by TDO-based chiral monolayers changed significantly. AFM data showed that the heights of grown silica in V-TDO, R-TDO, L-TDO and P-TDO chiral monolayers was ~3.87 nm, ~6.599 nm, ~4.12 nm and ~1.633 nm, respectively. The height of bare gold region showed no change. Samples before and after mineralization were naturally dried for AFM and SEM image acquisition.

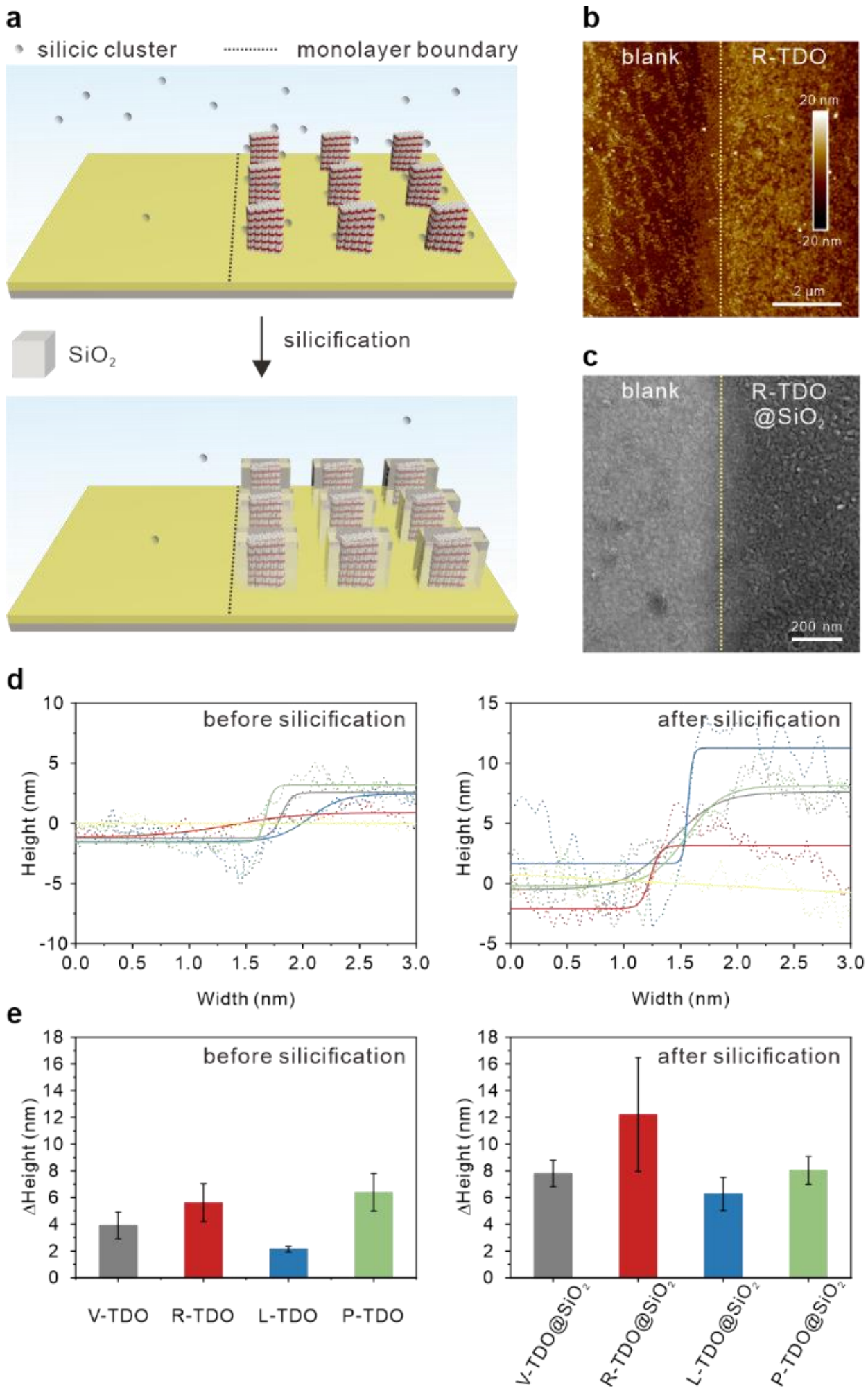

**Figure S4.** Boundary height difference of TDO-based chiral monolayers before and after silicification. (a) Schematic of the chiral monolayer boundary on the electrode surface (top); Schematic of the chiral monolayer boundary after silicification (bottom). The black dashed line on the electrode surface represents the boundary. (b) The AFM image of the boundary before silicification, the dotted yellow line represents the boundary. Scale bar was 2 μm. (c) The SEM image of the boundary after silicification, the dotted yellow line represents the boundary. Scale bar was 200 nm. (d) Representative data of the height of chiral monolayers before (left) and after (right) silicification. The gray, red, blue, and green dashed lines represent the experimentally measured heights of V-TDO, R-TDO, L-TDO, and P-TDO chiral monolayer, respectively. The solid lines corresponding to the colors represent their fitting heights. The height information is collected from a 3 μm × 600 nm region with boundary as its axis of symmetry. (e) Boundary height difference before (left) and after (right) silicification.

Spatial orientation of TDOs in silicified monolayer on electrode

SEM data revealed the details of TDO@$SiO_2$ particles. Image analysis was conducted using the ImageJ software. Monodisperse particles (highlighted in Figure S5e, right) within the size range of 512.69 $nm^2$ to 632.96 $nm^2$ and the roundness range of 0.50 to 0.76 were identified as TDOs oriented with the z-axis perpendicular to the interface. Conversely, monodisperse particles within the size range of 0 $nm^2$ to 512.69 $nm^2$ and 632.96 $nm^2$ to 750.46 $nm^2$, along with a roundness range from 0 to 0.50 and from 0.68 to 1.00, were regarded as TDOs oriented with the z-axis parallel to the surface.

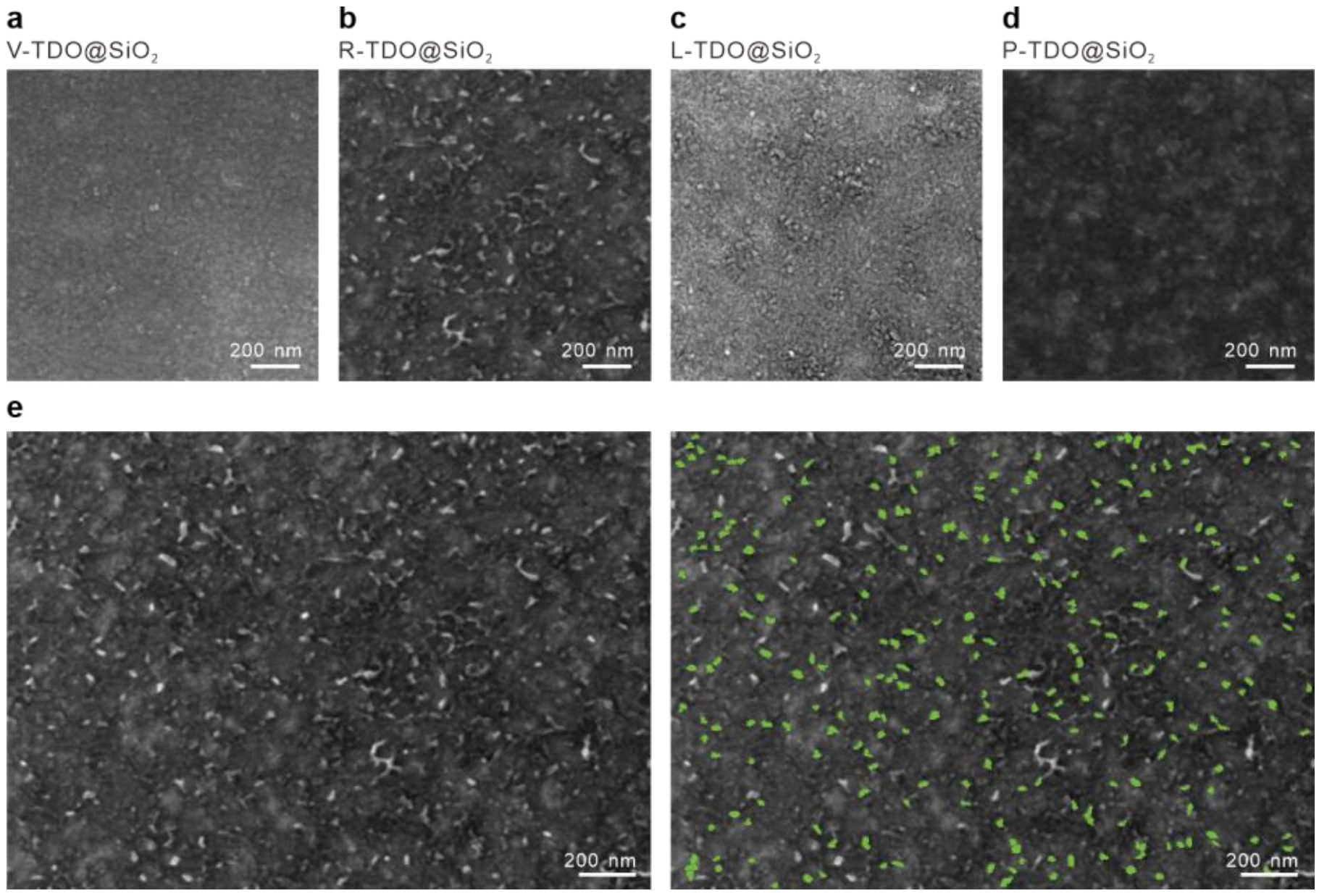

**Figure S5.** Details of TDOs in the silicified monolayer. (a-d) are the SEM images of V-TDO, R-TDO, L-TDO, P-TDO chiral monolayers after silicification on the electrode, respectively. (e) Silicified TDOs on the electrode surface (left), and corresponding particle analysis by ImageJ software (right). All single particles are marked and numbered in colors other than red for size and roundness statistics.

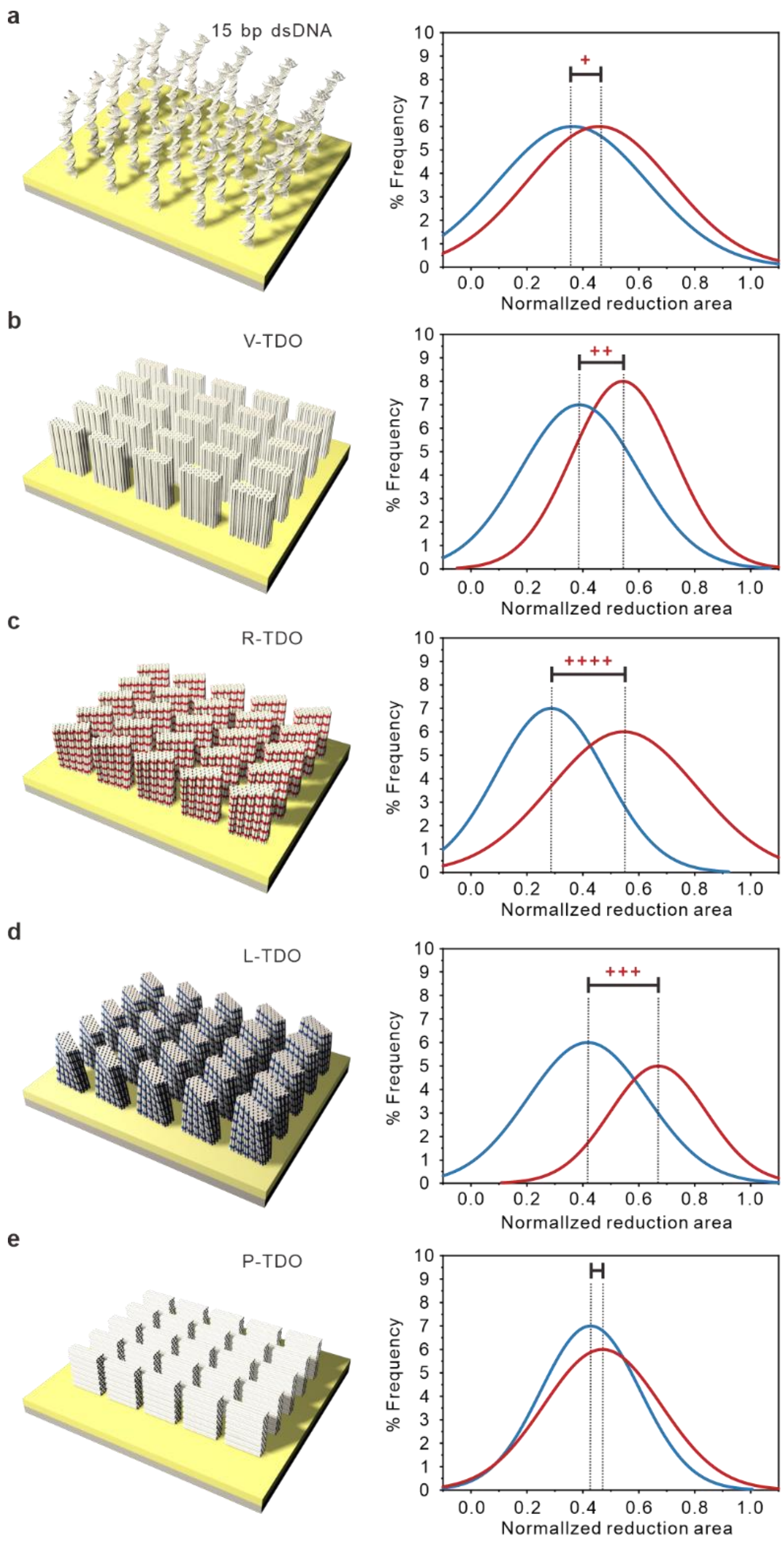

**Figure S6.** Representative CISS behavior of DNA chiral monolayers in different magnetic field directions. (a-f) Schematic models of blank electrode, 15 bp dsDNA monolayer, different TDO-based monolayers (left column), and corresponding electron transfer distributions in monolayers with different magnetic field directions (right column). The relative position of the peaks indicates the direction of spin filtering, with the + symbols indicating the degree of preference for the transformation of electron with up spin.

**Note S5. Design of twisted DNA origami**

Figure S7-10 shows the caDNAno diagrams of TDO. The numbers on the left indicate the DNA helices. The green strands are the extended strands that complements the linker strands.

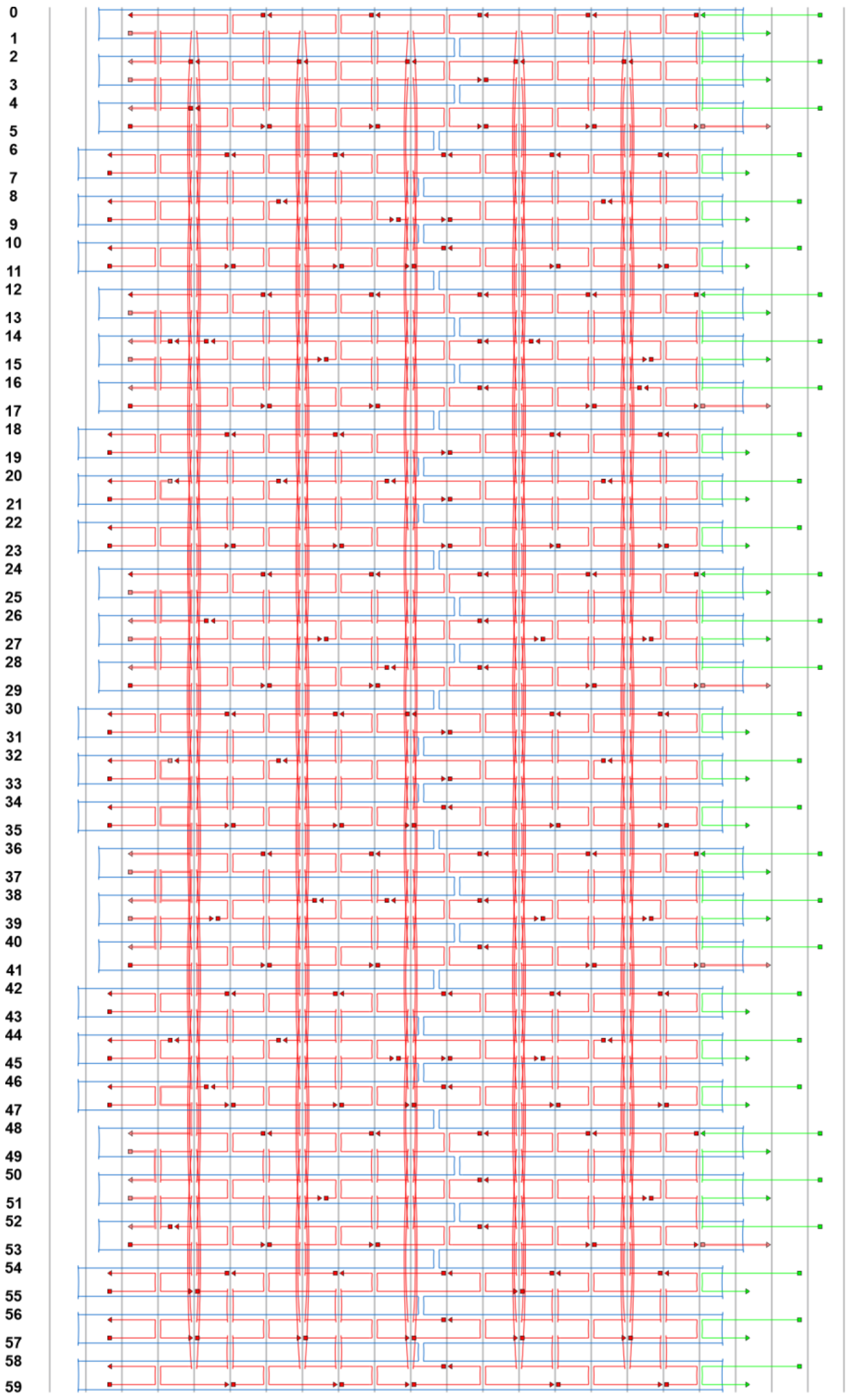

**Figure S7.** Strand diagram of V-TDO.

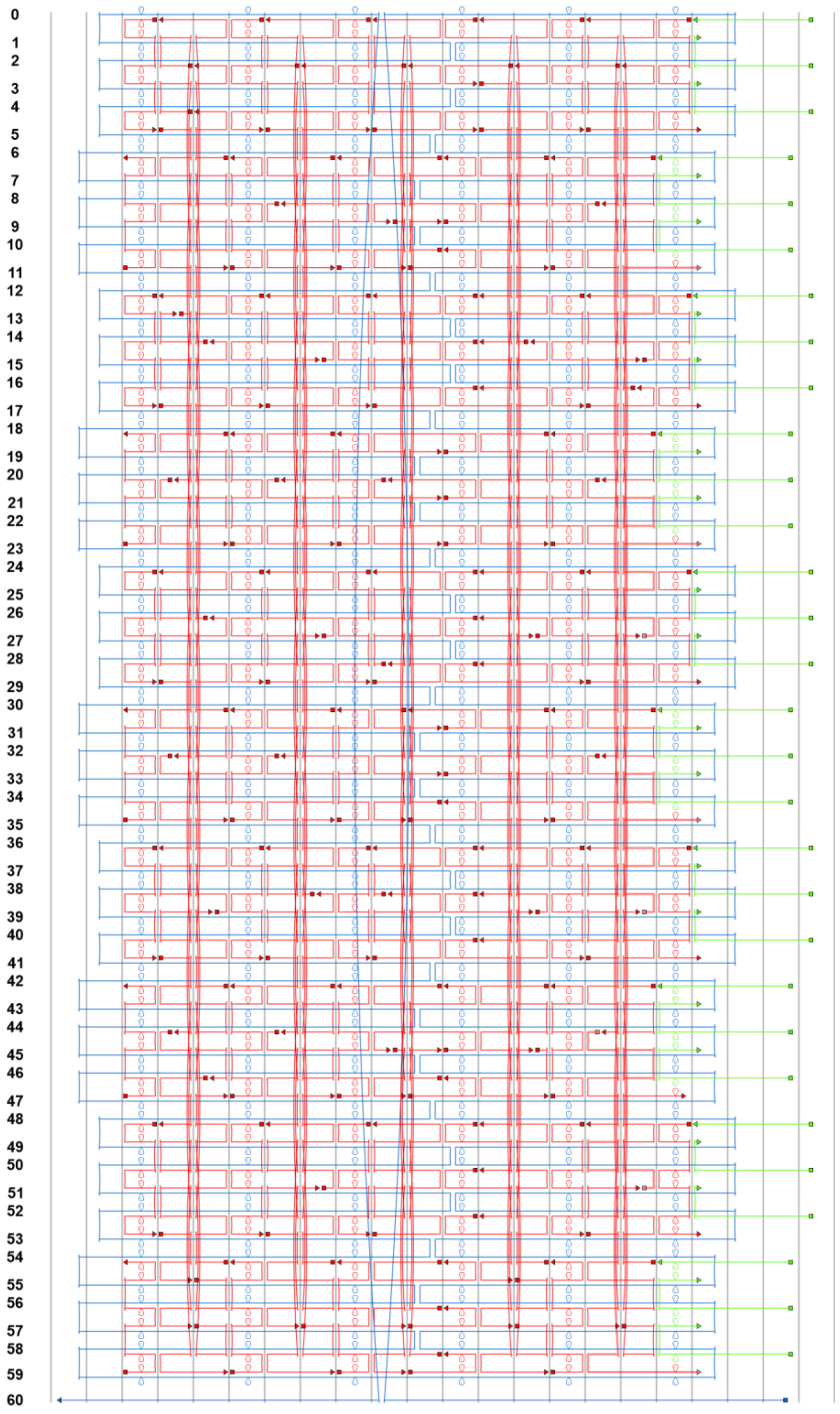

**Figure S8.** Strand diagram of R-TDO.

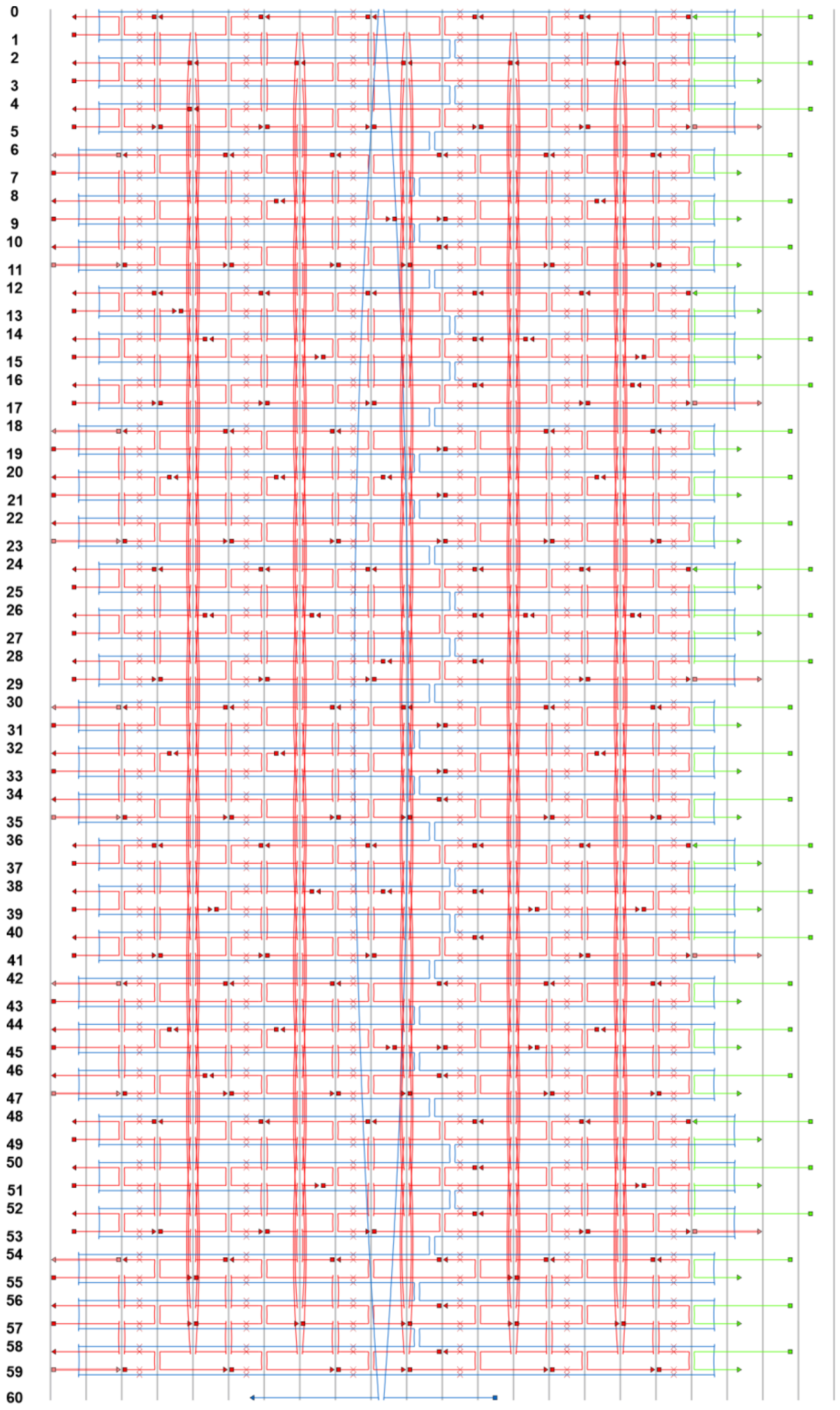

**Figure S9.** Strand diagram of L-TDO.

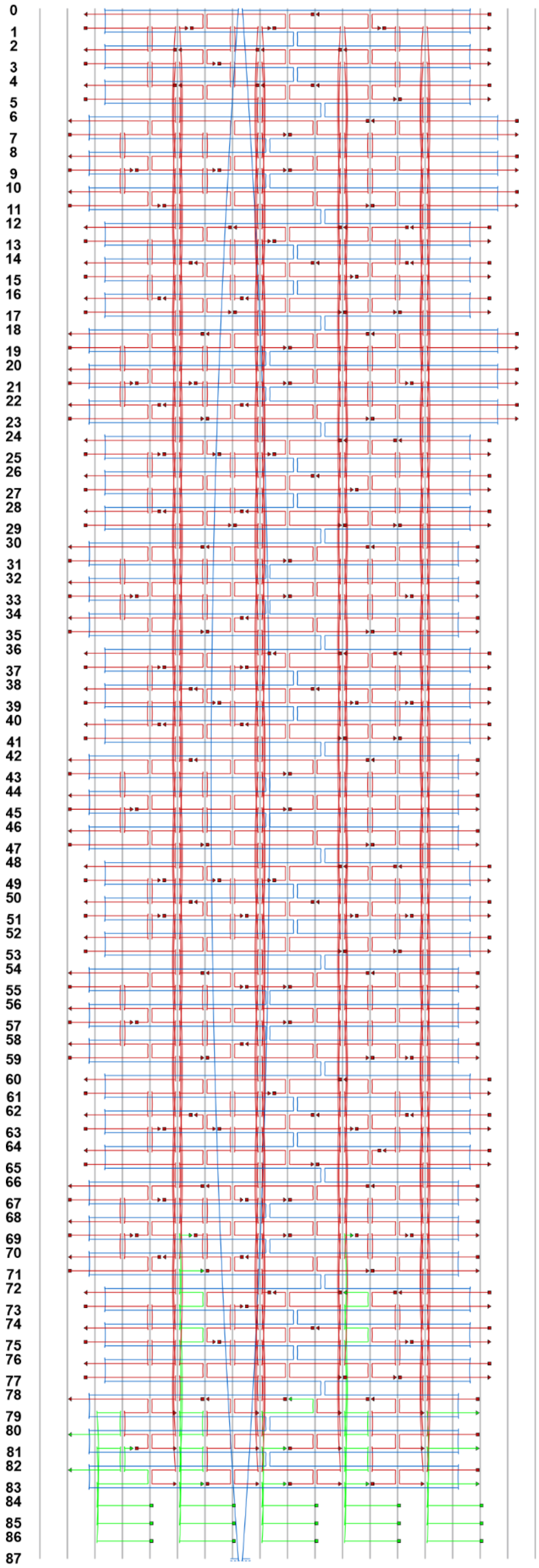

**Figure S10.** Strand diagram of P-TDO.

## Table S1. DNA sequences of V-TDO

| Strand name | Sequence | Color |
|---|---|---|
| VTDO-linker-1 | ATAATAATAATAATAAGTTGAGATTTTTTT | Green |
| VTDO-linker-2 | ATAATAATAATAATAAGAACTCAAATCGTCTGATTTTT | Green |
| VTDO-linker-3 | ATAATAATAATAATACTTTATTAGATTTTT | Green |
| VTDO-linker-4 | ATAATAATAATAATAAATGCATGTATTTTT | Green |
| VTDO-linker-5 | ATAATAATAATAATACGTGGGCGAATTTTT | Green |
| VTDO-linker-6 | ATAATAATAATAATAAGGCAGGTC | Green |
| VTDO-linker-7 | ATAATAATAATAATATCCAAATAAACTGAACACTTTTT | Green |
| VTDO-linker-8 | ATAATAATAATAATAGGGAACATTTTTTTT | Green |
| VTDO-linker-9 | ATAATAATAATAATAGCGTAAGGGCTTTTT | Green |
| VTDO-linker-10 | ATAATAATAATAATATTCTGTCCA | Green |
| VTDO-linker-11 | ATAATAATAATAATAATTACTCCTTTTTTT | Green |
| VTDO-linker-12 | ATAATAATAATAATAACCGAAGCC | Green |
| VTDO-linker-13 | ATAATAATAATAATAGCAAAAGTTGTTTTT | Green |
| VTDO-linker-14 | ATAATAATAATAATAAACCGCCTCAGGAGGTTGTTTTT | Green |
| VTDO-linker-15 | ATAATAATAATAATAGGTTTCATATTTTTT | Green |
| VTDO-linker-16 | ATAATAATAATAATATTTCAAACAATTTTT | Green |
| VTDO-linker-17 | ATAATAATAATAATACCTTTAGCGCACCACCGGTTTTT | Green |
| VTDO-linker-18 | ATAATAATAATAATAATAGCAAAACTTTTT | Green |
| VTDO-linker-19 | ATAATAATAATAATATATCATCATAAAATTATTTTTTT | Green |
| VTDO-linker-20 | ATAATAATAATAATAAATGGATTATACGTGGCATTTTT | Green |
| VTDO-linker-21 | ATAATAATAATAATATGAGAAATCCTTTTT | Green |
| VTDO-linker-22 | ATAATAATAATAATACCTGAACAAAGCTATCTTTTTTT | Green |
| VTDO-linker-23 | ATAATAATAATAATATAAATAAGGCGCTCAACATTTTT | Green |
| VTDO-linker-24 | ATAATAATAATAATATGTAGGAGCATTTTT | Green |
| VTDO-linker-25 | ATAATAATAATAATAGTAGGGCTTGTAAAGTAATTTTT | Green |
| VTDO-linker-26 | ATAATAATAATAATATGCTTTGAA | Green |
| VTDO-linker-27 | ATAATAATAATAATATCAGTGGCGGTTTTT | Green |
| VTDO-linker-28 | ATAATAATAATAATATACCGATCATTTTTT | Green |
| VTDO-linker-29 | ATAATAATAATAATACAGACAATA | Green |
| VTDO-linker-30 | ATAATAATAATAATATGCACGTAATCGCCTGATTTTTT | Green |
| VTDO-1 | GATCTAAAGTCAGCCCTCAT | Red |
| VTDO-2 | GATTAAACGTGCTTAAACGCTGACCAGTACCT | Red |
| VTDO-3 | CAGCAAACAGTAGCGTAGCGCGTTTTCACAAAATCCAAA | Red |
| VTDO-4 | CCTGTTTCTAGGCATTTTCGAATCAATAGATGATGCATTTCA | Red |
| VTDO-5 | GCGCGGGGCTAACTGAGCCGGGCTTTCCATCGGTGATTT | Red |
| VTDO-6 | ACGCGAGATCTTCTTAATTACTCCCGACAAGGCTTATAAAAAAGAG | Red |
| VTDO-7 | CATCACCTTAGGAGCACTAATTTTTTTATCAATCAGGTAACC | Red |
| VTDO-8 | CAGTACACAAGCCCACTCAGGAGGTTTAGCGGATACGGA | Red |
| VTDO-9 | AGCAGCGCGTAATGGGCAAAAGTCGAAACGAGGCGTCAAGAG | Red |
| VTDO-10 | TTATAATATAACATGCCTTGCAGGATTTTTGAAAGGAAC | Red |
| VTDO-11 | CAAATCAACCTTCACAGACGGGAATTTTGTTTCAGAGTGCCGATTCTGA | Red |
| VTDO-12 | CCACCAGGCCACGCATATCAAGGAAGGTAGATAATTTAAATC | Red |
| VTDO-13 | GCACCCAAACAGCCTTTT | Red |
| VTDO-14 | TGAACCCTCATTCACCATCTGACCAAGAAAGAGAGCTTGACGGGG | Red |
| VTDO-15 | GCAAAGAGAGGGAAGGTAATAATAAGTACTGCCTA | Red |
| VTDO-16 | TAAAGGCTTCGAGGAATTGTATTCAACACTGTATGTCGTCAC | Red |
| VTDO-17 | TTTACTGGTCTTTCCAGAGCCTTACCAAGCGAACCTAGAAAA | Red |
| VTDO-18 | CTGAATCTAATTTGTGAAAATATTAAAGTCAACCGCATCTTTACCCTCA | Red |
| VTDO-19 | ATCCCCGCCCGA | Red |
| VTDO-20 | ACCGTGTGATTTTT | Red |
| VTDO-21 | GATTGCAAAAGTTTGAGTAACCGAACGTTTAGAGCTACCGCC | Red |
| VTDO-22 | AAATCCGAAACTAGGATATTCAACCGTT | Red |
| VTDO-23 | CTTAGTGTCTCAGGATTAAGTAAGGGCGGGCACCGCATCTGC | Red |
| VTDO-24 | TAACCTGTTTGATACATTTC | Red |
| VTDO-25 | AAAGCCACCTCAGAGCCGCCAAAGGTGGCAATAATTTACCAG | Red |
| VTDO-26 | TACAATCGTATTGTAAAAATGTCC | Red |
| VTDO-27 | AGCCATCAATTAACCGTTGTAGAGTCTGGGAGGCC | Red |
| VTDO-28 | GCAACTACGAACGAGCGAGCTTCTGGCCCGCATTATGTT | Red |
| VTDO-29 | CAATGAGAGCCAGTACCAATTATTAAACAAA | Red |
| VTDO-30 | GAGCGGAATTTTTT | Red |
| VTDO-31 | TAAATCACAACGTCTCGTCGGTGACTCTAAGTGTC | Red |
| VTDO-32 | GATTAGCCACCC | Red |
| VTDO-33 | CAAAATAGCTACAAAGGTTTTAAACACCTTCTTACAGAC | Red |
| VTDO-34 | GAAAGCAAAGTCATCAGCAGATACTAAAATAAAAGACAAGGGAGT | Red |
| VTDO-35 | GTTCAACATGAGAACAATTATCATCCAATAAAACAAAGAACGGAA | Red |
| VTDO-36 | AGACGATCATTGACCCTCAGACCAGAGCTCAGACTGACAGAA | Red |
| VTDO-37 | CATTACCCATCGATTTTCGGTCATAGCCCCCTTGCGTCACCA | Red |
| VTDO-38 | TTTCGGAATAAACAGCTTGCTCGCTTTT | Red |
| VTDO-39 | ATGAAACATTAGCAATTATTCAGCAGCC | Red |
| VTDO-40 | AACAAACAACCGTGCTTCTGGTCCACACTACCGAGACGCTGGAGCCCGA | Red |
| VTDO-41 | TTAATGCGAACCCTGTCACACCATGGAAATCCAGATTCTTTG | Red |
| VTDO-42 | GTGGTTTTTCGGTTTGCGTA | Red |
| VTDO-43 | GAACGGTGTACAGAGTGAATGTAGTAACTTAGAGTTTTAATAAGA | Red |
| VTDO-44 | AATTCTGAAGTACGTTTGATATTTCAACAAGAACTAAAT | Red |
| VTDO-45 | TAATGCGGGAAGGGTGAAAGCATGGCTATAAAACA | Red |
| VTDO-46 | GTAACAACCCGTCAGGGGACCCAGCCAAAGC | Red |
| VTDO-47 | CTGGCTGACGTAACGGTTTAAAGAGGTCGAACCAGATAA | Red |
| VTDO-48 | AGTAAAAGCAATACACAATATCGTCAATTATCTAATTCT | Red |
| VTDO-49 | ACCCATGTACCGACCCGGAATAGGTGTAGGTTGATAACAACTTCGG | Red |
| VTDO-50 | CTAATAAAGACCTTAGATACCTTTAGTACCGACAATAACCATCCT | Red |
| VTDO-51 | TCAAGTTTGTTTTT | Red |
| VTDO-52 | AGGTGTCACATGAGCGCAAGAAACTCGGCTGTAATATCACAACAT | Red |
| VTDO-53 | GTCTCTGTGCCCGTACCTATTTCGAGAGTCACCGTAATAGGA | Red |
| VTDO-54 | TAGCATTCACCCTCCACCCTC | Red |
| VTDO-55 | TGAGACTATAAGTTAGTTGCGACCGATACGGCTACAGAGGC | Red |
| VTDO-56 | TAATCTTGACAAGCTTACTTAGCATATTGACGGAAAGGCCGG | Red |
| VTDO-57 | ATAAATCGCCTTTTCATTCAATAAAATGAAAAGAAAAGGGCATGA | Red |
| VTDO-58 | AACAGTTAATTCTACGCCATCCAGCTCATCGCCATCAGT | Red |
| VTDO-59 | AATGGGATAGGTTAATAGGAACTAGGTC | Red |
| VTDO-60 | GATAGGGACTATTAGGGGCCTCTATTTAGTCAACC | Red |
| VTDO-61 | CCCTCAGAGAGCCGAACGTAGAAGAACTTAAGCAG | Red |
| VTDO-62 | TTGCTTCTATCAAAATCATAAGTCAGGAACCCTGATACCAAG | Red |
| VTDO-63 | TCAGTTTTTAATATTTAAATATAATTTAATGAAAACAAATCGCGC | Red |
| VTDO-64 | GAATCAGGTTCGTCGCTTACATAAGCCAGTATAATGCACTGAACA | Red |
| VTDO-65 | GAGCCACCACCCCCTAATTAG | Red |
| VTDO-66 | AATAACACTGAGTTGGATTTTGCCGGAATCCGCGACCTT | Red |
| VTDO-67 | AGCCTAGTTTTAGTTAATTTCAAAACTTCTACCTTTAATCCT | Red |
| VTDO-68 | CCCTCAGAGCCACCCACAGATTTGTCGTGACCAAACAACGGCGTT | Red |
| VTDO-69 | TAGTTTGCATGTTTCGGATGGATTGGGCGTCAGGAAGATTTGTTTG | Red |
| VTDO-70 | CCTGGAGTGGGCACTAATGAGCCCCGGGAACATACCACATTA | Red |
| VTDO-71 | CCTCGAAAGTGT | Red |
| VTDO-72 | ATGTTAGGAACTATTATATCAAAAGAAGCCTCCATCAA | Red |
| VTDO-73 | TTCACGTTGGTGTACAAAGCGAAATTGTCATGGTCGAGAGTTTTGTTCC | Red |
| VTDO-74 | AGTAAAAACCATAAAGTCAGAGCAAACTTGCTCCTGTGTCTG | Red |
| VTDO-75 | AGGTCAATAAGC | Red |
| VTDO-76 | CTTTTTATAGCAATAGTCAGAAGAATTAGAAACGAATATTAT | Red |
| VTDO-77 | GAAAGGGGGACTTCGCTATT | Red |
| VTDO-78 | CACTCTCCGAGCTGACGTGGAACACGTAAAG | Red |
| VTDO-79 | GCGTTTCATGAGGCATACAGACGACGGAATCGTCATA | Red |
| VTDO-80 | TGTAACAGGAATCGGTTAGAACCCGCTTTAAAATACTGCGATAAA | Red |
| VTDO-81 | GAGGTGAATTAACATCTGGTCTCAACAGAGAAGTACAAACAA | Red |
| VTDO-82 | ACAATAGACGGGTATTAAACCGAAGGCACCATCGTCACCCTC | Red |
| VTDO-83 | AGACTTATTATA | Red |
| VTDO-84 | TTACATTATTAATTACGAGGAAGTGTAGCAATATTCGG | Red |
| VTDO-85 | GAGAGGGGGAGCAACCCCAAATAAAATTTTCCTGTCCGTGGG | Red |
| VTDO-86 | CCGATTTCGAAAGGACACCCGCGTATAAGGGATTTGCCACCG | Red |
| VTDO-87 | GGGATAGAACTACAAGTAAATTCAATCAGCCTGATGGCT | Red |
| VTDO-88 | TTCTAATTGAATCGACTCTGATCGTAATTATCCGCACTGCCC | Red |
| VTDO-89 | CAGTGAGCAGCAGGGGAGGATTAAACAGCAAGGCGAGAA | Red |
| VTDO-90 | TACCCAAAAAATACTTTATTTGCGACATGTGAATTGTAGCAC | Red |
| VTDO-91 | TTATGACACGACGGGTTTTCCTCAGGCTAACCAGGGATGGGC | Red |
| VTDO-92 | TTGCAAAACTAATGTTGAGATCGATTTTTTTAATCCATTACC | Red |
| VTDO-93 | TGGCCCACTAGAAAAACCGT | Red |
| VTDO-94 | CTCATTCACCAGGCACCGAACTCTTTCCAAAGGAATTGCTCAAAGAGGC | Red |
| VTDO-95 | AATTTACAAACCAACGTAGGACGCCCAATAAATCACTATTTT | Red |
| VTDO-96 | TAAAACAGTTAATGGAATGGAAAGCGCA | Red |
| VTDO-97 | TCGCTGAAGCTTGAAAGGCTCAGAATAGAGACGTTACGCCTG | Red |
| VTDO-98 | ATCGGCAAAACCTGTTTGAT | Red |
| VTDO-99 | AGAGGCGTTTTACATAGATTTTAATGGAGCAATTCTTTTGCG | Red |
| VTDO-100 | ACAAACACACCACCAACCGCCTCATAATTCGGCATAGCAGCA | Red |
| VTDO-101 | ATTACCTCAAATATACAGTAACAGCAGCAAATACCGATAGCC | Red |
| VTDO-102 | ATCATGACCCAACGCAAGAGAATGTGTTTAGGAGGCTT | Red |
| VTDO-103 | AATCCTCCTGCGCCAGGCCAGTGCAATCGATATAAATTAGACAGT | Red |
| VTDO-104 | TGCTGTAGCTCAAACCATTAAGCTATATTTCATCCGTTAATCGGG | Red |
| VTDO-105 | TTCGACAGAAACCAATTATCAGGTTATATAGATTACAGT | Red |
| VTDO-106 | ACGAACTAACAAGAAAAATC | Red |
| VTDO-107 | TTTCGAAAGACCACTACACATTCAAGAAGTTTTGCCCA | Red |
| VTDO-108 | AACCAAACAAAAGGACAGGTATATACCATTGAGATAAAGCTG | Red |
| VTDO-109 | CAAATCATTATTTCTGTAATAAGCATTATGGCATCGATTCCC | Red |
| VTDO-110 | CATGAAAGATCGGATTGAGAAAACGGATAAAAGGCCGGAATGCCG | Red |
| VTDO-111 | TGTAAAAACATT | Red |
| VTDO-112 | GACGACGACAAAAGAATTGAGAAGCCAACGTTAAAAATACCG | Red |
| VTDO-113 | GTGCAAAAGGCTAAAACAAACGGGATAACGCATAGCGAACTGGAT | Red |
| VTDO-114 | CCTCGTTTACGTAAGAGCAA | Red |
| VTDO-115 | TACCAAGAACGGATAACAGAACCATATCATTCCTGCCAGAAG | Red |
| VTDO-116 | AAGGAAACCATTGAGTTAAGCTCCAAGAATAAGTCGAACGCG | Red |
| VTDO-117 | AGATGTGACTTTGGATTATACAATATCTTTGCTGAAATAAAA | Red |
| VTDO-118 | GCGGGGAGGAAACGCAACATA | Red |
| VTDO-119 | GCCAGAGTCTTAGCTATCAAAAGG | Red |
| VTDO-120 | TTCACGTTGAAGGAATTGCG | Red |
| VTDO-121 | GTTCAGCATAAGAGACAACGCATACAAAGGAATCAGACCTAA | Red |
| VTDO-122 | ATGTGTAGGTTTTTAAATGC | Red |
| VTDO-123 | CAGTTTGGGATTCTAGCCAGCTTTTCATCAAGGCATCGAGCTAAGA | Red |
| VTDO-124 | CGGGAAACCTGTCGCCTGGCCCTGGGAGCTAAACATCCATCA | Red |
| VTDO-125 | CAAGAAACAAGAGAGAATAACATCCGGTCTCATCGTAATTTA | Red |
| VTDO-126 | CGTTTTACGCTAACGAGAACCGGATATTATTGTGACTTT | Red |
| VTDO-127 | CCGTAATATCACCAATCACCGTTTAACGGCGCATTCAAA | Red |
| VTDO-128 | AATCCAAGGTTTGATAAGAATGAAGCCTTAGCAAGAGAC | Red |
| VTDO-129 | TACATTTTTTGAATATCCTTGGATAGCTTAACTATTGATGCA | Red |
| VTDO-130 | TGAGAGATTTCAAATACATAT | Red |
| VTDO-131 | AGAGAATAAGATACATACCAGAACAATAAATCCTCATT | Red |
| VTDO-132 | AGAACCGCGGGGTTCAACTAAAAATCTCCCCA | Red |
| VTDO-133 | CTGAGTAGATTTTT | Red |
| VTDO-134 | ATAGCCGTAAGAGCTAATATCCAGGGAATCAAAAACCAGTTA | Red |
| VTDO-135 | GAACAAAACTCGTAACATTTGTGGTAATATACCTAACGAGCACCGCGCT | Red |
| VTDO-136 | ACCCAAAAAAATACCGATTTAACGGTCATACATGGCTTTTGATGA | Red |
| VTDO-137 | ACCAGAACGAAAGGCTTGCC | Red |
| VTDO-138 | GTGAAGAATCAGGTATAGAGGGGGTAAT | Red |
| VTDO-139 | AGATCGCACTGACGACAGTA | Red |
| VTDO-140 | AGTAGCGGTCACGCTAGAATC | Red |
| VTDO-141 | AAAGCCGCGAGAAACCGCTACCTATGGTACGCCAGAGTGTTT | Red |
| VTDO-142 | TATGAGCAAAAGAATATGTGA | Red |
| VTDO-143 | CTTTGCCATTATCAATCAATATTTAACCGTCAATAATGC | Red |
| VTDO-144 | GCTTTCCAGTCGCACTGTGTGCCACAACTAATAGATATTAAT | Red |
| VTDO-145 | GAAGTTTCATTCTAAGAGTACATCGAGG | Red |
| VTDO-146 | TGTGAATACAAGCCTTTTGAATTTGTAACAGAATTTACCGTTCCA | Red |
| VTDO-147 | CTGGTACCTCAA | Red |
| VTDO-148 | ATATAAAGACCGAAAATTCCCTTA | Red |
| VTDO-149 | GTGAGAAAATTTTTGTACCAAATACAGGTTGGGGCGTAGATT | Red |
| VTDO-150 | GGGACTATTTTGCAACAGGAATCCTCGTTGCGCGT | Red |
| VTDO-151 | TTGTCACCGACCAAAGAACCGGAAGCCGCCA | Red |
| VTDO-152 | GGCAGTTTTGTTTAGTATCATGTGAATTTGTAAATTAACGTC | Red |
| VTDO-153 | GCTAATTTTCAATTGCGTCGGGAGCAACAGTCAGAAGATTAGTCT | Red |
| VTDO-154 | AGAAAAATCTTTCCGCAAGCCGATATAGTTGCGGGTTTTATC | Red |
| VTDO-155 | TCAGAGCATAAGCAAAATTA | Red |
| VTDO-156 | CCTTGTGCTGGGCTTAAAAGCAACAAAGGGCCGTGAACCATCACC | Red |
| VTDO-157 | CTAAAACCAATTCTGGCCAACGCTAGGGCCTAAAGGGAGCCC | Red |
| VTDO-158 | CATCAAGGAAACAGATTAATTGAGAAGATCCGGCTACAAAGA | Red |
| VTDO-159 | CAAAGTCTTTGA | Red |
| VTDO-160 | TTGTGGCAGAATCAATACCGCCTGAAACAATTTACAAAAATTAAT | Red |
| VTDO-161 | GCATCGTGGCGGATATTCGCGGAAAAGGACATCCAACCG | Red |
| VTDO-162 | ATTGCGTATGAATCACAGCTGATTGCCCGCGGTCCCTCG | Red |
| VTDO-163 | GGGTGGGTAAGTTGGTGGAATATAAAGAACGAGTTTTTTGGGGTC | Red |
| VTDO-164 | ATTAGTACAGTGAGTAGACAGGAACGGTTGCTTTGCATT | Red |
| VTDO-165 | CAAATCATGGACTCAAAGAATTTTGCCCACGGGCAGGCCAAC | Red |
| VTDO-166 | TACCGCAATTCTAAGAACGTACCTGCTCCCTAAAAATGAAAC | Red |
| VTDO-167 | AACGAGGTTCCATTACTCATCTATCATCTAAGGGAGCATAGG | Red |
| VTDO-168 | GCCTAATGAGTGAGAGAGGCTTTTCAC | Red |
| VTDO-169 | ATCAAACATTAGAAAAGACAAGAGCAAGCTTCTGAATTCGCTCGC | Red |
| VTDO-170 | AATCCAACAGTAGTAGTCTTTTGCTATG | Red |
| VTDO-171 | TTATCCCAATTTTT | Red |
| VTDO-172 | ATTCATGCGCATGTTCTTCT | Red |
| VTDO-173 | AAGTGTTCATTTCAATAAATAACCCCGC | Red |
| VTDO-174 | CACTAAATCGGAACCGCTGGC | Red |
| VTDO-175 | GGGGGGTAATATCAATAGTTAGCACCGCCAGTGGCCTTGATATTC | Red |
| VTDO-176 | TTTTTGAGTAAGAATTTACATACGCTCAACTATCGCACTTGC | Red |
| VTDO-177 | GAGGTGCAGAGTCCTTGAGTGGCAGCAATTCACCGTGCCAGC | Red |
| VTDO-178 | CTAGCTGGAACGGTCCCGGTTGTTAAATAAAAATATGACCGT | Red |
| VTDO-179 | CCCACGCATACCGACAATGA | Red |
| VTDO-180 | GGGGGAACAAAGAGGAACTCAAATTCATATAAAAGATTTTTTGAGAGATCT | Red |
| VTDO-181 | AGCGTCCACAGTTCCATCAAATCAAAGCATTTTTGTAAATAT | Red |
| VTDO-182 | TTAAGACGCAGTATGAAAATTACCAGCGCTTGAGCTTAGAGC | Red |
| VTDO-183 | ATCTGTGCTACG | Red |
| VTDO-184 | ATTTAATTCGCAAGTAGGTTGGATGATGAGGGTTAGAAT | Red |
| VTDO-185 | CGTTTGCATTGAGGCACCACGTAACCCA | Red |
| VTDO-186 | AACATGAGCCTTGACTTAAACGGCTTGCGCATCGG | Red |
| VTDO-187 | ATTAAGCTAAAGATTGTTTGCCTGAGGGTGGACGACTTATGATACCGACAG | Red |
| VTDO-188 | AGAGCGAATAGCTGAATCATTGATAAAG | Red |
| VTDO-189 | GTAAGCGGGTCAGTAAGTATTGTACCAGGTACCGCATTTTCA | Red |
| VTDO-190 | AACCACCAGCGGGCAGAGATAGCGAACTGAACGAA | Red |
| VTDO-191 | ATCGCGCTTAAT | Red |
| VTDO-192 | CGCCACGCCATTAAAAATGAA | Red |
| VTDO-193 | TGCATTATGCGCTCTCACAATTGCCGGAGCGCAACAATTTTTGATA | Red |

## Table S2. DNA sequences of R-TDO

| Strand name | Sequence | Color |
|---|---|---|
| RTDO-linker-1 | ATAATAATAATAATAGCTCTCACGGAAA | Green |
| RTDO-linker-2 | ATAATAATAATAATAGGTAGAAAG | Green |
| RTDO-linker-3 | ATAATAATAATAATATAAACATCCCTTACACTGCGCGC | Green |
| RTDO-linker-4 | ATAATAATAATAATAGACGATAAAAACCTATTCATTGA | Green |
| RTDO-linker-5 | ATAATAATAATAATAAAAGACTTCAAAT | Green |
| RTDO-linker-6 | ATAATAATAATAATATAGCATTAA | Green |
| RTDO-linker-7 | ATAATAATAATAATAAGCGAGTAATT | Green |
| RTDO-linker-8 | ATAATAATAATAATAACTTTAATCAT | Green |
| RTDO-linker-9 | ATAATAATAATAATAGGCGCCAGGGC | Green |
| RTDO-linker-10 | ATAATAATAATAATATCGGTGCGGGCCTGCCAGTGCCA | Green |
| RTDO-linker-11 | ATAATAATAATAATAGTCCAGCAT | Green |
| RTDO-linker-12 | ATAATAATAATAATATGCACTCTGTGGTGAGGATCCCC | Green |
| RTDO-linker-13 | ATAATAATAATAATATAAAACACTCATCTTACTTAGCC | Green |
| RTDO-linker-14 | ATAATAATAATAATAATCGATGAACGGT | Green |
| RTDO-linker-15 | ATAATAATAATAATAGCAACGGCTACAGCAAAAGAATA | Green |
| RTDO-linker-16 | ATAATAATAATAATAACGAGGCGCAGAC | Green |
| RTDO-linker-17 | ATAATAATAATAATATACCGAGCTCGAA | Green |
| RTDO-linker-18 | ATAATAATAATAATACCCCTCAAATGCTAAGAGGAAGC | Green |
| RTDO-linker-19 | ATAATAATAATAATAGAACGAGTATA | Green |
| RTDO-linker-20 | ATAATAATAATAATATGTTAAAATTT | Green |
| RTDO-linker-21 | ATAATAATAATAATAAATCACCATCAATAGCAAACAAG | Green |
| RTDO-linker-22 | ATAATAATAATAATAGTGAGCTAATG | Green |
| RTDO-linker-23 | ATAATAATAATAATACGACGACAG | Green |
| RTDO-linker-24 | ATAATAATAATAATAGCGGCCTTTCA | Green |
| RTDO-linker-25 | ATAATAATAATAATAAACCGGATATT | Green |
| RTDO-linker-26 | ATAATAATAATAATAACGTGCCGGGC | Green |
| RTDO-linker-27 | ATAATAATAATAATAAATCCTGTT | Green |
| RTDO-linker-28 | ATAATAATAATAATATTTCAGAGGTGGATGGTGAAGGG | Green |
| RTDO-linker-29 | ATAATAATAATAATAGGATGGCTTCA | Green |
| RTDO-linker-30 | ATAATAATAATAATAGCGGGAGAAGCCTCCGGAGACAG | Green |
| RTDO-1 | TAATATCAACATCAAAGTGTTTTTATAAGGAGCTAGGCC | Red |
| RTDO-2 | CACCAGAGCCCTCAGACCCCTTATTAGCGACAGTAGCACCATTCATTA | Red |
| RTDO-3 | TGATGGTGTTGCCCCAGTGGTTTTAGGCGGTTCTCACATTCCTGGGGTGC | Red |
| RTDO-4 | TACATCGAGATTTTATTGTTTCATTGGCACATTCTAACAGGAGCGCTTA | Red |
| RTDO-5 | ATCGCGTCTTTTGAATTGCTGAGTCTGGATGATATCAGC | Red |
| RTDO-6 | TAATAACAATGAAAAGATAAAAACAGGG | Red |
| RTDO-7 | ACGTCAGATGAACCGCAATTCACCTGCCAGCACGCGTTGTTATC | Red |
| RTDO-8 | ATTGCGTGGAGAAAGAGCAAAATTCTGTTGTAATTCGGT | Red |
| RTDO-9 | TCCAAAGAAGTTTGGGGAATTAGCAAATCGG | Red |
| RTDO-10 | CAGGCGTTATATTAATTTAGATTAAGATAATCAGTTGGATCACCT | Red |
| RTDO-11 | TTCGCTATTAACAAATTGCGTTAACAGAGCCAAATGAAGAGAATT | Red |
| RTDO-12 | AATATCCAGCGCGTGGTCATCCCAATTACGCAGGCTGCGATCGCA | Red |
| RTDO-13 | CAGGTCCAAAAGGAAAGAGGGGG | Red |
| RTDO-14 | TTGTGTCAAGCGCGAACACCACTGGCTCAGCG | Red |
| RTDO-15 | TAACAGCTACGTCTTTCATAAGAAAAAACTTATATGTA | Red |
| RTDO-16 | AAGAGCAACTTTTGCAAATACTGCAAACGAGAAGCGGATTCGAGCTTC | Red |
| RTDO-17 | GGTCAATTTCATCACCAAATCAAAGATTTTAAACAATAA | Red |
| RTDO-18 | TTTCGGATTCAACGACGCTTCGCTCGCAACCTCATTGCAAAGGTTTCTTTGCTCG | Red |
| RTDO-19 | AATAGCCCCGCGCCTGGCCCTGACATCAGATCACTGTCCGGGCG | Red |
| RTDO-20 | TCATTCCAACAATACGACAATTCATTTCTACCTTTAAGAAACCGTT | Red |
| RTDO-21 | ACCAGTGCGGTCCAGTGTTCAAATGGGTAGGCGCT | Red |
| RTDO-22 | GCAAATAAAACTAGGCCTGAGAATATAAATTCCATGTTC | Red |
| RTDO-23 | AGAGGAATCGATGCGATTAAAACGGAGGACTGCGAAAGACAGCATCGGAACGAGG | Red |
| RTDO-24 | AATGGGGAAGCATCAAAAATAATTAATGCCTAATGTGCAATAAC | Red |
| RTDO-25 | GAACGCAGAAAAGCCCCAAAAACTACCCCGCTACAAATGTTTTA | Red |
| RTDO-26 | AGTAAAATTTCCTCCTGACCTTTGAATGAATTTTACTAA | Red |
| RTDO-27 | AATAATCAGAAAAATAAAGTAAGAAGATCATTTAAAATT | Red |
| RTDO-28 | GCCTTTACTAACGAGCAATTTTATTATAGAAGCCGCGCCCAACCAATC | Red |
| RTDO-29 | TAGGCCCGTATGCCATCCACCGGACCCAATATAGACGGAATAGCA | Red |
| RTDO-30 | AAAGAGGAGGCTTTAACTAACGTTGAGAATAACCCTCGTTTA | Red |
| RTDO-31 | AGTAGAAGAACTCTAACGGTACGCCAGAAAAGGGACCAGTAAACAT | Red |
| RTDO-32 | TTAAATATAATTACTGCGCAAGACAAAG | Red |
| RTDO-33 | GACATTCAACCGCTAAACGTAAGATTTT | Red |
| RTDO-34 | AAGGCTCAGAACGGTGTACAGACACTTTGATGATAAA | Red |
| RTDO-35 | GTTTATCAAGAACGAGCAAGCGACACCATAATAACGTTT | Red |
| RTDO-36 | AGAGGGATAGCAAGAGACGTTAGTTGTATCATCGCCAAGAGGAC | Red |
| RTDO-37 | CTCCAGCCTGGGCGCAGAACAAACTGGCCTTCCAGCTCATTTGTATAA | Red |
| RTDO-38 | ACGAAGGTTTCATGAAAAATCTTCAACTGCATAGT | Red |
| RTDO-39 | AACGCGAGTAAACACCTATCATATAGGCATTTACAAAAGGTAATATCC | Red |
| RTDO-40 | AACGTCAATAATTTGCTTGCTATTCTAAGAACATTTTCATTTTCCTTA | Red |
| RTDO-41 | AAGCTTCTGTTCAATAGCTAATAGAAGGAATCAGCAAA | Red |
| RTDO-42 | AATATGGCCCAACAGGTCAGGATACCAGACATAGTCAAGTGAAT | Red |
| RTDO-43 | AAATTGAGGGAGGGACGTCACAAGTGCCGGGTTTTAATGAATCCAAAAG | Red |
| RTDO-44 | ATACCACATACGTTAATTTAAGAAGAACGAGTCAAAGCTGATAGGCTG | Red |
| RTDO-45 | AAAGCCTTGAGGTAATACTTTAATGCAGGGA | Red |
| RTDO-46 | CATCCAATAATTCTACGATTTAGTTTGATTCCAGAGCTTATAAGAGGTCA | Red |
| RTDO-47 | CGCCATTCCAGCTGGCGTCACGACGGATAACCTGTGAGAGCAGCGGAT | Red |
| RTDO-48 | CAGAGGGTTAATATCAACCAGAAGAAACGCAACGGAATAATCACAATC | Red |
| RTDO-49 | AATAGAAATTATCACAGCAAAGTATCACTATTAAGTTCAGCGCGAATAA | Red |
| RTDO-50 | TGCCTGAGGGAGAGGGTAGCTCAAGTAATAAAAAGCCTGAAGTT | Red |
| RTDO-51 | AATAAGAAACAGCCATTGAAGCCTCGAACCTCCGAGAACAGGTATTAA | Red |
| RTDO-52 | CCAGCGCGGAAATTATTACCAATAAGTAAAGAGAATTGCTAAGAAAATC | Red |
| RTDO-53 | ACCCTCATTTTCATGGCGGATCATACCCTGACTATTCGGAAGCA | Red |
| RTDO-54 | CGCAAACTATCGGCCAGTCACATCAATACCTGATTAAAC | Red |
| RTDO-55 | TTGTACCAGGATAAAAAGATTCAATTCTAGCTAGGTCATTCATGTCAA | Red |
| RTDO-56 | ATGACAACGCTTGATAAGGAATTGGAGTGAGACACAGACATACAAACT | Red |
| RTDO-57 | CTTTTACCTTTCATAAAAAAATAGTGGCATCAAATCATTATGACCCTGTAATACT | Red |
| RTDO-58 | ATGTAGAAATAGCAGAAAATAATTACGCGAAACCAATCA | Red |
| RTDO-59 | TGAGGCTTTGTATCGGAAAAGGCTTTTCTGTAGTCTTTCCCCCAATAG | Red |
| RTDO-60 | CATCCTAATTTAAGAATATAAAAAGCGC | Red |
| RTDO-61 | AAAGCGCATTACCGTTGAACCTATCATGAAAGCGTACTCATACCGCCA | Red |
| RTDO-62 | AAAGAATCCTTGTTTAGGGAATCACCAACGCAGAGAGA | Red |
| RTDO-63 | CTTGATATAGTGTACTGTAACAGTGATTAGCGGTCGAGAGGAGCCACC | Red |
| RTDO-64 | GAGAACGTCAGCAGTGTCACTGGTCGCTGGTGTTCCGAAATCGGC | Red |
| RTDO-65 | AGCCCCCGAGCGAAAGGCTACAGGTCAGAGCGTCAGTGAGTCTTTGAT | Red |
| RTDO-66 | CCTTTTTAATCAAAATGAGTGAATGAAACAGTGCAGAGGCCTTTGAAT | Red |
| RTDO-67 | CAAACTTGTTGTGTATTGCCGTCTTCGCCCGTTTTCATTAATAACAGCT | Red |
| RTDO-68 | ATTCATCAGGAACAACATTGTGAAGAGATGGTTTCATTACAGAGTAATCT | Red |
| RTDO-69 | TATATTTTTAAATAAGCTTACCAGCGCCAACACCAGACGAGATAAGTC | Red |
| RTDO-70 | ACCACCAAAGTGTAAAAAATACTCAAAT | Red |
| RTDO-71 | ATCCGTATTAACTGATAACCACCACGCTAGGGGAACCCTAAAGGG | Red |
| RTDO-72 | ACAGTTGAATTAGAGCTGCCCGAACACCAGAAACTTCTGAATAAAGAA | Red |
| RTDO-73 | AACTGAACAAGAATTGAAGCAGATCCAAAAGATAAAAGAAATGGTTTA | Red |
| RTDO-74 | GATTCGCAACAGAAATAATGGATCGTCTAACCCTTGTTAGAAGCGCGTA | Red |
| RTDO-75 | CACGCTGAAGTATTAACAATATTTGAAAGCGTACATTTTGCAGGAAAA | Red |
| RTDO-76 | AAACGCTGGCCAGCAGTTGGGCGAAATTTCCCAGTCCCCAATAG | Red |
| RTDO-77 | AAGAGACAATCCCGAGGG | Red |
| RTDO-78 | ATCCTCATACATGGCTTTAATGCCGACTCCTCTAGCCCGGACCCTCAG | Red |
| RTDO-79 | GGAAACCACTGCAAGGCGATTAAGAAGTATGTACCCTCAATCAA | Red |
| RTDO-80 | ATAAAAAGCAGAAAATAGACTGTTAAATCTGTAGCTCAA | Red |
| RTDO-81 | TCTGACCTGTTTGAAATCAACAGTTGAGAATCATGTTCAGACGCGCCT | Red |
| RTDO-82 | CCGTCGAGCCAATCAGACCTGAATATCT | Red |
| RTDO-83 | TACAGCCGAAAGATTAGCAGTTACCGTGTGAAGTTAATGCTTAGG | Red |
| RTDO-84 | GAACCCAACCCTCAGGTTGATTTAGCAAAGCACCGTGATTAATTTT | Red |
| RTDO-85 | ATGTTTAGAAAATCAGGTCTTATAGTATGTAATAGCTCTTGCAA | Red |
| RTDO-86 | TAGTTAGCAACAGTAGGCTGACCCTGCCTCATAGCGCCG | Red |
| RTDO-87 | GAGTAAAATACGTAAGTACAACG | Red |
| RTDO-88 | CAATGAATTTACCTCCGTTCATCT | Red |
| RTDO-89 | CCAAACCCACACCCTGATTTGTTT | Red |
| RTDO-90 | GATTGCCCTTCACCTCAGTCGTG | Red |
| RTDO-91 | CAGCGGGGAGCTTACGACTTGTAGCACATCCTAGTGATGATAAAAAAAGC | Red |
| RTDO-92 | AAAATCCCGCAGCAAGAGACGGGCGAATCGGCCACTGCCCGCCGGAAG | Red |
| RTDO-93 | TTCGTAATGTAAAGAATTGCGTTGCGCTCAACGCGCCAG | Red |
| RTDO-94 | ATGCGCCGAGCGGGGCAGAAGCTAAAGCCAAATCA | Red |
| RTDO-95 | TGCTGAACCCGAACGAGCCCTAAATAAAAGGGAGATTCACCTTGCTGG | Red |
| RTDO-96 | TAGTAAATCACAAAGTTGAGAGATCCCTCAGGCCACCAAGAATGG | Red |
| RTDO-97 | CACGTAACTGATTGGAATTATAAACAACGCCATATGTTT | Red |
| RTDO-98 | TACAAGAAACCGGAACCGTTGAGG | Red |
| RTDO-99 | GCTTCTGGTAATAGGTCACGTTGGAACCCTTAAAGCTAAAATTA | Red |
| RTDO-100 | GGGTAACGCGGCGAAACGTACGAATCAGATCTTTACACAGCAGC | Red |
| RTDO-101 | ATTGCTCTTTAATTGCATCAAAACGTAAAGTAAATCCCAGCGTAAAACG | Red |
| RTDO-102 | TTTCCTGCATACGAGCTTTCCAGTCGGGAAACCTGGCACAATTC | Red |
| RTDO-103 | CAGTAGCGTTTAAACAGTTTGATGGCTTGCTATATATTCGGTCGC | Red |
| RTDO-104 | GGGGAAAGTGGCGAGATGCTTTGATAACGTGCGAGTCTGTAATTAACC | Red |
| RTDO-105 | CGGCTGGTGCAAATCGATCAGACGGCGGCGGGGTCCGTGACATAGCTG | Red |
| RTDO-106 | AGCAATAATAAGCTATATTTTCATTTTGCCTTACGAGAATGCAG | Red |
| RTDO-107 | CACACAATGTGAAATGCCTGTTTCCGGC | Red |
| RTDO-108 | CTGAAATGGTTAGGTAAATTTTTAGTGTAGAAGCTTTCCGCCATT | Red |
| RTDO-109 | GTTGTAGATTGCAAACGCTCAAAGGGTTCGGAACATTTA | Red |
| RTDO-110 | TGAGAGATGTTGATAATCCGAGC | Red |
| RTDO-111 | AATCGTATTTAAATAATTTTTTTCTCCGGCCGCCAGCCA | Red |
| RTDO-112 | GCTGACCCATAAGGCGACCTGCTCCATGTTTGACCTGGG | Red |
| RTDO-113 | CAGCGGTGATACCGGGGGTTTGATTTTAGAGCGCGTA | Red |
| RTDO-114 | GACAGGAGAGAGCCACTTTTCATATCGATAGCGGCCGGAAAAGGTAAA | Red |
| RTDO-115 | CCTCAGCAAAAGACTTCACCAACCATTATACCGAAATCCGGAACCGAA | Red |
| RTDO-116 | AAGCGCATATAAGAGCCGAAGCCCGACTCCTTCATACATAAAAAGGGC | Red |
| RTDO-117 | GTTAAAGGCCGCTTTTTCCATTA | Red |
| RTDO-118 | TACCTTTGCGGTCAGAGATAGGGTTGGC | Red |
| RTDO-119 | ATGAACCTTGCCAACGCTACCGACAAAATAAACGATTTACAAAGT | Red |
| RTDO-120 | AACTGCGCGAAATCCTTCGTCAATAGACGCTTATAACTTTTCAAA | Red |
| RTDO-121 | CATAAAGTCATGGTGCCTCCTAGTTAAAGCCTCCGCGGG | Red |
| RTDO-122 | GAGATTAGCTCAGTGGTCAGTTCGTAAA | Red |
| RTDO-123 | AGGCAAAGCGCGAGCTCATTTCGCGAAGTTTCTGCTGTAGTACCTTTA | Red |
| RTDO-124 | TCAGAGCCAACCGCCACATCGGCATTTAGCGTATTAGAGCCCGTCACC | Red |
| RTDO-125 | CAGTCAGGACTGCCCTGACGAGAAAACAAATGCCACT | Red |
| RTDO-126 | TCATATGAGGAAGATTTTTAACGGAATTTCACCGGGTCTCGTGGTC | Red |
| RTDO-127 | CTGTTTGCTACAACTAAAGTACGTAAATCAACTGGATATTATAC | Red |
| RTDO-128 | CTGACCACAGGCGCCTCATTCGAAGCAAATGACCAGTGT | Red |
| RTDO-129 | CCCTCAGTCACCAGGCCCTCA | Red |
| RTDO-130 | TTCGCACTGGTGGTGCGCTGGTCTCGCTGGCACGATGCTGACATCGAC | Red |
| RTDO-131 | CCAGCTGCACGGTCCCGGTGCAGCAACC | Red |
| RTDO-132 | GCAAGAGAGAGGTCCGTTTTTTCAAACAATCCAGGGTTGACCGT | Red |
| RTDO-133 | AATGCTGACGATAGCTTTCCCTTAATTAATTAGATGAAACGTACCTTT | Red |
| RTDO-134 | TAATTTTGAATTTCGCGTCATTAAAGCCGAACCAC | Red |
| RTDO-135 | AGCCCCCTGCTTAACGGGAGAGTTTTATAAATCAAAAG | Red |
| RTDO-136 | CATAACCGTTCGAGGTTTCACGTTACAACTTTCGTAACGATAACACTG | Red |
| RTDO-137 | ACTAAATCGCGCTGGCCACCCGCCGGCCGATTATCCTGAGCTTGCCTG | Red |
| RTDO-138 | TATCGGCCCTGCCAGTCAACCCGTCATCAACATCGCATTATGTAAACGTT | Red |
| RTDO-139 | TAGTAATCAGAACATTATTTAGGATTATGGAGCGGCAAT | Red |
| RTDO-140 | TCCAAAATTTATCAATACAGGTCACAAACAGCATT | Red |
| RTDO-141 | AAGGTGAAATTCATACGCAAACGTTTTTGCGAGGCTTAACAATATA | Red |
| RTDO-142 | AAAGCGATAGAGAGCTCAACAGGCTATCGATAAATTCGC | Red |
| RTDO-143 | GGTTATCTTAGGAGCAAAAGTTTGTCATTTTGAGAACCTAATTATTTG | Red |
| RTDO-144 | CGCAACAATTGGATTTAAACATAGTGCAAATCCAATCC | Red |
| RTDO-145 | CAGTTGACCAAGAAAGGTTATTTCCGTGCATTCAGGAAGCAACTGTTGGGAAGG G | Red |
| RTDO-146 | AACGTTGTAATCCGTGGTCGTAACAACGCAAAAAACATACAGGCA | Red |
| RTDO-147 | CATGTTTATGCTCATTTGTATACAGTAACAAAACATCAGTA | Red |
| RTDO-148 | GCCTTTTCGGTATTTCGCCAGTAATTAAACAAACCATCGCCCACG | Red |
| RTDO-149 | ATTGCTATTAGAGGTGAGGAAGAAATTTAGAGCTTGAC | Red |
| RTDO-150 | TTGGGTTAGAGAAGAGAAATCGTCATTTGAATAATTACCTCAATAACG | Red |
| RTDO-151 | ATAACCCTCAGAGCACATATATT | Red |
| RTDO-152 | ATACATAAACGGACGTTGGGAAGAGGAAGTTGCGGGATCGTCAC | Red |
| RTDO-153 | CTATGGTAAGGAAGGGCGGTCGAGCCAGTGAGGAA | Red |
| RTDO-154 | TAACACAGTTGCTGCGGCGGGGAGTCTTTTC | Red |
| RTDO-155 | ATTTTGCACCGAAAAGTAGTTAAGACCGCCTCCGCCGCCAAATAA | Red |
| RTDO-156 | TATTGACCAAAGACAAGGTGGATCATTAGCTTATCTAGG | Red |
| RTDO-157 | TGAAAAATATAAAACAGTCTTTAAAGAGATAGGAAATGGAATATTACC | Red |
| RTDO-158 | AACCGCCTGTACCGTCTAAAGTTTTGTCTGGGATTGGAT | Red |
| RTDO-159 | GCCAGCCCAATACTGCCACCG | Red |
| RTDO-160 | CTGAACAGGCTGTCCGTAGGACAACATAACTGGCATAAT | Red |
| RTDO-161 | AGTTTCGAACCGCCAATAGGTATCACCAGAATCAAGGAA | Red |
| RTDO-162 | TATCTGGTACATTTGACGACAACTATCATATTTAATCCTGCAGGTTTA | Red |
| RTDO-163 | CCGAAGGGTGTTAGATAGAAAAGGCGAGAGGCACTATCTTTAGGA | Red |
| RTDO-164 | ATCAAATGAGCCGGGTGCCGGGT | Red |
| RTDO-165 | GTCGGCGGATTTTCCCAGAAAGGGCACCGTCCAATCCGTGCCCTG | Red |
| RTDO-166 | CGGTTGCGTAATGCCAACGGCAGGGATGTGGGCAAAGCGGCACC | Red |
| RTDO-167 | AACGGGCAGTTTTACGCAGACGATTGGC | Red |

## Table S3. DNA sequences of L-TDO

| Strand name | Sequence | Color |
|---|---|---|
| LTDO-linker-1 | ATAATAATAATAATAAGATACATTTTAACATCCTTTTT | Green |
| LTDO-linker-2 | ATAATAATAATAATATGCCTGTCATTTTTT | Green |
| LTDO-linker-3 | ATAATAATAATAATAAGGCCGAGAATTTTT | Green |
| LTDO-linker-4 | ATAATAATAATAATATCGCGTCTGGTAATGGGATTTTT | Green |
| LTDO-linker-5 | ATAATAATAATAATACAGGGAGTTAAAGACTTTTTTTT | Green |
| LTDO-linker-6 | ATAATAATAATAATATTGCGGGAG | Green |
| LTDO-linker-7 | ATAATAATAATAATAAATAAATCAGTAATACTTTTTTT | Green |
| LTDO-linker-8 | ATAATAATAATAATATAGGTCACGAGCTTTCCGTTTTT | Green |
| LTDO-linker-9 | ATAATAATAATAATAGGTCTAATCATTTTT | Green |
| LTDO-linker-10 | ATAATAATAATAATAGGCGCAACGATTTTT | Green |
| LTDO-linker-11 | ATAATAATAATAATATTCATGAGGCATCTTTGATTTTT | Green |
| LTDO-linker-12 | ATAATAATAATAATATAAGAGCAAAGTAAAATGTTTTT | Green |
| LTDO-linker-13 | ATAATAATAATAATAGCACCGCTT | Green |
| LTDO-linker-14 | ATAATAATAATAATATGTAGAATGCTTTTT | Green |
| LTDO-linker-15 | ATAATAATAATAATAGAGGACTCTATTTTT | Green |
| LTDO-linker-16 | ATAATAATAATAATATTTAGACTG | Green |
| LTDO-linker-17 | ATAATAATAATAATAGCGGATTTCATTTTT | Green |
| LTDO-linker-18 | ATAATAATAATAATACAGACCGAACTTTTT | Green |
| LTDO-linker-19 | ATAATAATAATAATAAAAGGTGGCGTTTTT | Green |
| LTDO-linker-20 | ATAATAATAATAATACAACGCGCGCCCTGAGAGTTTTT | Green |
| LTDO-linker-21 | ATAATAATAATAATAGGTTTGAGATTTTTT | Green |
| LTDO-linker-22 | ATAATAATAATAATAGACAAATCTTTTTTT | Green |
| LTDO-linker-23 | ATAATAATAATAATACAGCTTTTATTTTTT | Green |
| LTDO-linker-24 | ATAATAATAATAATAGATAGGGTT | Green |
| LTDO-linker-25 | ATAATAATAATAATAAGTTGCAGCATAGCCCGATTTTT | Green |
| LTDO-linker-26 | ATAATAATAATAATAAACGGAACAGAGGCATAGTTTTT | Green |
| LTDO-linker-27 | ATAATAATAATAATACCCCCAGCG | Green |
| LTDO-linker-28 | ATAATAATAATAATACCACAGCATTTTTTT | Green |
| LTDO-linker-29 | ATAATAATAATAATAGTAAAAAAGTTTTTT | Green |
| LTDO-linker-30 | ATAATAATAATAATATGTATAAGATTTTTT | Green |
| LTDO-1 | TAAATCAGGGAGGTAACCCCGCGCTAGACAGCCAGA | Red |
| LTDO-2 | ATAAATAGAAAATCGCGCACCTTTAAACAGTTTGGA | Red |
| LTDO-3 | TTTTTAACGATAAGATTTTT | Red |
| LTDO-4 | TTTTTCCTAAGGAACTTTTT | Red |
| LTDO-5 | TTTTTGGTACGGAACTTTTT | Red |
| LTDO-6 | GAACCACTGAAACAACAGGCTCATGCCAGCTTAACC | Red |
| LTDO-7 | TCGATAAATCAGGGAACCCCGGAAGCAGGTGCCTTG | Red |
| LTDO-8 | CCCTGAATTTACCGCTGCCTATTGTTGAAAATCTGTTGCG | Red |
| LTDO-9 | GAATAAAGATGACATTTCGATGAACGTAGACATCAA | Red |
| LTDO-10 | GGCGAATAAGAATCACCTAGCGACCGCCACCCTCAGAGCCACCAC | Red |
| LTDO-11 | AGTAACAGAGGCTGCCTCAGATAATCAATTTATTTAGGG | Red |
| LTDO-12 | AAGCCTGACCCTTACAGGGTAGCATCGCAATTTAGT | Red |
| LTDO-13 | CCAGAAGTCATCGGCCACCAGTAGCACC | Red |
| LTDO-14 | TTTTTTGAATTACC | Red |
| LTDO-15 | TTGACCATTTTTTT | Red |
| LTDO-16 | TGTAATTCGTTATACCGGAATGCACGTATACATCGGTCAATACTAA | Red |
| LTDO-17 | CCCGCTGACCACGGGTAAAAGAATGTTAGCGTACAAACTACAACG | Red |
| LTDO-18 | GGTGTACTCATTCAGTGTATAAAAACTTCCAGATTAGAGG | Red |
| LTDO-19 | GTTGTACTGTTTCCTGTAGGTATAACAAAGCGA | Red |
| LTDO-20 | GCTGCGCCCCCCGATTTTTGATAGCCCTCCACGCT | Red |
| LTDO-21 | TAACCCAGATAGCCACGCAATGGTTGAGCCGCCTCAGACTCCTGCTCAG | Red |
| LTDO-22 | AAGCCCACTTTTTACCAAAAGCGCCGCCACCCTCAATGAAAGAGTGCCG | Red |
| LTDO-23 | TTTTTTTTTAACTTGAATCAATAAATCAAAAATGGA | Red |
| LTDO-24 | TCATCGCACACCAACCTAATTTTGTCACACTGAGTTTCGT | Red |
| LTDO-25 | AATCAACGCTGCCAAAGGCCAGCA | Red |
| LTDO-26 | ATTATAAACACTAAGTTTAGGACTAAAGGCGGTCGC | Red |
| LTDO-27 | TACCAGGTTAGTACAGAATCACACCAATACCGTCA | Red |
| LTDO-28 | AAAAGAGTCTGTTCTTTCCTCGTTAGAACGAGCACGAAAAACAGAT | Red |
| LTDO-29 | CCTGAGTAGTTTTT | Red |
| LTDO-30 | TGATGATCAAACAAGACAGGAAATAACGTACATAAGTCA | Red |
| LTDO-31 | CGACTTGCATGCGCATCAGGAAGATCACCATAGGTAAAAACGCAA | Red |
| LTDO-32 | TTTTTAGTAATAAGGAACAAGAATTTTT | Red |
| LTDO-33 | ATCAAATTACCCTCAAAGCGGAAGGAATATCTCAAC | Red |
| LTDO-34 | TAACCGAAATTTCTGGAGCCTTTAATTGATAGAAAGGCT | Red |
| LTDO-35 | CATATTTGTATAAAGCGTTAAGAAATTGTATACAGAGGA | Red |
| LTDO-36 | GAGGCGTCTTACTTGCCAAGAGAGGGAGAAGAGCAA | Red |
| LTDO-37 | GCTCAACAGTAGATCCGACCGGTATGTC | Red |
| LTDO-38 | ATGTTTTGAGTAGAATGGTCACCCCAAACTGGAGCCGGC | Red |
| LTDO-39 | GGATAGATGGGCCTGCATGCTGCAATAAATCTGTTCCAGTTTGGA | Red |
| LTDO-40 | TTTTTAGACGGATTATTTTT | Red |
| LTDO-41 | CCGACTCGATTGTGTCACAATACAGAATACAGAAAA | Red |
| LTDO-42 | TAGAAGAGAACGCGGCTGTCAGCTAGACGAAATCGC | Red |
| LTDO-43 | ACTCACACAGTCGGCCAGGGTGGTTTTTCTTTGCACTGCC | Red |
| LTDO-44 | GCACAAAGGCAGTCAAATCGCACTCCGGAAACTATTAC | Red |
| LTDO-45 | GATTAATTTTTAACTAAAATCGTCGCTA | Red |
| LTDO-46 | CAGTCCCAGGAAGGGACTCCAACGTCCA | Red |
| LTDO-47 | TTTTTGCGAGAGAACTTTTT | Red |
| LTDO-48 | ATTAGTAGAAGTGTGATTAAA | Red |
| LTDO-49 | ATCCTTATTTATCTGATGTTAGTTTAAACACAAATT | Red |
| LTDO-50 | CGGTACTCAAATGCAAAAA | Red |
| LTDO-51 | AGGCATAACCTCGAGCTTGACTATACATTATTTATTCGATTCAA | Red |
| LTDO-52 | ACAAAGATCTGAATATTATTTCATAATTTTTTCAAATAG | Red |
| LTDO-53 | AACGTGGAGATGGTGGTTCGCCAGGGGAAGGGCCATTCAG | Red |
| LTDO-54 | ACAAGAGGCAAACCCCAGACAGCTTGGGCGGAAACC | Red |
| LTDO-55 | AAGGAGCGCGAGAATTCTGACCCAGCAG | Red |
| LTDO-56 | AATCGTAAAGAAACCAGAAGATACATTCATGTTGGG | Red |
| LTDO-57 | TTGCCATTAATTCATCATCAATCACAGAGGGGTCCAATACCATAA | Red |
| LTDO-58 | GAAAACAGAAGCATCGCGTCAGGATGGCTTCAACTA | Red |
| LTDO-59 | AATGACCAGTGACAATATTTAGAGGTGCCGTAAAGCAC | Red |
| LTDO-60 | AAGTACGAGGCAAAATACCAACGGCGTCACCACGCA | Red |
| LTDO-61 | AAGGTTTCGCATCGGAACGGAATAATGAATTTT | Red |
| LTDO-62 | ATCGGCCAGTGGGTAACCGAAATCGTCCACTATTAAAG | Red |
| LTDO-63 | TTTTTTTTTCATCG | Red |
| LTDO-64 | AATAGCTACCGCTGTAGACTGTTTAAGGTACCAACA | Red |
| LTDO-65 | AGCCCATAGGTGGGTTGAATTATTTGCCCCTTCCAG | Red |
| LTDO-66 | AAGCGCGGCCTCGTAACCCCGTGGCATCAAATTTTT | Red |
| LTDO-67 | CCTCATTCATACATGTATAAACAGTTAACTGAAACGAAC | Red |
| LTDO-68 | GCTGCGGTGGGGACGACGAAACCGTTGCAATGCTAGAACC | Red |
| LTDO-69 | AACTTTAATCTACGTTACAGGAATTGCTCAAACTCAATAGTACAAG | Red |
| LTDO-70 | TTTTTAAGGAAACCTAAAAGAAATTTTT | Red |
| LTDO-71 | AAACGAACTTTTTT | Red |
| LTDO-72 | TTTTTCCTGTAAAAGTTTTT | Red |
| LTDO-73 | AAGCCTTTAGTTTCCAAATTTTTTTTGAGCGAGAGA | Red |
| LTDO-74 | AGCCGGAGACGGAGAGTAGAACCGGGGCTTAATTTC | Red |
| LTDO-75 | AGTTACCAGTTTTT | Red |
| LTDO-76 | ATCGTTAGTAAATAATTTT | Red |
| LTDO-77 | AAAGCGCAGTCTGAAGCCGCCCTAATAAGGCTTGTAAGAA | Red |
| LTDO-78 | GTTGTAGGAGGCCACGGGAGCTAAACAGGCGTACTCCTA | Red |
| LTDO-79 | AAAAATTACAAATGGAATAGCGAT | Red |
| LTDO-80 | CGAAGCCATAATAATTAACTGTCTGTCCAATGCAGTCTTCTGACTA | Red |
| LTDO-81 | CGCTAGCGTTAGAAAATGGGCGACTCCTGAATTTTAGCTCAGATA | Red |
| LTDO-82 | CTGGCAAGGGGAAAAATACGTTAAAAATTAACACC | Red |
| LTDO-83 | TCCTACAAAACAAAATTGCATCACAACAGTGAAAACAT | Red |
| LTDO-84 | ATTAGAACAAAATCATATGTATGTATGATTTCTTAC | Red |
| LTDO-85 | TTTTTAACTACTAACTTTTT | Red |
| LTDO-86 | TTTTTTAATGCGCG | Red |
| LTDO-87 | CTTAGCAATAGCTAAAGACTCACCAGAACAGAGCCGGAACCTTATAAGT | Red |
| LTDO-88 | TTTTTCGCAAAGACGACGGAAATTTTTT | Red |
| LTDO-89 | GATAGCGGTAATCACTATAATTACACATTATTAATA | Red |
| LTDO-90 | ATTACCCTGTAGCCTTATCACCCTCCACCAGAATGG | Red |
| LTDO-91 | ATATTCAACAGGAGTGCCTTG | Red |
| LTDO-92 | AAGGGTGGAGACTATCAGGAGAGTAACAGGAAGCAA | Red |
| LTDO-93 | GAGTGTAAAAGAAAGCGGGCCTGGGGGAGACATTAA | Red |
| LTDO-94 | ATGTGTCAATATTTTGAGAGAATCATAATCAACGTT | Red |
| LTDO-95 | CATAGGGACATGAGGAAGAACCTCAACATCACAATATACCTTAGA | Red |
| LTDO-96 | TTTTTCAAGATAAATTTTTT | Red |
| LTDO-97 | TAACCAAAATTGTAAGAAAAGATAACCTTTCTACTCAAC | Red |
| LTDO-98 | GAAACAATGAAAGGAAGCGCAAATTTGA | Red |
| LTDO-99 | CCCATGTACCGTAGTCTTT | Red |
| LTDO-100 | TGGTTTGATCCCTTAGGCGATCTCTTCGCCAGGCA | Red |
| LTDO-101 | TTTTTCTCAGAACC | Red |
| LTDO-102 | TTTTTGATTGCAGACTTTTT | Red |
| LTDO-103 | AAATCTGATAAATAAGAGG | Red |
| LTDO-104 | AAACGTAGTTTGTTTTCAGCCGCCAGCATTATAAAT | Red |
| LTDO-105 | AAAACCAAGTAATCGCAAGAGTCAGAGAACATCATTACGGTTTTG | Red |
| LTDO-106 | ACAGAGCAACGAGCCCTTCTAAGAACGC | Red |
| LTDO-107 | TGAATCGGCTTTTT | Red |
| LTDO-108 | CCTGTAGACAGCCAACAGGTGAGATATCGGTGCTTT | Red |
| LTDO-109 | TTTTTACCTACCATCGGATTCGCTTTTT | Red |
| LTDO-110 | TGCCGACGATTTCAAATAAAGCGGCTAAATCAAATTTTCTGAGTA | Red |
| LTDO-111 | TTTTTTTTTAACGGGAAGGATTATTTTT | Red |
| LTDO-112 | AGCAGCTATTGATATTCCAGTATCCATTCGCGATCGGT | Red |
| LTDO-113 | AAATCCTGTTTAGGTTCACCAGTGAGGTCATAGAAACGACTGCCAGT | Red |
| LTDO-114 | GTAAGCACAAGAATAAGTCAGCCGACAAATCAACAATATATTCAAA | Red |
| LTDO-115 | TTTTTACCCGACCACTTTTT | Red |
| LTDO-116 | AGTAGAGGCGGTAAATGCAAAATCAACCAAGAAGCAAAGAACCTC | Red |
| LTDO-117 | GCCACCCTCAGAGGGTTTTCAAGAGGTCAGTGTACT | Red |
| LTDO-118 | AATCATATTCGCATATGATGGCAATTTTTTCAGTGTG | Red |
| LTDO-119 | AGCGGAACTGATTGAAATAAAATAAGAAAATTTCAAACG | Red |
| LTDO-120 | GAATCCGTTAGCGAGAGGCACTTTGATGTGTCGATTTGTA | Red |
| LTDO-121 | CCTGATTATCAGTATTCGACAACGTGAAATTGTTCGTTGC | Red |
| LTDO-122 | TTCACTCACCACCCTTCGGTCCATGAAC | Red |
| LTDO-123 | ACCTTATTCAGGACCAGTTGATTGCGGATTAGAGAAAGGTGGATTA | Red |
| LTDO-124 | TATAATTATTTGAAACAAAATATCTCAGTATACCGAAC | Red |
| LTDO-125 | AGGTTTTAATCTCGTTTGTTTTGCTAAGGGAGCTCCATCGAAACA | Red |
| LTDO-126 | GGGATTTTTAATGCCGTCTGAGTCACACATCTTTAATCT | Red |
| LTDO-127 | GTCGAGCTTGACGTGTAGCAGGGCGAGGCCTTTTAT | Red |
| LTDO-128 | GAGTAAAATTTTTGTTAAATGTTAAAATGTACCTTCATTT | Red |
| LTDO-129 | TTTCGAGCCTTTTT | Red |
| LTDO-130 | TCAAATTACCGTTGGGTACTACCTCATTCCAGCTTATC | Red |
| LTDO-131 | GCCAGCGGGATGGGTCGATCCCCGAAGCATGCCTGG | Red |
| LTDO-132 | TTTTTAAATAATATCCGTTTTTATTTTT | Red |
| LTDO-133 | TGCCGGGTAAAACTAGCTTTTTACAAGTTGGCAACACGAC | Red |
| LTDO-134 | ATACTAATTTATCCCAAGCTATTTGAAGGTAAATTATCGAAACCA | Red |
| LTDO-135 | AATTGAGCGACAATTTAGACGAATAACATTACGCAGGTT | Red |
| LTDO-136 | CGCTTTCTTAATTGATCCGCTTGTGAGC | Red |
| LTDO-137 | AAGTACGTGATTCCCTATATTCCGGTTGGATGAACGGATTCTGTGC | Red |
| LTDO-138 | AAGATACTGTTGGTTTTCC | Red |
| LTDO-139 | GGTAATAAGTTTTT | Red |
| LTDO-140 | CGAGGTGTATATTCCGCTTTTGTAAATTGATATTCCAAA | Red |
| LTDO-141 | TTTTTAAGAACTCATTATTTACATTTTT | Red |
| LTDO-142 | TTTTTTATCAGCTAATTTTT | Red |
| LTDO-143 | TGAGGCTTGTTTTT | Red |
| LTDO-144 | GGTGCCTCCAGCTGGGCGGTTTGCGTATGATTGCCAGCT | Red |
| LTDO-145 | GCTTGATCATCGCCCCTCAGCGAGAAACACGTAACCTAA | Red |
| LTDO-146 | TTTTTTTGGCAGATATTAGTCTTTTTTT | Red |
| LTDO-147 | AAGAGATCTAAAGAATTACCAAAATATAGTCGAGAATGACTGCGG | Red |
| LTDO-148 | AAAACTAGACCACATTCAAAAAGCTGCAGACCACACTACG | Red |
| LTDO-149 | AGCGAGCCGGGGTACCGCTTCACCTCCACGC | Red |
| LTDO-150 | TAAGCGTAAAGCCACCAGAGCAACTGGCTAGCAAACCTTTACGTTA | Red |
| LTDO-151 | TTTTTCATTTATTATTTTTT | Red |
| LTDO-152 | AGGGTTAGATTTTT | Red |
| LTDO-153 | GGATAAGGTTGTAGCAAACATCAAGTTTAGCAATTC | Red |
| LTDO-154 | AAAATTATTTACGAGCAACTCATCATAGTGAGAAAACAACAGTAC | Red |
| LTDO-155 | TTATACTAACCACCATTTTAAGCTGGTAGCTCAATGCCGCTACGGTCAC | Red |
| LTDO-156 | TAAGAGGAAGTTTCGGCTT | Red |
| LTDO-157 | AGGCGGAAACCCAGTTGAAGAAGTTTAAATCATATT | Red |
| LTDO-158 | TCATATGTAGGCAGTAAAGTAAGGGTAAGTTTAACAGGT | Red |
| LTDO-159 | TGTCGTGAATGAGTAACATACCAGCTTTGAACAAAAAAC | Red |
| LTDO-160 | CCTCCCAACGCGAGCTGAAGTACCTTGAAAGACAAAAACC | Red |
| LTDO-161 | AACTGAATGGCTTCACCAAATGGAAACTATCACTTG | Red |
| LTDO-162 | TATAATCTTATCATCCTTTGCTATTACCGGAAATAATGGTTGTAGGGCG | Red |
| LTDO-163 | CATCTGACCATTTTGCACATAAATCAGTTCA | Red |
| LTDO-164 | GACAAGCAATGACAACAACACCGATACCAAAAA | Red |
| LTDO-165 | CAACTTTCCTCATAACACTAACCAAGCGGTTACTT | Red |
| LTDO-166 | TGCGAACAAATATGAGAGCTTTAGAAAGATAACGCATTA | Red |
| LTDO-167 | CTTACCAAACAACGAAGTAATAACACCCATAGCAGCGTA | Red |
| LTDO-168 | TTTTTCACCAAGAGCTTTTT | Red |
| LTDO-169 | AAGGTGAATATTACCACGAACATAGAGGAAGAACAA | Red |
| LTDO-170 | GCCTGCCTTGCTGGTTATGATAATTTATTAAGAAGG | Red |
| LTDO-171 | TTTTTCTGATTGCTAATTTCATTTTTTT | Red |
| LTDO-172 | TTGACAACAACAACCCGTCGGTAATCAGAGGGTAATAAAG | Red |
| LTDO-173 | AAAGGCTGAATTGCAGGGTAGGTAATGCGGCG | Red |
| LTDO-174 | TTAAATCTAGCTGATAAAGATATCTGCAAAAACAGAGGTG | Red |
| LTDO-175 | AAGCCCTAATTGCTCCTACGGTTTGAGGCTTAGTGATTAA | Red |
| LTDO-176 | CAACAATAATGGTATTAATAGGTCAATTTTCTGTGAGT | Red |
| LTDO-177 | ATATTTATAGGAACCTGT | Red |
| LTDO-178 | AGGTGCACCCTGCGGGACGCGCCC | Red |
| LTDO-179 | AATATTTTCAGCTCCATTAAACACAATTAATCATGACGGGCACAGGCGA | Red |
| LTDO-180 | AAAAATAATTTTTT | Red |
| LTDO-181 | CGCCATGGCACAAATAAATTTGACATATCCTCTTTG | Red |
| LTDO-182 | GCGGGCTAAGTTGCCAAGATTCGTCCACACGAGCTA | Red |
| LTDO-183 | GAGAGCTGAAAAGGAGCAATAGATAGTAACTGCGGA | Red |
| LTDO-184 | CTCATACTAGAGCATAAAGATTGCATCTTTAAAATTCATT | Red |
| LTDO-185 | CGACCTACCGAAAGGCTGAAATCAACCAGATGAATT | Red |
| LTDO-186 | TTTTTACCGTGCAGCTTTTT | Red |
| LTDO-187 | CCTCATGGAGGTCGGATATATTAAGTGCCCGGCTTT | Red |
| LTDO-188 | GAAAATCAATTGCCATCCCTTTAGGTACTCATTTCAGGGATAGCA | Red |
| LTDO-189 | AGCTTACTGAGAAGACAAAAAACTACTAGATTAGTA | Red |
| LTDO-190 | CACCAGTAACGAGGATTTCTAAAGCCAAAATAAACA | Red |
| LTDO-191 | CCGACTAGCTACAGCCATATGAAATGAACATGAGTT | Red |
| LTDO-192 | TTTTTGGATTAGCGACCGCCACCTTTTT | Red |
| LTDO-193 | CCCAAAAACGTGGGGCGCCTTTGATCAGAGCCGAGT | Red |
| LTDO-194 | GAATGGCTCATTATACCAGGCGATTTCCCTGACAGCGAAA | Red |
| LTDO-195 | TTAATTTGAGAGTATATAACCTAAAATAAGGCCAAC | Red |
| LTDO-196 | AGAAATCAACTCAATCACAAAAGACCTTGCTTCTGTTT | Red |
| LTDO-197 | CTGTATGTCTAAAGAACGAAAAACGGAGAAATCCG | Red |
| LTDO-198 | GCCCATCACGCAAACATTGCAACTCGTAATTAGACAACG | Red |
| LTDO-199 | TTTTTGCAAACAGCATTTTT | Red |
| LTDO-200 | AAGAAAAATCATTGACGAGTAGCGGGATCTACAGAGGAACAATGCTAAA | Red |
| LTDO-201 | TTTTTTATTCATTA | Red |
| LTDO-202 | TCGAGAGTATCACCCGTCAGAATTAGCATTTGGGA | Red |
| LTDO-203 | TTTAAGGAATTTCTGGCGCGTAAGGCCGGCGTCAAGTTTTTTGGG | Red |
| LTDO-204 | TACATAGCCCCGCGTTTCCCGGAAATAGGAA | Red |
| LTDO-205 | GGGGTATCCATATAACAGTGTGTCTGGTCATTTGATTTAG | Red |
| LTDO-206 | AATCAGTCAATACTAGAACAACCGAACGACATTTGTAAC | Red |
| LTDO-207 | ATAATTTAATAACATGTTCTTTCCGCCT | Red |
| LTDO-208 | TAGGAAAGCAAGCCCATCAGTCCTAGAATAAGGCAT | Red |
| LTDO-209 | ATCCTGAATAACATCGGCCTTAAGTTTGTAGAGCCGGAG | Red |
| LTDO-210 | CGCAACCAATAATAAACAATTTTAATTCAACTGAGCCAAGGCCGG | Red |
| LTDO-211 | CTGGTGCCAGCCTTGGTGTTGACCGCCTTCGCCATC | Red |

## Table S4. DNA sequences of P-TDO

| Strand name | Sequence | Color |
|---|---|---|
| PTDO-linker-1 | ATAATAATAATAATACGTTAGTAACTTTCGTACTCAGTCGAGACCAG | Green |
| PTDO-linker-2 | ATAATAATAATAATAGTAGCAACGTTTTT | Green |
| PTDO-linker-3 | ATAATAATAATAATACCTAAAACGTTTTT | Green |
| PTDO-linker-4 | ATAATAATAATAATATTTATCAGCTTGCTAGCCTTTCAA | Green |
| PTDO-linker-5 | ATAATAATAATAATAAAAGGAAAATTGTATCTTTTT | Green |
| PTDO-linker-6 | ATAATAATAATAATATGAATTT | Green |
| PTDO-linker-7 | ATAATAATAATAATAGCGAATAAAGGCTCAGAGAAGCTATTATAACGGGG | Green |
| PTDO-linker-8 | ATAATAATAATAATATTATCGT | Green |
| PTDO-linker-9 | ATAATAATAATAATATGACAAC | Green |
| PTDO-linker-10 | ATAATAATAATAATAAAGTTTTAGTGAGAATTTTTT | Green |
| PTDO-linker-11 | ATAATAATAATAATATATAAACGGGTAAATACAAAC | Green |
| PTDO-linker-12 | ATAATAATAATAATATACCGAT | Green |
| PTDO-linker-13 | ATAATAATAATAATAATGCCACAGACTTTGTTACTTTTGAAAGTTG | Green |
| PTDO-linker-14 | ATAATAATAATAATAAAAGACAAGGGAGTCAGGCGCTTACCCA | Green |
| PTDO-linker-15 | ATAATAATAATAATACATAACCGATTTTT | Green |
| PTDO-1 | CATATTCTTAAAAGATCCTTTCAATAGACTAACAACTGGTCA | Red |
| PTDO-2 | AATGCGATTATTTATGCGGTGCCGGTGCTGAAGGGTTCTCATTTGCCGC | Red |
| PTDO-3 | TTTTTTCAAAAAGACAATAAAGCCTCTTTTT | Red |
| PTDO-4 | CGCATCGCCATTTGAGGACAACTCATACAGTAAA | Red |
| PTDO-5 | TGCACCAATACCAGACCGGAAAGACCAAAGAACAAAAAC | Red |
| PTDO-6 | TTTTTAAACAAATAGATGATACATTTTT | Red |
| PTDO-7 | TTTTTCACGCTGAGAGCCAGCAGCAATCAACTATCTTTATTT | Red |
| PTDO-8 | CAGCAGTGCGGCCTGGTCCGTACGTGCCATGCCAAGTCCAGC | Red |
| PTDO-9 | TGCCCAATCCAAGAGAGAGAGAGAATTGATAACCCACAAGACAATGAAACC | Red |
| PTDO-10 | CAATAGTAATGTATAAGCAAAAACCGTTGTGTAGGATCCAAT | Red |
| PTDO-11 | GAGGGTCAGACCGACAGAGACAGGAGCAGTCTGGCGGATATT | Red |
| PTDO-12 | AGCAAACGTAGAGAAACGGAGCCTAACAATTTGAACCTC | Red |
| PTDO-13 | TTTTTCCGGAATAGGTGCCACCCTCATTTTTTTT | Red |
| PTDO-14 | TTTTTTATATTCGGCTTCATCAAGAGTTTTT | Red |
| PTDO-15 | TGGATGCTTTAGAAACATCAGTGAGAACCGG | Red |
| PTDO-16 | CCAGACCCTCATAGTACCGGATTTTTTTCACG | Red |
| PTDO-17 | GTTGTACTTAGCAAGTAGTAGTCATTTGTTAGATAACAGTTG | Red |
| PTDO-18 | TTTTTTGCGGTATGAGCCAGTCGGGAAACCTGTCGTTTTTT | Red |
| PTDO-19 | TTTTTGCGCGCCTGTGCGGGGTTTCTGCCAGCACGCTTTTT | Red |
| PTDO-20 | GGTTTACATATAAAAAAATACTAACGGAGAAGGAAATAGCAA | Red |
| PTDO-21 | CAGCAGACGATATTAAAGGGGTTGACTCCTCACAAAAGGTTCGAGG | Red |
| PTDO-22 | CTAATCAGCGGGTCGTCTAGTTAGCGTAACGATCTTTTT | Red |
| PTDO-23 | CCACATAACGCCAAGACGACGGCTGAATCCTTTTGTAT | Red |
| PTDO-24 | TGACGAACCAGAGCCGTGCCCGAACAGGTTTGAAATAACCGG | Red |
| PTDO-25 | TTTACCATTCAACCCAATAAATATTCATAAATTGGAAACGTAACAAAGC | Red |
| PTDO-26 | TTTTTTTGCTTCTGAAGAGTCAATTTTT | Red |
| PTDO-27 | AGGCGCTCCTGCGGTTGCGCTAAAGCCTGCGGTCCCGGCAAA | Red |
| PTDO-28 | TTTTTGCCCCCTTATTAGCGTTTGGAACCGCCCA | Red |
| PTDO-29 | GATCAAATGGTGAACCATGTTGGGAACGGTAAAACGGCGATT | Red |
| PTDO-30 | CCACTACCAGTCAAACGCTTAGTGACCCCTGCCGGCCAG | Red |
| PTDO-31 | TGAGCGATTCGCGTTAAATCAATTGTAACAGAAAAAATCGTA | Red |
| PTDO-32 | TTCAACTTCTACGTTAATATTATCAAGTATCACCAAATTCGA | Red |
| PTDO-33 | CAAAACAATAATTAAGAATCATAGGTATAAAAAC | Red |
| PTDO-34 | TTTTTCTTTACAAAGCGGAACAATTTTT | Red |
| PTDO-35 | TTTTTACAAAAGGTAAAGTAATTCTGAACAATGTAGAACACCCTGATA | Red |
| PTDO-36 | CGATCGAACTGGGCTGACTCGCTGAGCCCACG | Red |
| PTDO-37 | CTTAAACCAAAAAAATAATTTGCTAAACAAATGAACATTCCA | Red |
| PTDO-38 | AAACTAGCTGAGAGTCTACAATTGACCAGGG | Red |
| PTDO-39 | TTTTTGTTTTTTGGGGTAACGTGGCGAGATTTTT | Red |
| PTDO-40 | GATTGCCTCAGCGTGGTGCTACCAGCAGACAGACAAACGAAT | Red |
| PTDO-41 | AACCAGCGCCGGAGGGAAGGTAAACCATTTGCATTACCTCA | Red |
| PTDO-42 | CTTTCAAATATTAATGGTTTGAAAAATTACTACAAATTTAT | Red |
| PTDO-43 | TTTTTTAATGGGATCTCTCACGGAAATTTTT | Red |
| PTDO-44 | AATTAATCCTGTCAGATGATGGCAATAAAGTTAAACGTTGAA | Red |
| PTDO-45 | TTAAGCTCATTACGACAGTAGATGGTTCTCCGCTTAAAT | Red |
| PTDO-46 | TTTTTCCTTATTACAATAAGTTTTTTTT | Red |
| PTDO-47 | TTTTTAAGGAAGGGAAGGCTTAATGCGCCTTTTT | Red |
| PTDO-48 | GCCAGGCAAAAAGGTTTCTTTTAGAACGGTTCCGGCCCGGAA | Red |
| PTDO-49 | TTTTTAGAAACCACCAGAAGGTCATTTTCAA | Red |
| PTDO-50 | CCAGTCAACCTGAATGAATGGTACCGAAGGCGGTCAAATCTA | Red |
| PTDO-51 | ATATTCAATAGGCTACCAACTAGCCGGAGCCTGATATTATAC | Red |
| PTDO-52 | TTTTTCACGCAACCAGCTTACGGCGCCGGGCGCGGTTTTTT | Red |
| PTDO-53 | ATTCCCACAACTAATGTAGCTACCCTCGTTTTGCCATACCAGATG | Red |
| PTDO-54 | TTTTTACCCTCAGAGATATTCACTTTTT | Red |
| PTDO-55 | TTTTTATTAGAGAGAGCTGAAAAGGTTTTTT | Red |
| PTDO-56 | CCATATACATTTCGCTATTTTGATATTCTAT | Red |
| PTDO-57 | TAATTTGAGTCCTTCTGCACGACCGAAATACCTACATTTTGCTGGAGT | Red |
| PTDO-58 | TCGCCATTTGAATTACCTTTTTACATTTTACAAAAATTGGTC | Red |
| PTDO-59 | TTTTTCCTACCATATCAAACAGTACCTTTTTTTT | Red |
| PTDO-60 | TTTTTAACAAGAGTCCACTATTAAAGAACGTTAG | Red |
| PTDO-61 | TCACATTAGAATCAGAGCGGGTAACGTGGCGGTCA | Red |
| PTDO-62 | CGCGTACCTTTACTCCAAAGAC | Red |
| PTDO-63 | TTCGTACAGCGGGGATAGAGGTCACAACGGCGTTAAATG | Red |
| PTDO-64 | CACCCTCGGAAGTTTCCATTGGCCACCCAGTACCACTTTAAT | Red |
| PTDO-65 | TTTTTCTAAGAACGCGAGGCGAGGCTTAGCA | Red |
| PTDO-66 | GCCGGAGCAGAACGCGCCTGTCAACATGTAGGCAGCCGCCAT | Red |
| PTDO-67 | AATTAAGCGTACTTGAGTATTGACTTCATAT | Red |
| PTDO-68 | TCAGTGCGTTCCAGGGAAAGCGGTTGAGAGAGCCGAGCCGCC | Red |
| PTDO-69 | GGTTCCGAAATACGCTGGTTGCCCTGCCAGGG | Red |
| PTDO-70 | GTTGGCAAAATGAAAGTATTATTTATAAATCACGCAAGGAGCGGGGAAA | Red |
| PTDO-71 | TTTTTTTGCAACAGGAAAAACATATTACAGT | Red |
| PTDO-72 | TTAAGCGGAAAATAATGGAAGGGTTAGAATTTTT | Red |
| PTDO-73 | CCTCTGGCCTTGCCGCCACACCCTCAGAGCCACCTTTTT | Red |
| PTDO-74 | TCACGGTTCCTCACTTTCCTGAAAGTGTCACTGCCCGCGCGG | Red |
| PTDO-75 | TTTTTATCGGCGAAACGTCTCGTCGCTGGCAGCCTCTTTTT | Red |
| PTDO-76 | CCTAACAGTGCCCGTATGCTTGAGATGGTGAGAG | Red |
| PTDO-77 | TTTGTGTAATCAGAGGTGGAGGCCGGAAGGGGACGTTTTAAC | Red |
| PTDO-78 | TTTTTAGAGCATAAAGCGCCTTTATTTTTTT | Red |
| PTDO-79 | TTTTTTCCAGCCAGTCACCGGAAACATTTTT | Red |
| PTDO-80 | TCAATCCTGGAGGTCGGCAGCCGACGTTGATAACCCTTTCCGCTC | Red |
| PTDO-81 | AGAATCTGTCCTCAGTGACAGGAGGCCGATTAGGTTGCTACCACAC | Red |
| PTDO-82 | TTTTTGGCCTCTTCGTAACGCCAGGGTTTTT | Red |
| PTDO-83 | TTTTTCAGTAGGGCTTATATAAAGTACCGTTTTT | Red |
| PTDO-84 | TTTTTGCCAGCTGCATTAATGAATTCTTTTCACCAGTTTTT | Red |
| PTDO-85 | TTTTTGACAGTCAATATTTTGTTAAATTTTT | Red |
| PTDO-86 | AAAAGCCCAATGCGGGAGTTTTGCACCCAGCTATT | Red |
| PTDO-87 | TTTTTTTTTCCCAGTCAACCGTCGGTGGTGCCATCCTTTTT | Red |
| PTDO-88 | TTTTTTGAAACCATCGATTTTCGGTCATATTTTT | Red |
| PTDO-89 | ATTATTCAAGATTACAATACAGGCAAGGTTCAAATGACCTGACTATTAT | Red |
| PTDO-90 | TTTTTTGAGACGGGCAACAGCTGATTTGCCCGAAGCATTGTGAAAAACG | Red |
| PTDO-91 | TTTTTGCTAACGAGCGTTTTTGTTTAACGTTTTT | Red |
| PTDO-92 | TTCTGCATCAATATGCGAGAGATAGACTGCGCATCTCCGGATTCTCCGT | Red |
| PTDO-93 | GAGTAATCTAGCTGATAAATCAACATGTATCATATACCTGCC | Red |
| PTDO-94 | AAGTTGGGCTATTAAAGGGCGTGATAATACG | Red |
| PTDO-95 | AATTAATTTAATGGTTAGAATAGGC | Red |
| PTDO-96 | CAAACGCTGAGTAAATCGATGTGAGTGAATAACCTTTTT | Red |
| PTDO-97 | AACTTTTAGGTTGGAGACTACTAGCGATTTTTCCCAAACAGT | Red |
| PTDO-98 | CCAGGATTGAGAAAGACAAAAGGGCATCGCGTTTTAAGGAAA | Red |
| PTDO-99 | AAGCATCTCAATATCTAATAGTCGC | Red |
| PTDO-100 | TTTTTAAGAAGATGATGAAACAAATCAATATTCGCTATAAGA | Red |
| PTDO-101 | CCCGTATCTTCTGATAGGAACGCCATCATCATATAAAGACCCTGTAATA | Red |
| PTDO-102 | ACCCTTGAGTGCCTATTTCGGAACGATTAGGAAGTGCCGGA | Red |
| PTDO-103 | TTTTTGCTACAGGGCGCGTACTATAAGGGATAAGTGTTACAC | Red |
| PTDO-104 | AATACCGTTGTTCCTGAGTTTAGACAGGAACGGTTTTTT | Red |
| PTDO-105 | TAAAACTACGTGAACCATCACCCAAATCAATTTTT | Red |
| PTDO-106 | TTTTTCCACACAACATACGAGCCGCAGCAGGCGAAATTTTT | Red |
| PTDO-107 | CCGACTTAGCAAGCCGTAGGATTCCAAGAATAATCCAATAGATAA | Red |
| PTDO-108 | TTAAATCACCAGGTGAGAACTAATAAATTAAGTTAAGAGAGC | Red |
| PTDO-109 | TTTTTAAGTAAGCAATTAAGACTTTTTT | Red |
| PTDO-110 | ACTAATGCAGATACGGAACAAGTCGAAATCCGCAGTTCGTCACCTGACC | Red |
| PTDO-111 | AAAGCCATAAATCAAAAACCCTCAAATAGCGTAAAGAAGTTTA | Red |
| PTDO-112 | ACATAAACATCAAGATTATTCAGGAGCATAATACATTAAAAACTA | Red |
| PTDO-113 | TTTTTTGGCCAACAGAGTGCGCGAACTGATTTTT | Red |
| PTDO-114 | CGTCCCGGGTTATAGCTGAGTTGAGCCCGAGAGGACTCCAACGTCA | Red |
| PTDO-115 | TTTTTCGGCCAGAGCACATCCTCACACTGGTGTGTTTTTTT | Red |
| PTDO-116 | TTTTTATGTACCCCGGTATCGGTGCGTTTTT | Red |
| PTDO-117 | TTTTTTGAATTATCACCAAACGTCACCAATTTTT | Red |
| PTDO-118 | GAACAGAGCCATATCACCAACAGTTAGGAATT | Red |
| PTDO-119 | TGGCGAAAGGGGATTCAGGCTAAACAGGAAGATTTTTCAGCTAATCGAT | Red |
| PTDO-120 | TTTTTATTCGCATTAAAAGATCGCACTTTTT | Red |
| PTDO-121 | TCGTCAATTATAGCGAGAGGCAGAAAAATTAATCAAGGACAG | Red |
| PTDO-122 | TTAATTCGAGCACGACGACAATAAATTATCAAGGCTGTCTAGACGGATA | Red |
| PTDO-123 | GGAGAGGCGGTTTTAATTGCGCTGGTAATGGGTGATGAACCTCAATTGC | Red |
| PTDO-124 | TTTTTTCAAAAATGAAAAACAAAGTCAGATTTTT | Red |
| PTDO-125 | TTTTTAAACGAGAATGACGGATTGCATTTTT | Red |
| PTDO-126 | CCCAGCGAAATTGTCATTATTGTTGGGATTT | Red |
| PTDO-127 | TTTTTATCCTGTTTGATGGTGGTTGAGTGTTGTGAGCCCATACCGACTC | Red |
| PTDO-128 | ATTTAGAATCGGAAATCGGCCTTGACGCAGATTCA | Red |
| PTDO-129 | CAAACAAGAGAATGAGGGTAGCAAATGGTCAATGATAAGCCCAATCATT | Red |
| PTDO-130 | AAGGGCGGGTCAAAAGAATAGGATCCCCTGGGCGGGCCGTTT | Red |
| PTDO-131 | AACAAACGTCACGCCTGAGAGGCGAAAAACAA | Red |
| PTDO-132 | GCTTCAAGTCATTTTTGCGACAAAAACACATATTACCTGCCT | Red |
| PTDO-133 | ACCCTCACCATCTTCGTTTTCGTTACCAATACCCACCTTTACAAT | Red |
| PTDO-134 | GGTTGAACCGCCCATGTACCGTAACCCTGTAGTTTTCTG | Red |
| PTDO-135 | TTTTTCTTTCATCAACAGATTGACCGTTTTT | Red |
| PTDO-136 | ATTTCATTAACTAAAGATAGAACCCAAAATAAGTAACAA | Red |
| PTDO-137 | AGCCCAAGTACACGAGCAGAAAAATAATATCCCATTTTT | Red |
| PTDO-138 | TTTTTAATCCAATCGCAAGACTGTAAATTAT | Red |
| PTDO-139 | TTTTTATTTTGTCACAATCAAACCACGGGCA | Red |
| PTDO-140 | CAGACAGTAGGAACCACCCTCCACCACCGCAGGTCACCGTAAATT | Red |
| PTDO-141 | GCCGGCGCGAGGTGCAGAACAGCT | Red |
| PTDO-142 | AGTCAGAGAAGCCCGAAGCAAAATTGCTATAATGCAGTACGG | Red |
| PTDO-143 | ATTCAACGCCCTTATCAATCATTATTTATCCAGTTACATGGC | Red |
| PTDO-144 | CTTTTGCAATTTTTTCAAAAGTCAATATTGAGAGATCTGGAG | Red |
| PTDO-145 | TAGCTATTCAGAGAAACTGAAACCAATCAACGGGTATCGCCACTT | Red |
| PTDO-146 | GTAGTGTAGCGTTCATAATCAAAATCCCTCAGCCG | Red |
| PTDO-147 | AACCATCGGCTTGCGCATCGGGGACTAATACGAAGACACTAA | Red |
| PTDO-148 | TTAGTTGCCTGTAATATCCCG | Red |
| PTDO-149 | TTCGATTTAGATCTAAAAAGTTGAAAGGAATTGATTTTT | Red |
| PTDO-150 | GTATTGGCATGGATAGCCAAGCCCTTTTTAAGAATTTTT | Red |
| PTDO-151 | TTTTTTAGCCCTAAAACCTGCAACAGTGCTTTTT | Red |
| PTDO-152 | ACCAGTCTGAGGTTATATTTGCACGACTTCTGTTATCAT | Red |
| PTDO-153 | CTTAGAGATAAGAGCAAGAAAATTGAGTAGCGCATTTTTAAT | Red |
| PTDO-154 | TTTTTTAATCTTGACAAATAAGGCTTTTTTT | Red |
| PTDO-155 | TACGCATTGGCTCAATCGTCTGAAAGGGTACCGAGCTCTCCC | Red |
| PTDO-156 | GCAGATCCAGCGCACAGTGGGCGGAACAGCG | Red |
| PTDO-157 | AACTAACCACCGGAACCGCCTCACCGGACTTTAGCAAGATGG | Red |
| PTDO-158 | ATCAGCGGGGTCAACAGCAACAGTGCCAAGCTTGTAACAATTTCGATGT | Red |
| PTDO-159 | TTTTTGCCCTGACGAAACAGTTCAGATTTTT | Red |
| PTDO-160 | TACAACGACTGAGTAACCGCCCAAAACG | Red |
| PTDO-161 | ACATGCAATAAATACATAGTAGCACGGAATTAATTT | Red |
| PTDO-162 | TTTTTAACACTATCATACAACATGTTTTTTT | Red |
| PTDO-163 | TTTTTTTAAATATGATTCTGCGAACGTTTTT | Red |
| PTDO-164 | TAAAAAAAACCGTCTACGATGGCCCGCACTAAGCTTGACGGG | Red |
| PTDO-165 | TTTTTTCTTTGATTCGCCAGCCATTTTT | Red |
| PTDO-166 | AGTTTGAACCAGGCAAAGCTTGAATACCAGATTTTCGATGCT | Red |
| PTDO-167 | GCATTGCTTTGAAAACACTTTTTAGCGTTATAGA | Red |
| PTDO-168 | TTTTTGGCATCAATTCTAAGGCCGGATTTTT | Red |
| PTDO-169 | TGAAGAGTAGTTGAATCCTCAGGTC | Red |
| PTDO-170 | AGCGAAATCTCAGCTTGA | Red |
| PTDO-171 | TGGTCAGTATCAAACCCTCAAACCTTGCTAAAACAGACTAAC | Red |
| PTDO-172 | TGTGGTGCTGATCAGACTCAGATGATAAACACTGTTGCTTCGCAC | Red |
| PTDO-173 | TTTTTAAGAGGCTGAGATATAAGTATAGCTTTTT | Red |
| PTDO-174 | GCTGCAAGACGGCCCGCAAGAGGACTTGGCT | Red |
| PTDO-175 | TTTTTTCAACGCAATCCTGTAGCCAGTTTTT | Red |
| PTDO-176 | TTTTTTCAGGGATAGCAAGCCCAACCCTCATTTCCAGA | Red |
| PTDO-177 | TTTTTGGAGTGTACTGGTAATGGCTTTTAAT | Red |
| PTDO-178 | TGTCTGGAAGTTTACTTAATTATAAAAACCAAAGCACCAGAG | Red |
| PTDO-179 | TGCTCATCCAGAACTTACCTTTCAGGACACAGGTAACATTCA | Red |
| PTDO-180 | TCCACAAAGACTAGAAAAGGAAATTATTCATTAAAGGTTTTT | Red |
| PTDO-181 | CATGAGTAATAAAAGGGACATTCTTTTT | Red |
| PTDO-182 | TTTTTAAAATCCCGTAAAAAAAGCCGCAGTGTCACTTTTTT | Red |
| PTDO-183 | AGGATTTTTGTCTGGCCTGGATAAAGGGAGAATAAATCG | Red |
| PTDO-184 | AGTTGCGCGCTTTTGCGGGGAGGGTTTTGCCC | Red |
| PTDO-185 | TTTTTGGGTAATTGAGCGCTAATACTTACCGGAACAAAATCGGCATAG | Red |
| PTDO-186 | TTTTTAAAGAGGCAGAAACAAAGTACTTTTT | Red |
| PTDO-187 | TTTTTGTGCCTGTTCTTCGCGTCCGTTCCAGTTTGGTTTTT | Red |
| PTDO-188 | GGGAACAGTTGGTGTATCGGCGCACCGCTGTTGGGCGCCAGC | Red |
| PTDO-189 | TTTTTTAAATAAGAATAGCCAACGCTCAATTTTT | Red |
| PTDO-190 | TCGTAATCTAATGAGTGAGGTTGGCAAGGCCCCCG | Red |
| PTDO-191 | AAATCGCGGAACCTGTTTAGCATAAGAGAGCGAACTGCGGAA | Red |
| PTDO-192 | TTTTTCAGCAAATCGTTTTGTTATCCGCTCACAATTTTTTT | Red |
| PTDO-193 | TTTTTGCTACAGAGGCAGACGGTCAATTTTT | Red |
| PTDO-194 | TTTTTACGCCAGAAAGCAATACTTTTTT | Red |
| PTDO-195 | GAAATTGAGAATTAAACGTTTTTAAACGATTCTT | Red |
| PTDO-196 | TTTTTCGAGAACAATCCGGTATTTTTTT | Red |
| PTDO-197 | AAGATTTTCATAAATCAGGAATCATTACCGACGCGAGAA | Red |
| PTDO-198 | CCGCCGCAAAGCGAAAATTAAACATCACTCTTTAAATAGAACAACA | Red |
| PTDO-199 | TTATAACTATAAAAGAACCGTGTGATAAATAAGGCGTTTTTT | Red |
| PTDO-200 | TTTTTAAGAGACGCAGATTGTGTACATCGACATAAATTTTT | Red |
| PTDO-201 | TAAAGTTGCTAGTTTTGAAGCCTTATTTTAAATGCAAGCGCCTGTTCCTA | Red |
| PTDO-202 | TTTTTTCCTAATTTCGCACTCATTTTTT | Red |
| PTDO-203 | TGGTTTTCGGCCAACGCTTTCCGGGTCATCCCTTATAACGGATTT | Red |
| PTDO-204 | TTTTTGGAAGGTTAAGTATTAGATTTTT | Red |
| PTDO-205 | TTTTTTACATCGGGAGATTACCTGAGCAATTTTT | Red |
| PTDO-206 | AATCAACAGTTAATGCCGGTGTACAGACTAAAGGCCCGACAA | Red |
| PTDO-207 | CCAAAGGAATTACGAGGGAATACCGAAAGATTATCATCACG | Red |
| PTDO-208 | CATAAGTTTTTCTGAAACATGAAAGTATTTTTTT | Red |
| PTDO-209 | GTCCTGTCCAGCAGTAATTAATTAAAGCTTAGACGGATTGATGAATGTAT | Red |
| PTDO-210 | TTTTTGGTCATTGCCATGTCAATCATTTTTT | Red |
| PTDO-211 | TTTTTTAGTGAATTGCTGATGCATTTTT | Red |
| PTDO-212 | AATTATTTTAGTTAATTGTAACCGTGCATCTAGA | Red |
| PTDO-213 | GCAGAAAGAACGTTAGCAAGGCCGGGTCACCGCATA | Red |
| PTDO-214 | CGCTCGTAACCTTGACGAGCACGTAAGCTAAAGGCCACCGAGG | Red |
| PTDO-215 | CAAGCGCAAAGAATGCACCAA | Red |
| PTDO-216 | TTTTTTGGCTCATTAGAGGGGGTAATTTTTT | Red |
| PTDO-217 | CGCTGCGAGGGCGCAAAAGAGGAACTCAATATTTTAGCGTAAATTA | Red |
| PTDO-218 | TGAGATACTTTCCTCGTGCGTATTGGGCTCACCGCCTGGCCC | Red |
| PTDO-219 | TTTTTTCATAAGGGAACTTTAAGAACTTTTT | Red |
| PTDO-220 | ATTCTGATTAATTGTTTGGATTATTAA | Red |
| PTDO-221 | TTTTTAGTAAAATGTTTCAGGTCAGGTTTTT | Red |
| PTDO-222 | TTTTTGAGATTTAGCATAGTAAGAGCTTTTT | Red |
| PTDO-223 | GAACGGTGCCCCAAGCGCAACTTCTGGTCCG | Red |
| PTDO-224 | AGGCGCTTTGAAACGAGG | Red |
| PTDO-225 | ATCCCTTATAAAGATGGAGTTGCAGCAAGGGGTGCCATGGTCACCTGCA | Red |
| PTDO-226 | ATGAATCTCGACCTGCTCCATTTCATGAAGCAGCG | Red |
| PTDO-227 | TTTTTAACGGAGATTTGTCATCAGTTTTTTT | Red |
| PTDO-228 | ACCGATATAGATTTTAGCTATCCTGAATCTTACCAACTTTTT | Red |
| PTDO-229 | TTTTTAGTAGATTTAGTAGGCTATCATTTTT | Red |
| PTDO-230 | AACACTCATCTTAGATACGTA | Red |